\documentclass[aps,pre,twocolumn]{revtex4}
\usepackage[utf8]{inputenc}
\usepackage[T1]{fontenc}
\usepackage{lmodern}
\usepackage{mathptmx}
\usepackage{amsmath,amssymb}
\usepackage{xcolor}
\usepackage{graphicx}
\usepackage{physics}
\usepackage{float}
\usepackage[colorlinks,linkcolor=blue,urlcolor=blue,citecolor=blue]{hyperref}
\usepackage{dcolumn}% Align table columns on decimal point
\usepackage{bm}% bold math
\usepackage{epsfig}
\usepackage[exponent-product=\cdot]{siunitx}

\usepackage{color}
\usepackage{soul}

\newcommand{\e}{\mathrm{e}}
\DeclareMathOperator{\erfc}{erfc}

\begin{document}

%\preprint{}

\title{ 
Optimal mean first-passage time of a Brownian searcher with resetting 
in one and two dimensions: Experiments, theory and numerical tests}

\author{ F. Faisant, B. Besga, A. Petrosyan, 
S. Ciliberto}\email[E-mail me at: ]{sergio.ciliberto@ens-lyon.fr}

\affiliation{Univ Lyon, ENS de Lyon, Univ Claude Bernard, CNRS,
Laboratoire de Physique, UMR 5672, F-69342 Lyon, France}

\author{Satya N. Majumdar}

\affiliation{LPTMS, CNRS, Univ. Paris-Sud, Université Paris-Saclay, UMR 8626, 91405 Orsay, France}

\date{\today}

\begin{abstract}
We study experimentally, numerically and  theoretically the optimal mean time needed by a Brownian particle, freely diffusing either in one or two dimensions, to reach, within a tolerance radius $R_{\text tol}$, a target at a distance $L$ from an initial position in the presence of resetting. The reset position is Gaussian distributed with width $\sigma$. We derived and tested two resetting protocols, one with a periodic and one with random (Poissonian) resetting times. We computed and measured the full first-passage probability distribution that displays spectacular spikes immediately after each resetting time for close targets. We study the optimal mean first-passage time as a function of the resetting period/rate for different target distances (values of the ratios $b=L/\sigma$) and target size ($a=R_\text{tol}/L$). We find an interesting phase transition at a critical value of $b$, both in one and two dimensions. The details of the calculations as well as experimental setup and limitations are discussed.
\end{abstract}

\maketitle

\tableofcontents
\newpage
%\begin{figure}[t]
%\includegraphics[width=\linewidth]{text_\text{f}igure_1.pdf} % Here is how to import EPS art
%\caption{\label{fig:brownian} A typical trajectory of a Brownian motion $x(t)$ during the time interval $[0,T]$, starting from $x(0)=0$. The global maximum $x_{\max}$ occurs at time $t_{\max}$ and the global minimum $-x_{\min}$ at $t_{\min}$. The final position $x(T)$, measured with respect to $-x_{\min} \leq 0$ is denoted by $x_{\rm f}$. The total time interval $[0,T]$ is divided into three segments: $[0,t_{\max}]$ (I), $[t_{\max}, t_{\min}]$ (II) and $[t_{\min}, T]$ (III), for the case $t_{\min}>t_{\max}$.}
%\end{figure}

\section{Introduction}

The time needed by a freely diffusing Brownian particle to reach a fixed target in space
is a widely studied problem not only for its fundamental aspects but also for its
numerous applications from chemical reactions, astrophysics, all the way to animal foraging
in ecology. This time
is a random variable and its probability distribution is called the first-passage time
distribution~\cite{Redner,pers_review}. The first moment of this distribution, when it exists,
is called the mean first-passage time (MFPT) which is often used as a measure of
the efficiency of a search process. The smaller the MFPT, the more efficient is the
search process.
The MFPT can be optimized using various stochastic search algorithms (such as simulated annealing) which 
speed up the search process~\cite{VAVA91,LSZ93,TFP08,Lorenz18}. 
More recently it has been demonstrated that the MFPT for a single particle (searcher) 
to a fixed target can be minimized using  
the reset protocols (for a recent review see ~\cite{reset_review}). 
Searching a target via resetting is an example of the so called
intermittent search strategy~\cite{BLMV11} consisting of a mixture of
short-range localised moves (where the searcher actually performs the
search) with
intermittent long-range moves where the searcher moves to
a new location and starts a local search in the new place.
 
Consider, for example, a fixed target at some point in space
and a particle (searcher) starts its dynamics from a fixed initial position
in space. The dynamics of the particle is represented
by a stochastic process, e.g. it
may just be simple diffusion~\cite{EM2011_1,EM2011_2}.
Under resetting protocol, the motion of the searcher is interrupted
either randomly  at a constant 
rate $r$~\cite{EM2011_1,EM2011_2}
or periodically with a period $T$~\cite{PKE16,BBR16}, and then
the particle is sent back to its initial position (usually instantaneously) and
the search restarts again from the initial position
after each resetting event. {For a given fixed distance between the target
and the initial position of the searcher, if one plots the MFPT 
as a function of the resetting rate $r$ (or $T$ for periodical
resetting), the MFPT typically displays a unique minimum at some optimal value
$r^*$. This led to the paradigm that resetting typically makes
the search process more efficient if the resetting rate is
chosen to have its optimal value $r^*$~\cite{EM2011_1,EM2011_2}}.

While this `optimal resetting' paradigm has been tested and verified in
a large number of recent theoretical and numerical 
studies~\cite{EMM13,MV13,EM2014,KMSS14,RUK14,RRU15,CS15,KGN15,PKE16,NG16,Reuveni16,PR17,CS18,Belan18,EM18,MPCM19a,MM19,DLLJ19,PP19} 
(see also the review \cite{reset_review}), it has been been verified only in a 
few experiments limited to the one dimensional (1D) case \cite{Tal2020,us_1D_Phys_rech}. 
In Ref.~\cite{us_1D_Phys_rech} we have demonstrated the emergence
of new physical effects when the
resetting position is not fixed to be exactly the initial position, but
has a distribution of finite width $\sigma$ around an average initial position.
A finite value of $\sigma$ led to interesting metastability and phase transition
in the MFPT as a function of resetting rate/period. Very recently~\cite{FP_2021}, we have
shown that a finite $\sigma$ in the initial condition also affects, rather profoundly, 
the first-passage probability distribution of the searcher, 
even in the absence of resetting. 

The purpose of this article is to extend these results to the two dimensional (2D) case, showing 
both theoretically and experimentally that {the new physical
effects induced by a nonzero $\sigma$ discussed above are not restricted 
just to 1D, but are also observed in 2D.} Our article is organized as follows. In the next 
section we describe the experimental set-up and the difficulties that one encounters in 
performing such experiments. In section \ref{sec:1D} we recall the main 
experimental results in 1D, whereas the theoretical predictions for 1D are summarized in the 
Appendices \ref{apx:A}-\ref{apx:C}. Although the 1D results have been recently
published in a shortened form without details in Ref.~\cite{us_1D_Phys_rech}, we 
think that it is useful to recall them with details in this article, 
in order to have a full comparisons 
with the 2D results described in Section \ref{sec:2D}. {The details of the 
2D theoretical 
computations, which are rather nontrivial, are provided in the 
Appendices \ref{apx:D_2d} and \ref{apx:E_2d}. We finally 
conclude in Section \ref{sec:conclusions} and some further details
are relegated to the rest of the Appendices.}

\section{Experimental set-up.}\label{sec:exp_set}
 
\begin{figure}
	\includegraphics[width=0.45\textwidth]{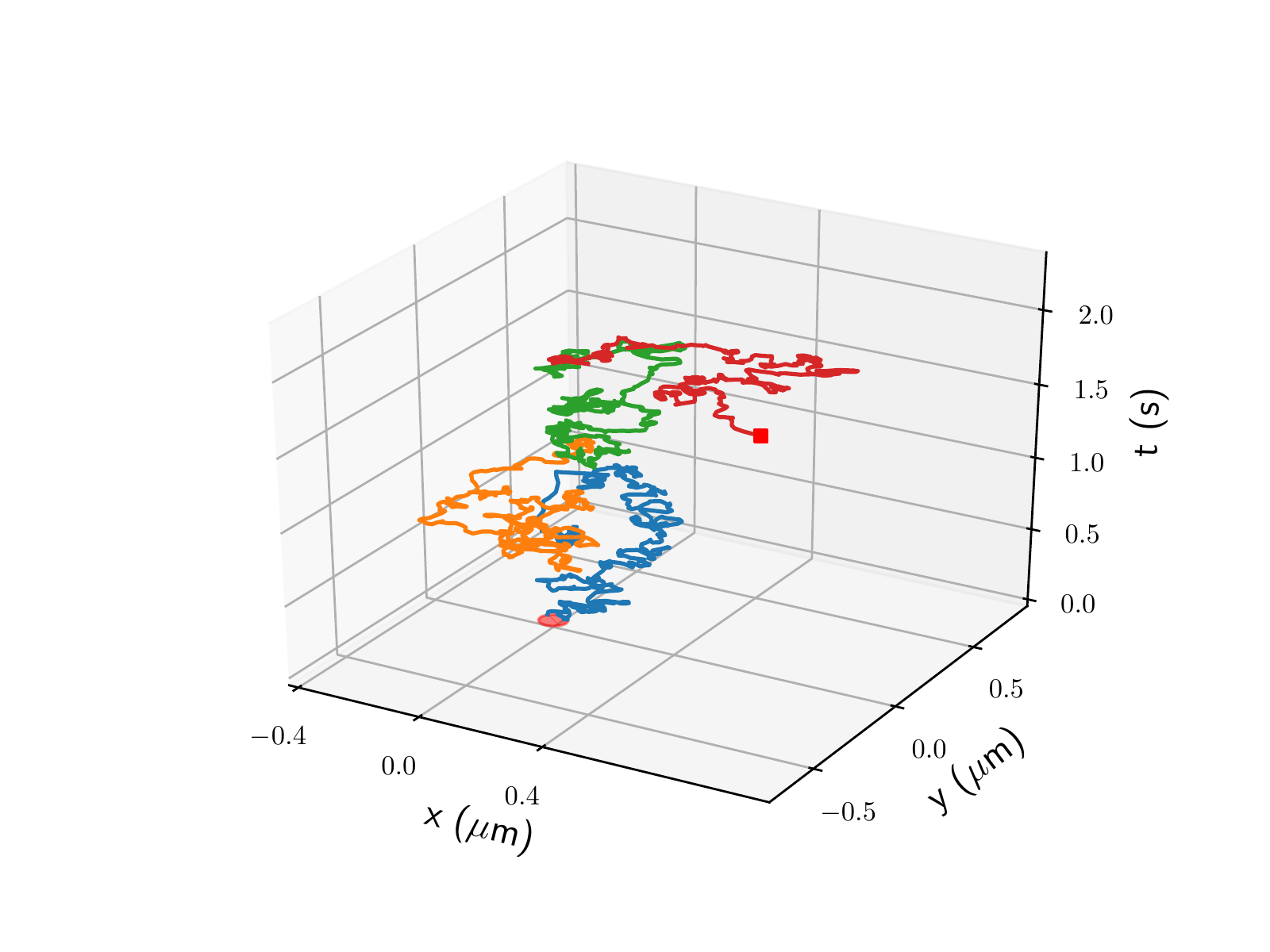} 
	\includegraphics[width=0.45\textwidth]{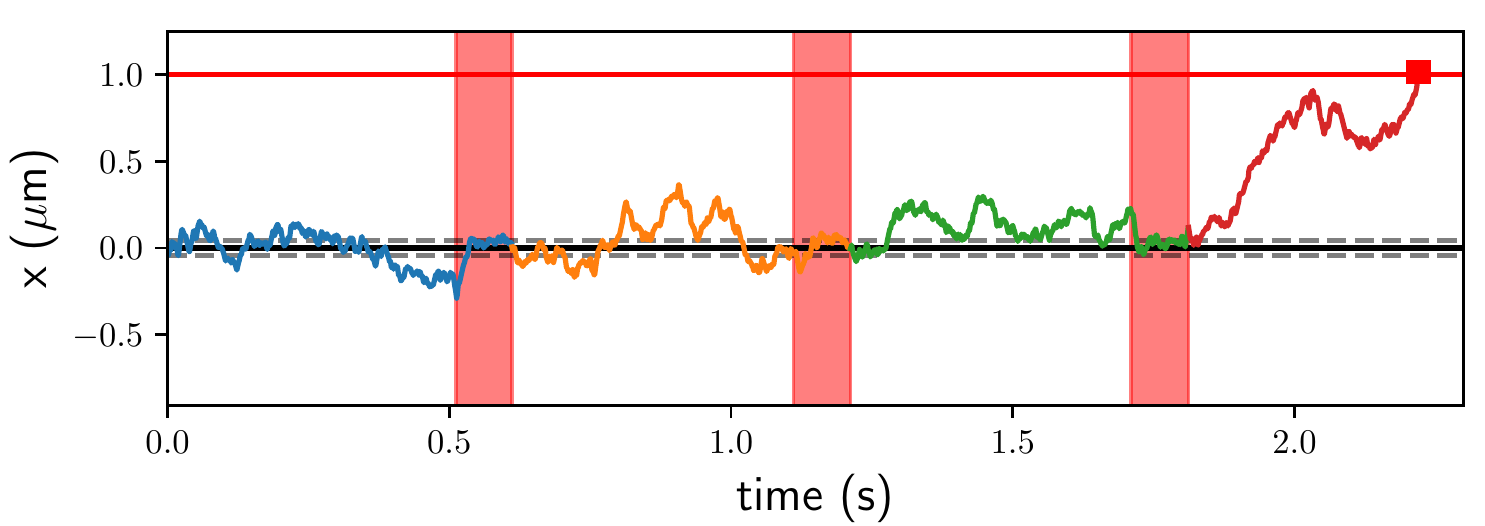} 
	\caption{\textit{Top panel :} Brownian trajectory in the 2D plane $(x,y)$ over time $t$ with resetting.
		\textit{Bottom panel :} Corresponding 1D Brownian trajectory with periodic resetting 
		after $T = \SI{0.5}{\s}$ (red colored area). The equilibrium standard deviation of the trapped particle $\sigma_x = \SI{43}{\nano\m}$ is shown (dotted lines). The target (red square $\SI{1}{\micro\m}$ away from the center of the trap) is reached when $x$ crosses the red line for the first time.}
	\label{fig_setup} 
\end{figure}

The optimal search protocol is implemented in an optical tweezer 
experiment~\cite{Berut2016}. The Brownian searcher is a silica 
micro-sphere of radius $R=\SI{1}{\micro\m}$ ($\pm 5\%$) immersed in pure water. The 
fluid chamber is made of two parts : The colloids are inserted in the thicker part 
of the chamber (reservoir) and the measurements are performed in a second part 
(thickness around $\SI{50}{\micro\m}$) with very few particles which allows us to 
take long duration measurements (tenths of hours). The optical trap is implemented 
with a near infrared laser with wavelength $\lambda = \SI{1064}{\nano\m}$ focused 
into the chamber through an oil immersion objective (Leica $63\times$ and $1.4$ 
NA). The particle is trapped in the transverse plane $(x,y)$ by a harmonic 
potential with stiffness $\kappa_{x,y}$. The stiffness is controlled by changing 
the optical power (typically hundreds of mW) in the chamber directly by laser 
current modulation (up to 10 kHz) or faster using an electro-optical modulator 
(EOM). The position of the particle in the $(x,y)$ plane can be tracked (see Fig. 
(\ref{fig_setup})) either by a white light imaging on a camera or from the 
deviation of a second laser on a quadrant photodiode (QPD). The camera allows us to 
extract the position of the bead at maximum speed around 1000 frames per second over 
a broad range ($> \SI{100}{\micro\m}$) and the center of the bead is extracted in 
real time with an accuracy of $\SI{2}{\nano\m}$. On the other hand the QPD allows 
for a fast position detection (around $\SI{1}{\mega\Hz}$ bandwidth) with a smaller 
range ($< \SI{1}{\micro\m}$) and a better accuracy (below $\SI{1}{\nano\m}$).

We implemented different search protocols (see below) involving a free diffusion 
phase where the particle actually searches for a target (with duration $T$) 
and a second phase resetting the particule to an initial distribution ${\cal 
P}(x_0,y_0)$. The resetting is implemented by turning on the optical trap during a 
time long enough for the particle to equilibrate in the trap. 
{During this equilibration period, we do not make any first-passage measurement.}
The first passage 
time is measured when the position crosses, {during the diffusive phase}, 
the target line in 1D for the first time 
(see Fig. (\ref{fig_setup}) bottom panel) or when the particle enters for the first 
time the tolerance radius around the target in 2D (see Fig. (\ref{fig:capture_2d})
in Section \ref{sec:2D}).

We point out here a few difficulties encountered in such experiments. First the 
free diffusion phase cannot be too long (typically $T < \SI{1}{s}$) to prevent the 
particle to diffuse too far from the center of the trap, as it needs to get 
trapped again in the next resetting phase. It is also important to make 
measurements close to the bottom surface of the chamber to avoid errors in position 
reading due to sedimentation during the free diffusion phase. Finally the 
acquisition rate has to be high enough to detect fast events (see discussion in
Section \ref{sec:1D}).

We will first discuss periodic and poissonian protocols in the 1D case in
the next Section, where we 
focus only on the $x$-component of the particle's position, i.e., one 
dimensional trajectories. Later, in Section \ref{sec:2D}, we will 
analyse the full 2D trajectories in detail.

\section{The optimal resetting protocols in one dimension}\label{sec:1D}

\subsection{ Periodic resetting protocol.} The experimental protocol leads us to study
a model of diffusion in (effectively) one dimension subjected to periodic resetting.
We consider  an overdamped particle in thermal equilibrium
inside a harmonic trap with potential $U(x)= \kappa\, x^2/2$, 
where the stiffness $\kappa$ is proportional to the trapping laser power.
The particle is allowed to equilibrate in the trap, and once it has
acheived thermal equilibrium, the time is switched on.
This means that the initial position
$x_0$ is distributed via the Gibbs-Bolzmann distribution which is
simply Gaussian in a harmonic trap: 
${\cal P}(x_0)= \e^{-x_0^2/2\sigma^2}/\sqrt{2\pi \sigma^2}$
with width $\sigma= \sqrt{k_B {\cal T}/\kappa}$, where ${\cal T}$ is the temperature. 
We also set a a fixed target at location $L$. 

At time $t=0$, the trap
is switched off for an interval $T$ and the particle undergoes free diffusion
(overdamped) with diffusion constant $D=k_B {\cal T}/\Gamma$, where
$\Gamma$ is the friction coefficient. 
At the end of the 
period $T$, the particle's 
position is reset (see Fig. (\ref{fig_setup})). Performing the resetting of the
position poses the real experimental challenge. 

In standard models of resetting one usually assumes 
instantaneous resetting~\cite{reset_review}, which is however
impossible to achieve experimentally. In our experiment, the
resetting is done by allowing the particle to relax to its
thermal equilibrium in the trap. This poses two problems.

\begin{itemize}

\item
First, there is always a finite relaxation
time needed for the particle to reset or equilibrate (it is never instantaneous
as in the theoretical models).
There have been recent theoretical studies of resetting
protocols with a finite `refraction'
period needed to reset the particle to its initial
postion~\cite{Reuveni16,EM2019,MPCM19a,PKR2019,GPP2019,BS2020,MBMS20,GPKP21}. 
In our experiment, the analogue of this refraction period is
the relaxation time to equilibrium once the trap is switched on
after the period $T$ of free diffusion. However, since
the theoretical calculations are much easier for
the instantaneous resetting case~\cite{reset_review}, it would
be easier to compare the experiments with theory if we can 
devise an experimental set-up that mimics
the `instantaneous' resetting. To devise such a set-up, we first note
that the relaxation inside the trap can, in principle, be made
arbitrarily fast using, e.g., the recently developed
"Engineered Swift Equilibration" (ESE) technique~\cite{ESE1,ESE2,RMP19,PGTP19,GPP2019}.
In our experiment we determine {the characteristic relaxation time inside the 
trap  $\tau_c=\Gamma/\kappa$}. During the period
$[T, T+\tau_{\rm eq}]$ with $\tau_{\rm eq} \simeq 3\, \tau_c$, we do not make
any measurement. In other words, 
even if the particle encounters the target during this relaxation period, 
we do not count this as a first-passage event. Thus in this  
set-up, the thermal relaxation mimics the instantaneous resetting.

\item Secondly, while the set-up described above achieves `instantaneous'
resetting, it still has one important ingredient that differs
from theoretical models of instantaneous
resetting. Using this relaxation technique we do not reset it to exactly 
the same initial position, rather the new `initial' position, at 
the end of the time epoch $T+\tau_{\rm eq}$, is drawn from the Gibbs-Boltzmann
distribution ${\cal P}(x_0)$. Thus it is important to modify the theoretical
results for instantaneous resetting by taking into acount a finite spread
$\sigma$ of the resetting position. One of our main results in this
paper is to provide new analytical calculations that take into acount 
the finite $\sigma$ effect on the optimization of the MFPT.

\end{itemize}

To complete our protocol using the two caveats mentioned above,
at time $T+\tau_{\rm eq}$, we again switch off the trap
and we let the particle diffuse freely for another period $T$, followed by 
the thermal relaxation over period $\tau_{\rm eq}$. The process repeats periodically.
During the free diffusion, if the particle finds the target at $L$, we
count it as a first-passage event and measure this first-passage time $t_\text{f}$. 
Note that the 
first-passage
time $t_\text{f}$ is the total
`diffusion' time spent by the particle before reaching the target 
(not counting the intermediate relaxation periods $\tau_{\rm eq}$).
Averaging over many realizations,
we then compute the MFPT $\langle t_\text{f}\rangle$, for fixed
target location $L$ and fixed resetting period $T$. 
As mentioned before,
we find that a finite $\sigma$ affects the results for MFPT as a function
of the period $T$ in a significant way, leading
to new physical phenomena such as metastability and dynamical phase transition which
do not exist when $\sigma=0$ strictly.

\vspace{0.3cm}

\noindent{{\bf Mean first-passage time.}} We start with the MFPT, the
central object of our interest.
The MFPT for this protocol can be computed analytically, as detailed in 
Appendix \ref{apx:B}. Our 
main result can be summarized in terms of two dimensionless quantities
\begin{equation} b= 
\frac{L}{\sigma}\, ; \quad {\rm and} \quad c= \frac{L}{\sqrt{4\,D\,T}}\, . \label{bc_def} 
\end{equation} 
The parameter $b$ quantifies how far the target is from the center of the trap in units of the trapped 
equilibrium standard deviation. The second parameter $c$ tells us how {frequently} we reset the particle 
compared to its free diffusion time. We show that the dimensionless MFPT $\tau$ can be expressed as a function 
of these two parameters $b$ and $c$ (see Eq. (\ref{mfpt_gauss.1}) in Appendix \ref{apx:B}) 
\begin{equation} 
\tau=\frac{4D\langle t_\text{f}\rangle}{L^2}= w(b,c)\, , 
\label{tau_bc} 
\end{equation} 
where 

\begin{equation} 
w(b,c)=\frac { \int_0^1 \ dv \ \int_{-\infty}^{\infty} \  du
\ \e^{(-u^2/2)}\  {\rm erf } 
\left( \frac{c}{\sqrt{v}}\ |1-u/b| \right)}{c^2 \  
\int_{-\infty}^{\infty} \ du 
	\ \e^{(-u^2 / 2 )} \erfc\left(c\ |1-u/b|\right) } 
\label{wbc_def} 
\end{equation} 

While it is hard to evaluate the integrals explicitly, 
$w(b,c)$ can be easily plotted numerically 
to study its dependence on $b$ and $c$. Furthermore,
one can also evaluate the function $w(b,c)$ explicitly in different limiting cases.  

\begin{figure}
\includegraphics[width=0.40\textwidth]{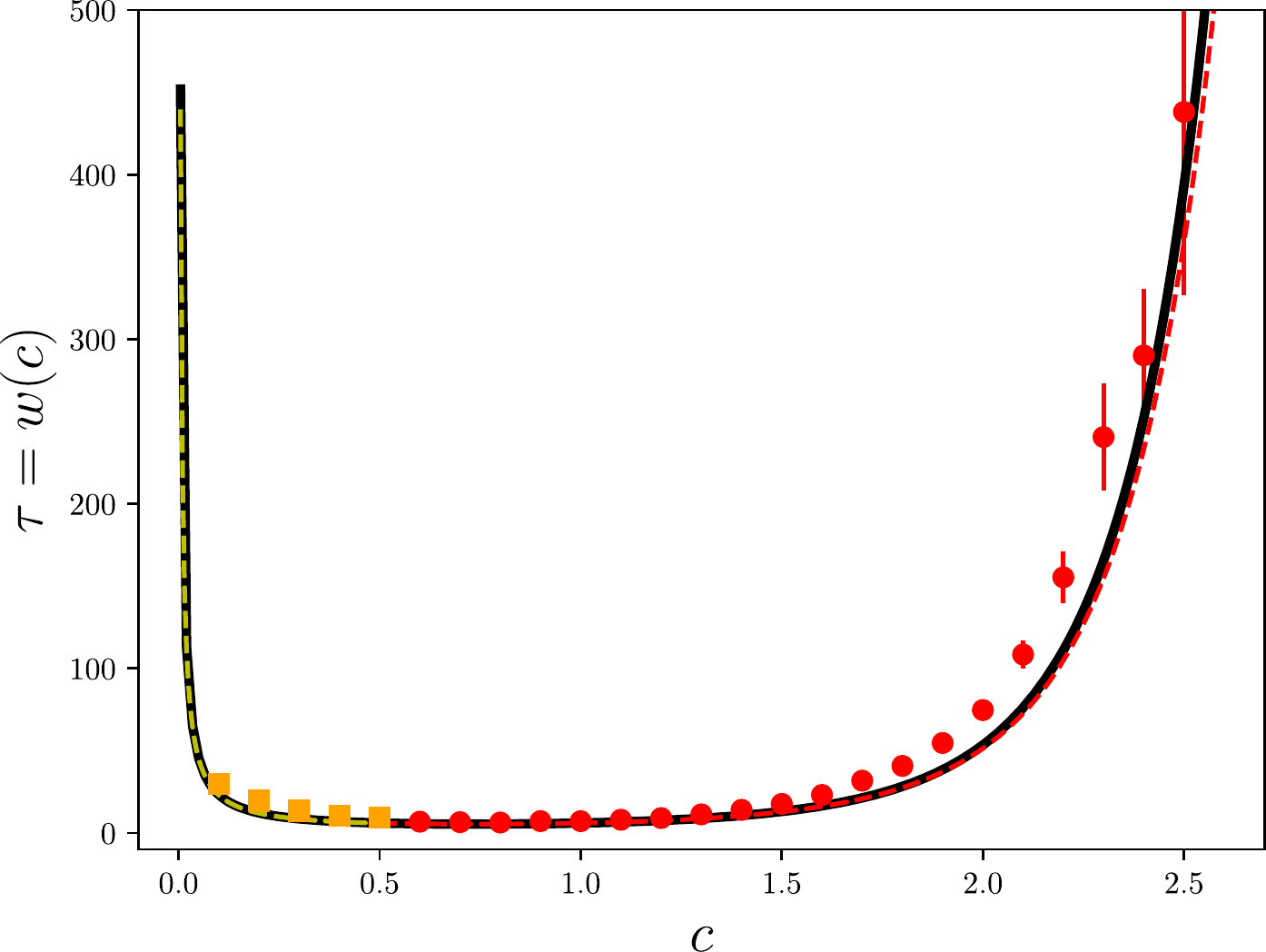} 
\caption{$w(c)\equiv w(b\to \infty,c)$ vs. $c$ curve (see also 
Fig.~(\ref{fig_wbc})). The solid black line represents the
theoretical formula for $w(c)$ in Eq. (\ref{delta_reset.1mt}), while 
the dots denote the experimental data with $b = 30$ (red dots) and $b=5$ (yellow 
square). The dotted lines shows
$w(b,c)$ with $b=5$ for small $c$ {($c \leq 0.6$)} in yellow, and with 
$b=30$ for higher $c$ in red.
The error bars are given by the standard deviation of the MFPT 
distribution divided by the square root of the number of events.}
\label{fig_wc.1} 
\end{figure}

To analyse the MFPT, we start with the limit 
$b\gg 1$ of Eq. (\ref{wbc_def}), i.e., $L\gg \sigma$. This 
limit corresponds to the case when the target is much farther away 
compared to the
typical fluctuation of the initial position. In this case, taking
$b\to \infty$ limit in Eq. (\ref{wbc_def}) we get
\begin{equation}
w(c)=w(b\to \infty,c)= \frac{{\rm
erf}(c)+2c\left(\e^{-c^2}/\sqrt{\pi}-c\, {\rm erfc}(c)\right)}{c^2\,
{\rm erfc}(c)}\, ,
\label{delta_reset.1mt}
\end{equation}
where ${\rm erf}(c)= (2/\sqrt{\pi})\, \int_0^c \e^{-u^2}\, du$ and
${\rm erfc}(c)=1- {\rm erf}(c)$. In Fig.~(\ref{fig_wc.1}), we plot
$w(c)$ vs $c$ and compare with our experimental data and find good 
agreement with no adjustable parameter.
Typically, to have a good estimate of the MFPT,  
we follow the particle for a few hours which allows us  
to detect between 1000 and 10000 first passage times, 
depending on the values of  $b$ and $c$. 
The standard deviation of first-passage times is of the same order as the MFPT. 
We see a distinct optimal value around $c^*=0.74$,
at which $w(c^*)=5.3$. Our results thus provide a clear 
experimental verification of the optimal resetting paradigm.
Let us remark that the authors
in Ref.~\cite{PKE16} studied
periodic resetting to the fixed initial position $x_0=0$
and obtained the MFPT by a different method than ours.
Our $b\to \infty$ limit result in Eq. (\ref{delta_reset.1mt}) indeed
coincides with that of ~\cite{PKE16}, since when $\sigma\ll L$, our protocol
mimics approximately a resetting to the origin.

\begin{figure}
\includegraphics[width=0.45\textwidth]{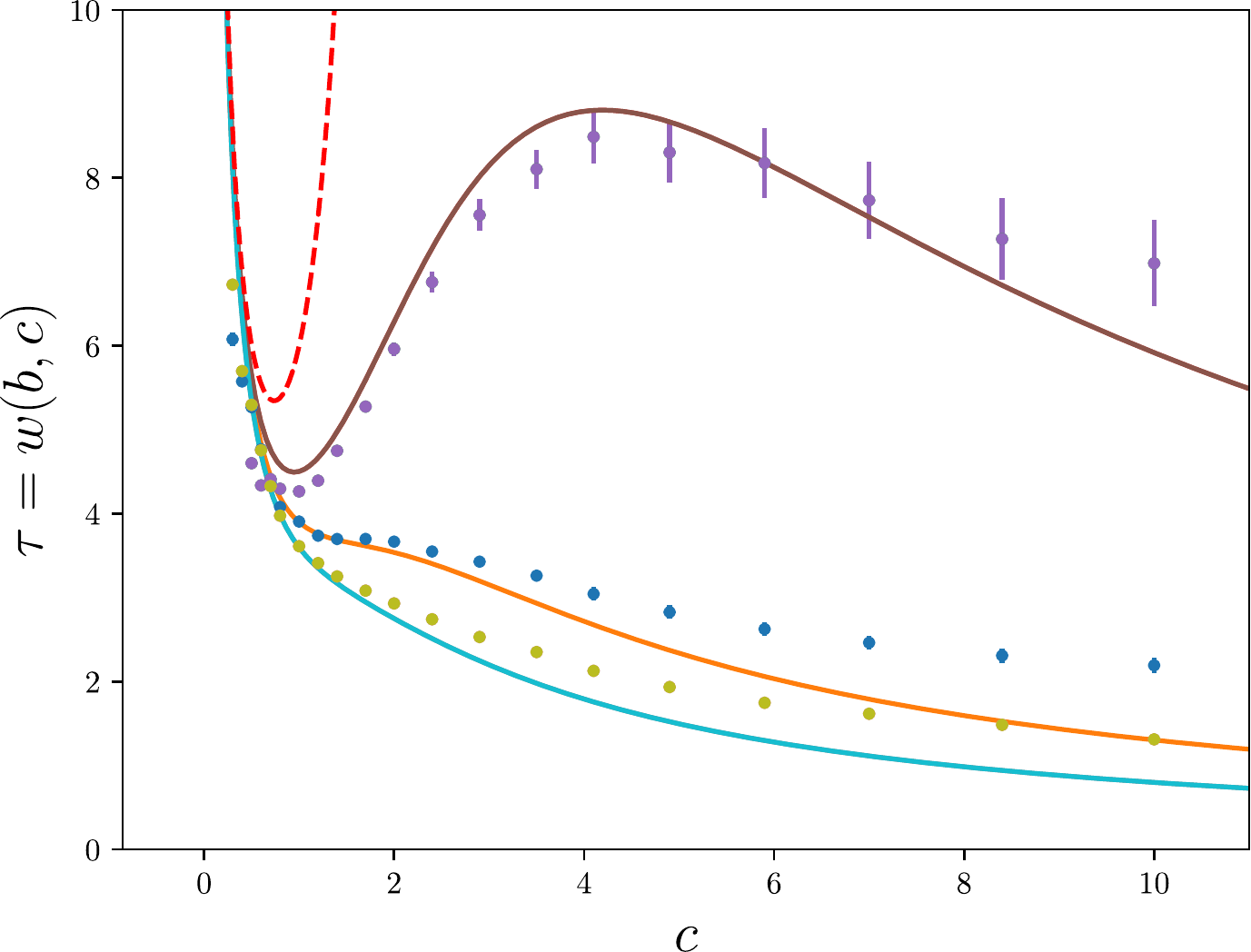} 
\caption{The scaled 
MFPT $\tau = w(b, c)$ vs. $c$ curves and experimental data for $b=3$ {(brown curve and 
purple dots), $b=2.3$ (orange curve and blue dots) and $b=2$ (blue curve and yellow 
dots) } , for the periodic resetting protocol. {The dotted curve recalls the $b\to 
\infty$ limit.} The theoretical curves are obtained from Eq. (\ref{wbc_def}).
The scaled MFPT has two different behavior depending on how far is the target. 
If $b > b_c\approx 
2.3$ the MFPT $\tau$ exhibits a local minimum, whereas it decreases 
monotonically for $b$ lower than the 
critical value $b_c$. }
\label{fig_wbc}
\end{figure}

What happens when $b=L/\sigma$ is finite? Remarkably, when we plot $w(b,c)$
in Eq. (\ref{wbc_def}) as a function of $c$ for different fixed values of $b$,
we see that $w(b,c)$ decreases as $c$ increases, indeed achieves a minimum 
at $c_1(b)$, then increases again, achieves a maximum at $c_2(b)$ and then decreases
monotonically as $c$ increases beyond $c_2(b)$ (see Fig. (\ref{fig_wbc})).
When $b\to \infty$, the maximum at $c_2(b)\to \infty$ and one has a
single minimum at $c=c_1(b)$ as in Fig.~(\ref{fig_wc.1}).
However, for finite $b$, the ``optimal'' (minimal) MFPT at $c=c_1(b)$ 
is thus actually a `metastable'
minimum and the true minimum occurs at $c\to \infty$, i.e., when the
resetting period $T\to 0$. Physically, the limit $T\to 0$, i.e., $c\to \infty$
corresponds to repeated (almost
continuous) resetting and since the target position and the initial location
are of the same order, the particle finds the target very quickly simply by resetting,
without the need to diffuse. This explains the second minimum at $c\to \infty$. 
Interestingly, this metastable minimum at $c=c_1(b)$
exists only for $b>b_c \approx 2.3$. When $b<b_c$, the curve $w(b,c)$
decreases monotonically with $c$ and there is only a single
minimum at
$c\to \infty$, or equivalently for $T\to 0$. Thus the system
undergoes a `dynamical' phase transition as the parameter $b=L/\sigma$ is tuned across
a critical value $b_c\approx 2.3$, from a
phase with a metastable minimum at a finite $c=c_1(b)$ to one where
the only minimum occurs at $c\to \infty$. 
This phase transition is well reproduced by the experimental data points. 
The deviation from theoretical prediction for high values of c is due to 
the limited experimental acquisition rate (here $50000$ Hz) that prevents us from 
detecting very fast events and thus leads to an 
overestimation of the MFPT as confirmed by our numerical simulations.

This phase transition was rather unexpected and came out as a surprise.
In a recent paper~\cite{FP_2021},
we have shown that a similar dynamical phase transition already occurs in
the first-passage probability density of a diffusing particle {\em without
resetting}, when averaged
over the initial condition drawn from a distribution ${\cal P}(x_0)$ 
with a finite width $\sigma$. Indeed, the transition that we observe
in the $w(b,c)$ curve in Fig.~(\ref{fig_wbc}) also has its origin in
the finiteness of $\sigma$. 
It can be traced back to the fact that there are two time scales in the
system when $\sigma$ is finite, namely $T_1\sim \mathcal{O}(L^2/D)$ and
$T_2\sim \mathcal{O}(\sigma^2/D)$. In terms of the parameter $c=L/\sqrt{4DT}$, they correspond
respectively to $c_1(b)\sim L/\sqrt{4D T_1}\sim \mathcal{O}(1) $ and
$c_2(b)\sim L/\sqrt{4D T_2}\sim L/\sigma \sim b$. When $b\gg 1$ is large,
$c_1(b)\ll c_2(b)$. However, as $b$ decreases, the two times scales approach
each other or equivalently, as a function of $c$, the minimum at 
$c_1(b)$ approaches the maximum at $c_2(b)$ in Fig. (\ref{fig_wbc}).
Finally, at a critical value $b_c$, the two time scales merge with each other
and for $b<b_c$, the function $w(b,c)$ decreases monotonically with
increasing $c$ with a single minimum at $c\to \infty$. The phase transition
is `dynamical' in the sense that it arises when two time scales merge with each other.

\vspace{0.4cm}

\noindent{{\bf The full distribution of the first-passage time.}} Going beyond the first moment $\langle t_\text{f}\rangle$ and computing the full probability 
density function (PDF) of the first-passage time $t_\text{f}$ is also of great 
interest~\cite{pers_current,Redner,Louven_review,pers_review}. Indeed, we computed the 
PDF $F(t)$ of $t_\text{f}$ (see Appendix \ref{apx:C}) and the result is plotted in 
Fig. (\ref{fig_stat}). For small $b$, we found striking spikes in $F(t)$ just after each 
resetting event. We show in Appendix \ref{apx:C} that setting $t=n\, T+ \Delta$ with 
$n=0,\,1,\,2\ldots$ and $\Delta\to 0^+$, the first-passage density $F(t=n\, T+\Delta)$ 
displays a power law divergence (the spikes) as $\Delta\to 0^+$
{
	\begin{equation}
	F(t=n\, T+ \Delta)\simeq A_n(b)\, \Delta^{-1/2}
	\label{spike.1}
	\end{equation}
}
with an amplitude $A_n(b)$ that can be computed explicitly (see Appendix \ref{apx:C}). 
We find that as $b$ increases, $A_n(b)$ decays rapidly and the spikes disappear for 
large $b=L/\sigma$ (see the inset of Fig.~(\ref{fig_stat})). Instead for large $b$, 
$F(t)$ drops by a finite amount after each period (as seen in the inset).  In 
theoretical models of resetting to a fixed initial position ($\sigma=0$ or $b\to 
\infty$), these spikes are completely absent and hence {they are characteristic of the 
finiteness of the width $\sigma$ of the resetting position ${\cal P}(x_0)$}.

\begin{figure}
\includegraphics[width=0.45\textwidth]{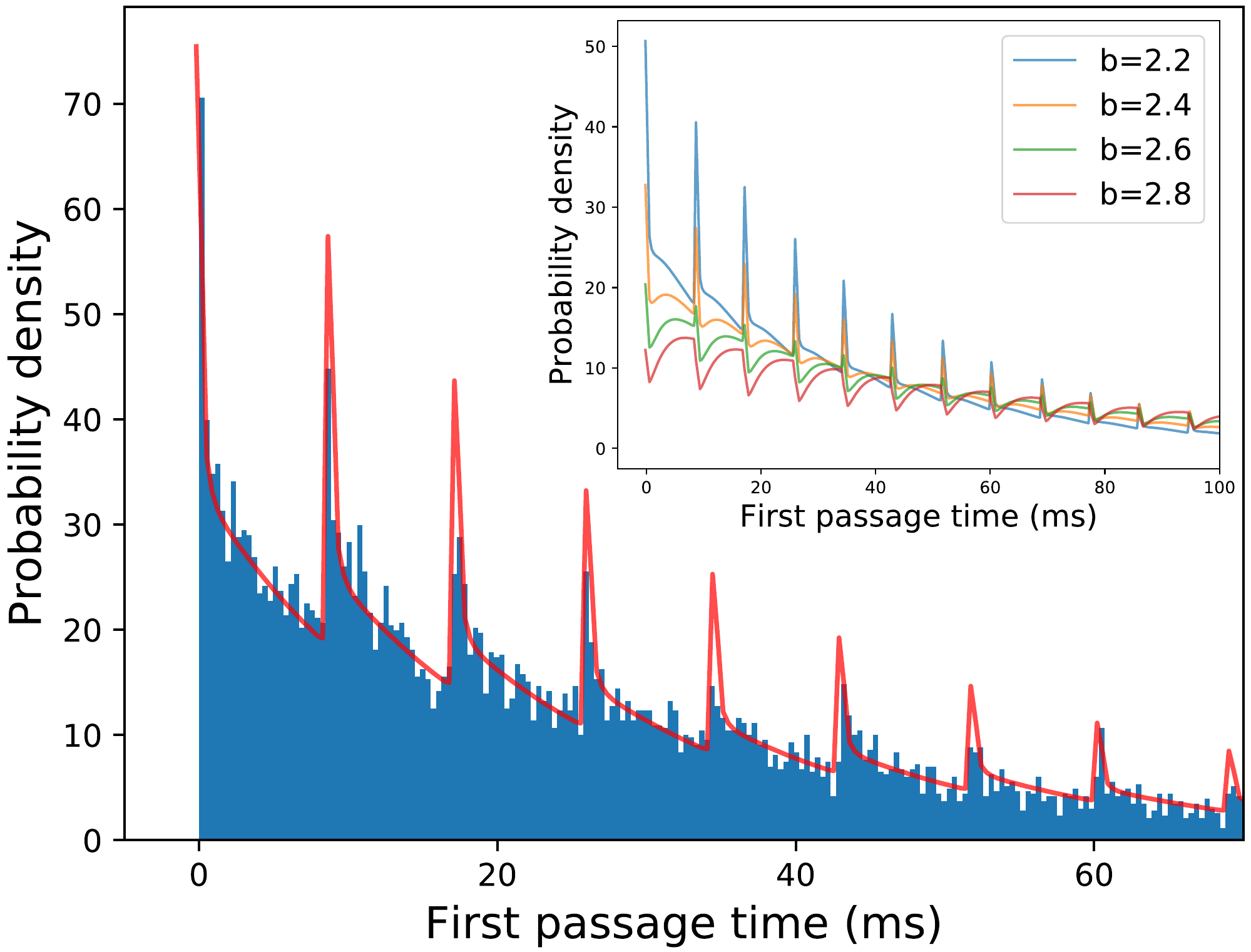} 
\caption{Experimental (blue histogram) and theoretical (red line) PDF $F(t)$ of the
first-passage time. We measured $1.2 \times 10^4$ first-passage times 
for $b=2$ and $c=1$ ($T=8.6$ ms).
The spikes occur just immediately after $t=n\, T$ where $n=0, \, 1,\, 
2, \ldots$.
\textit{Inset:} The PDF $F(t)$ vs. $t$ (theoretical) for $b = 2.2$ to 
$2.8$ at $c=b/2$ (i.e $T$ constant) that demonstrates
that the spikes disappear rapidly with increasing $b$.
The PDF $F(t)$ is normalized such that $\int_0 ^\infty F(t) d{t} = 1$ with  $d{t} = 3.5 \ 10^{-4}$s.}
\label{fig_stat} 
\end{figure}

In order to test experimentally these results we realized a periodic resetting 
protocol and measure the statistics of first-passage times. The diffusion coefficient 
(typically $D \simeq \SI{2e-13}{\m^2/\s}$) is measured during the free 
diffusion part and the width of the Gaussian (typically $\sigma \simeq 40$ nm) when 
the particle is back to equilibrium. These independent and simultaneous measures allow 
us to overcome experimental drifts {which may appear}. In Fig.~(\ref{fig_stat}) we 
show the experimentally obtained PDF of $10^4$ measured first-passage times for $b=2$ 
and $c=1$. We also compared with our theoretical prediction (see Appendix 
\ref{apx:C}). We observe a very good agreement with no free parameter.

\subsection{Poissonian resetting protocol.}
It turns out that this metastability and the phase transition is rather
robust and exists for other protocols, such as Poissonian resetting
where the resetting occurs at random times (as opposed to
periodically in the previous protocol) with a constant rate $r$. Here, the associated
dimensionless variables are 
\begin{equation}
b=\frac{L}{\sigma}\, ; \quad {\rm and}\quad c=\frac{\sqrt{r} L}{\sqrt{D}}\, .
\label{parameters_Poisson} 
\end{equation}
The scaled MFPT $\tau= 4 D\langle t_\text{f}\rangle/L^2$ again becomes 
a function $w_2(b,c)$ of $b$ and $c$  
(analogue of Eq. (\ref{wbc_def})). In this case, we get
a long but explicit $w_2(b,c)$ (see Eq. (\ref{mfpt.2p2}) in Appendix \ref{apx:C} 
for details).
In Fig. (\ref{fig_wbc_Poisson}) we plot $w_2(b,c)$ vs. $c$
for different values of $b$ 
together with experimental data. We have a good agreement between theory and experiment and here again the 
deviation at high $c$ comes from limited experimental acquisition rate. Once again we see that there is a 
metastable minimum that disappears when $b$ decreases below a critical value $b_c\approx 2.53$. When $b\to 
\infty$, there is only a single minimum at $c^*=1.59362$ where $w_2(\infty, c^*)= 6.17655$. Thus this 
phenomenon of metastability and phase transition in MFPT seems to be robust.

\begin{figure}
	\includegraphics[width=0.45\textwidth]{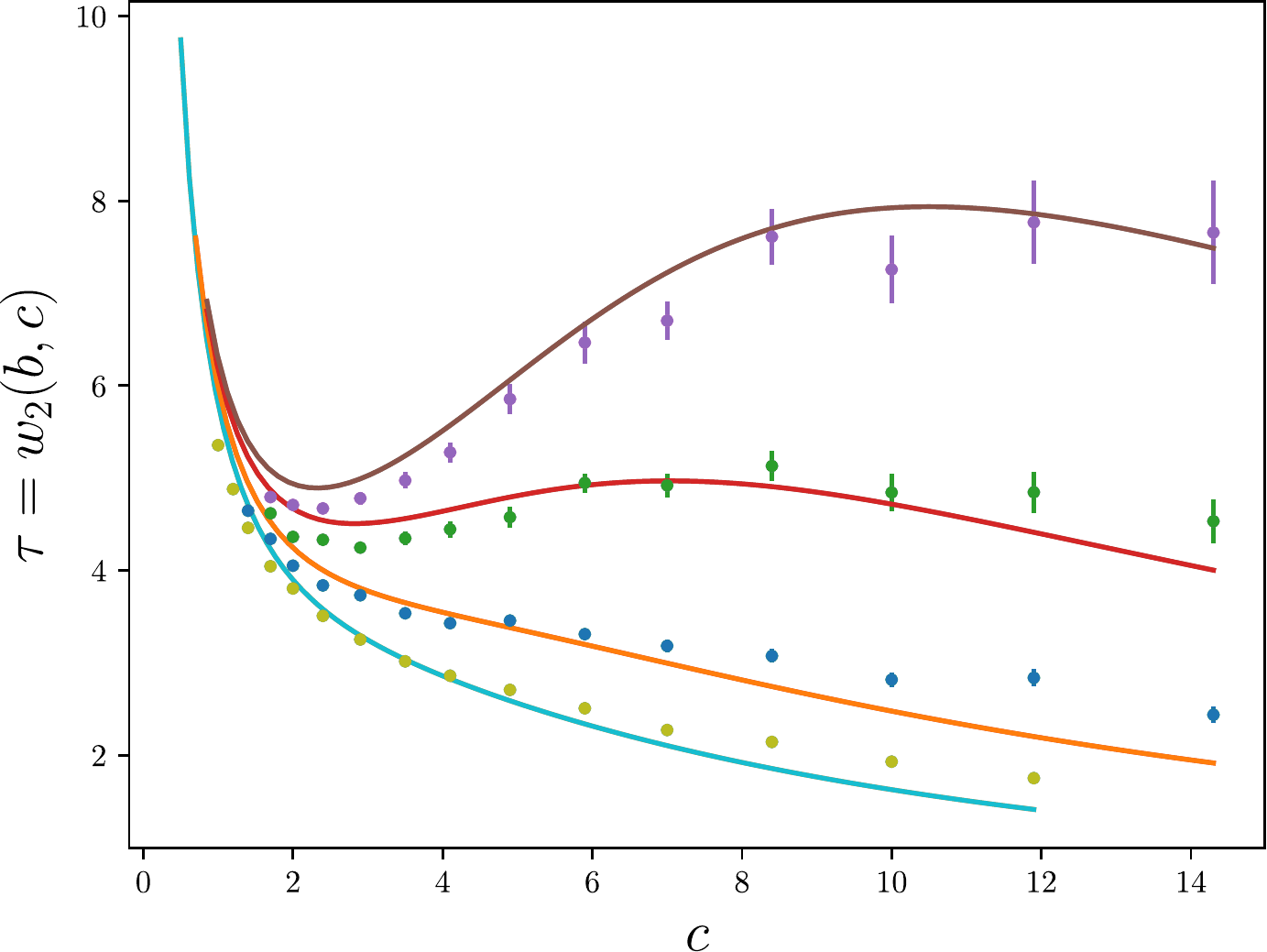}
	\caption{Scaled MFPT $\tau = w_2(b,c)$ vs $c$ curves and experimental data points for $b = 2, 2.3, 2.7$ and $3$ (from bottom to top)  in the case of Poissonian resetting. The metastable minimum disappears for $b<b_c\approx 2.53$.}
	\label{fig_wbc_Poisson}
\end{figure}

\section{Optimal resetting in 2 dimensions} \label{sec:2D}

\begin{figure}
	\centering
	\includegraphics[width=0.4\textwidth]{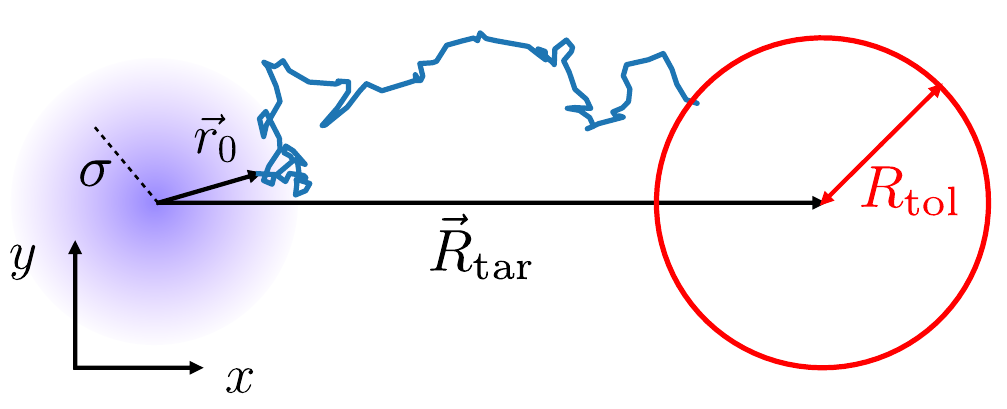}
	\caption{We consider a searcher in $2$-dimensions starting from an 
initial position $\vec r_0$ {drawn from} the initial distribution centered at the 
origin, {which here is a Gaussian distribution of width $\sigma$}. 
A fixed point target is located at $\vec R_{\rm tar}$, 
with a tolerance radius $R_{\rm tol}$. If the searcher comes within a distance 
$R_{\rm tol}$ from the target, the target is reached.}
	\label{fig:capture_2d}
\end{figure}

In the previous Section we have studied the MFPT with resetting
in 1D, both for the periodic and the Poissonian resetting protocols. 
We now consider a Brownian searcher in 2D, starting initially at 
the position ${\vec r}_0$. 
This position is random, and is drawn from a Gaussian distribution of width $\sigma$.
An immobile point target is located at 
${\vec R}_{\rm tar}$ with a tolerance radius $R_{\rm tol}$ (see 
Fig.\ (\ref{fig:capture_2d})). As the target and the initial position distribution 
are rotationally invariant, only the target distance $L=|\vec{R}_\text{tar}|$ matters. 
When the searcher comes within the distance $R_{\rm tol}$ 
from the target at ${\vec R}_{\rm tar}$ for the first time, the target is detected and the time at which this occurs is recorded as the first-passage time. A finite tolerance radius is needed in 2D, since
a point particle will never find a point target in 2D -- making it necessary
to have a finite size (cut-off) of the target. This introduces an
additional length scale and consequently the analytical calculations
are much more involved (see Appendices \ref{apx:D_2d} and \ref{apx:E_2d}
for the detailed calculations in 2D). In the 2D case,
we therefore introduce three dimensionless parameters, to characterize
the MFTP $\tau$. Two of these parameters read
\begin{equation}
a=\frac{R_{\rm tol}}{L}<1\, ,
\quad\,\, b= \frac{L}{\sigma}\,,
\label{eq:def_a_b}
\end{equation}
where $a$ keeps into account the tolerance radius with respect to the target 
distance, and $b$ quantifies how far is the target with respect to the spread of the initial distribution as in the 1D case.
The third parameter $c$ depends on the protocol: for the
periodic protocol with period $T$ and the Poissonian protocol with
rate $r$, the parameter $c$ reads respectively  
\begin{equation}  c= \frac{L}{\sqrt{4\,D\,T}} \quad\, {\rm and} \quad\,
c=\frac{\sqrt{r}\, L}{\sqrt{D}}\,.
\label{eq:c_def}
\end{equation}

{The MFPT $\langle t_\text{f}\rangle$ for this
model is computed analytically in Appendix \ref{apx:E_2d}.
For this, we first computed the survival probability up to time $t$ of a resetting
Brownian motion in 2D (with the resetting position chosen
from a Gaussian distribution of width $\sigma$) by imposing an 
absorbing boundary condition
on the perimeter of the circle of radius $R_{\rm tol}$ which is taken
as the target. The MFPT $\langle t_\text{f}\rangle$
is subsequently computed from this survival probability.}

As in the case of 1D, we are able to express the scaled MFPT $\tau = 4 D \langle 
t_\text{f}\rangle / L^2$ as a function of the three parameters $(a,b,c)$. This 
functional dependence of the MFPT on the three parameters for the periodic protocol 
$\tau=W_\text{periodic}(a,b,c)$ is not easy to obtain explicitly (see
Eq. (\ref{tau_b_final.1} in Appendix \ref{apx:E_2d}), though different 
limiting forms can be computed (see Appendix \ref{apx:E_2d}). The corresponding 
function is more explicit in the case of Poissonian resetting, where 
$\tau=W_\text{Poisson}(a,b,c)$ is computed explicitly in Eq. (\ref{tau_poisson.1}) 
in Appendix \ref{apx:E_2d}.

\subsection{Periodic resetting in 2 dimensions}

Here the particle is reset after a free diffusion period $T$. The theoretical 
prediction of MFPT for the periodic resetting is given in Eq.~\ref{tau_b_final.1}, 
which does not have an explicit expression and is numerically integrated (see 
Fig.~(\ref{fig:tau_per}) in Appendix~\ref{apx:E_2d}).

\subsubsection{Numerical simulation of Langevin dynamics}

\begin{figure}
	\centering
	\includegraphics[width=0.9\columnwidth]{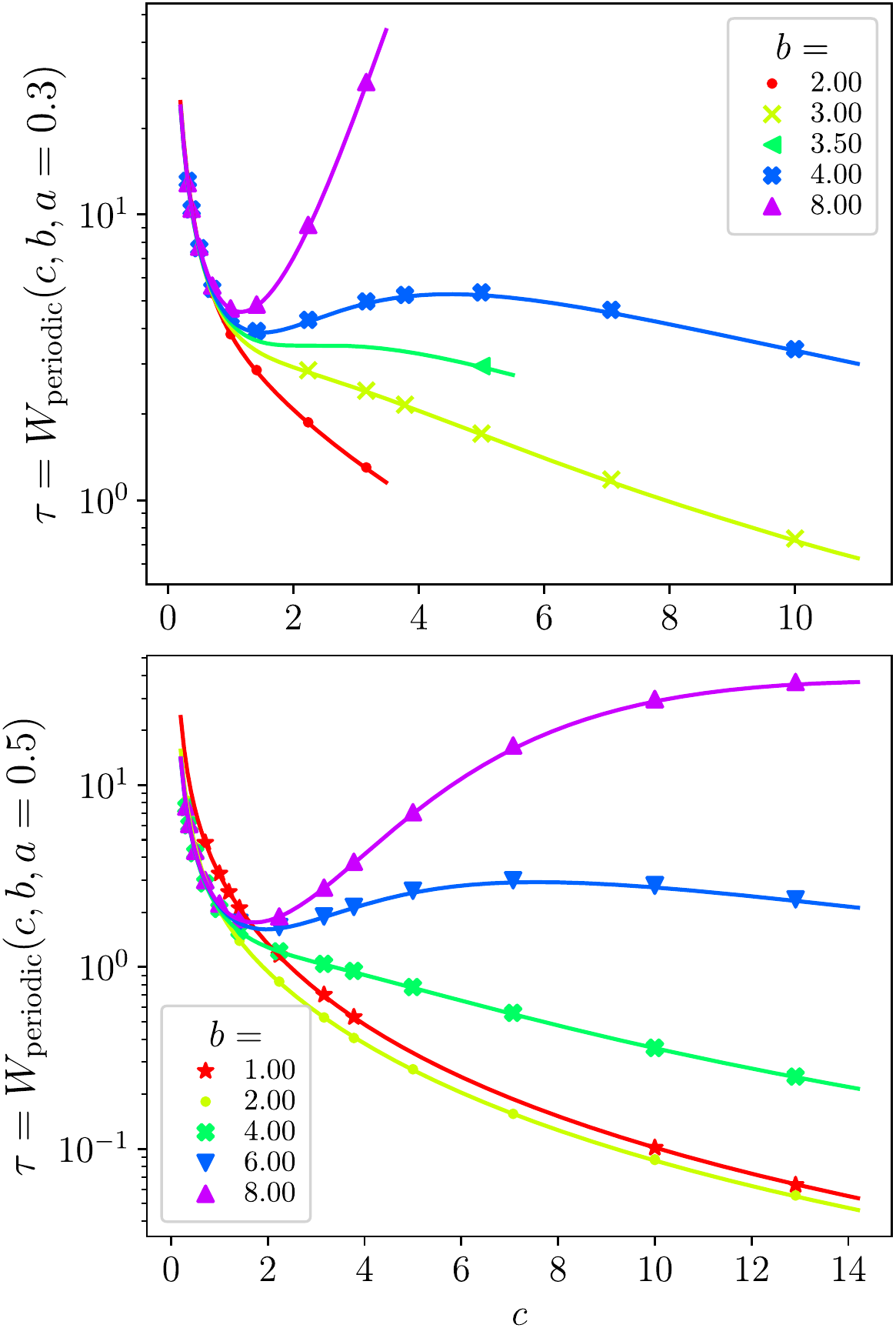}
	\caption{MFPT versus $c$ for the periodic resetting protocol in 2D for finite $b$ and $a=0.3$ (top panel) and $a=0.5$ (bottom panel). Continuous lines are the the theoretical predictions of Eq.~\ref{tau_b_final.1} and the symbols corresponds to the results obtained from the numerical integration of the Langevin equation. In the bottom panel ($a = 0.5$), the curves for $b=6$ and $b=4$ illustrate the two different regimes, whereas the phase transition appears at lower $b$ for $a=0.3$ (top pannel).}
	\label{fig:b_2d_numerical_period}
\end{figure}

\begin{figure}
	\centering
	\includegraphics[width=0.9\columnwidth]{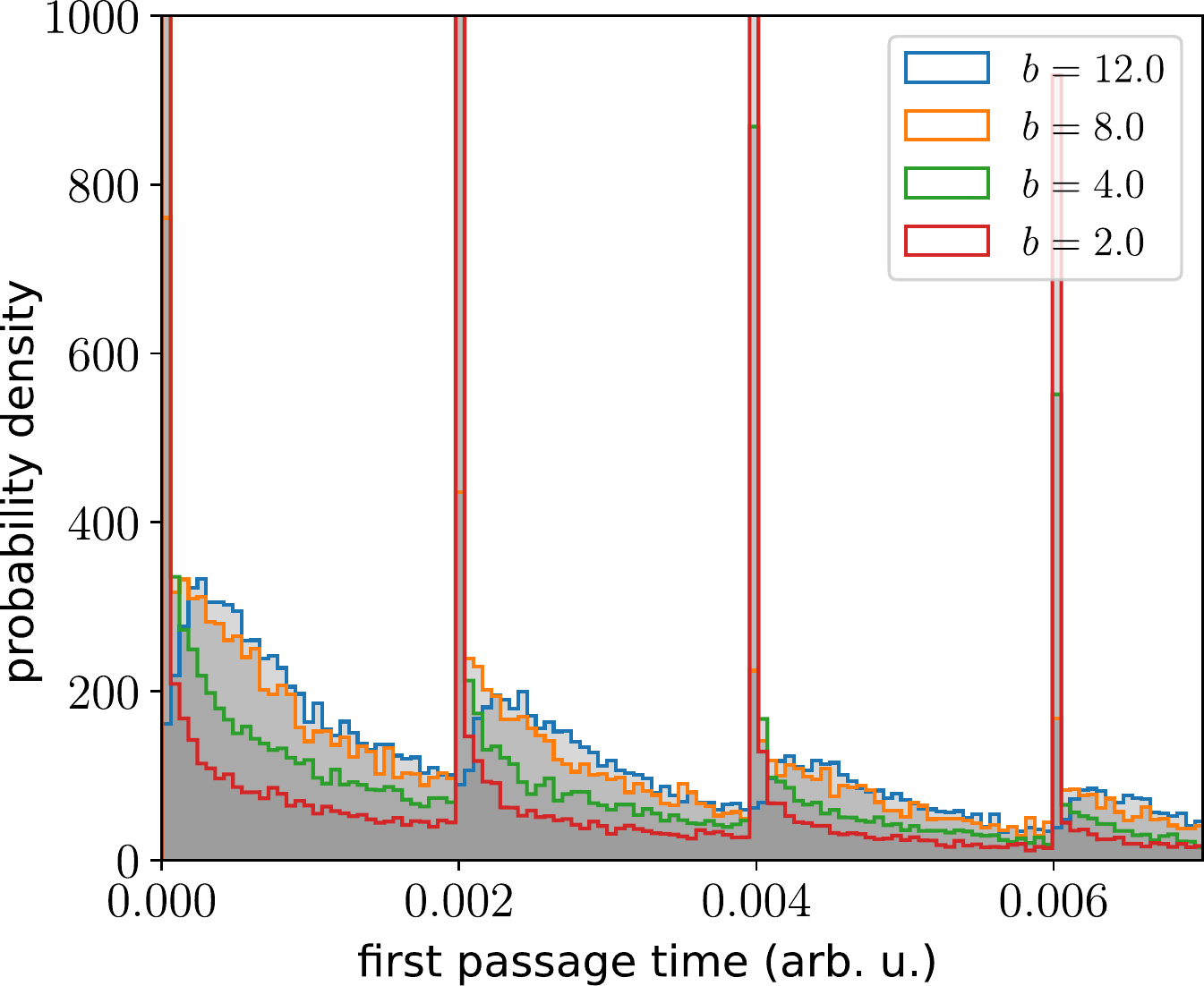} 
	\caption{Distributions of first passage times $t_\text{f}$ obtained from numerical simulations of the Langevin dynamics wih periodic reset at $c=2.23$, $a=0.75$ and various $b$'s.}
	\label{2D_FPT_distribbutions}
\end{figure}

The comparison with the numerical simulation obtained from the numerical 
integration of the Langevin dynamics is plotted in 
Fig.~(\ref{fig:b_2d_numerical_period}) for two different $a$ and various $b$'s. The agreement is excellent and the presence of a metastable minimum for $b > b_c$ and a monotonous decrease for smaller $b$ confirm the presence of the dynamical phase transition in 2D for the periodic resetting. First passage times distributions has been numerically evaluated and the results at $c=2.23$, $a=0.75$ and various $b$'s are plotted in Fig.~(\ref{2D_FPT_distribbutions}). Notice the presence of the spikes of period $T$ as was already observed in 1D. 

\subsubsection{Experimental results} \label{sec:res-exp2d}

\begin{figure}
	\centering
	\includegraphics[width=0.9\columnwidth]{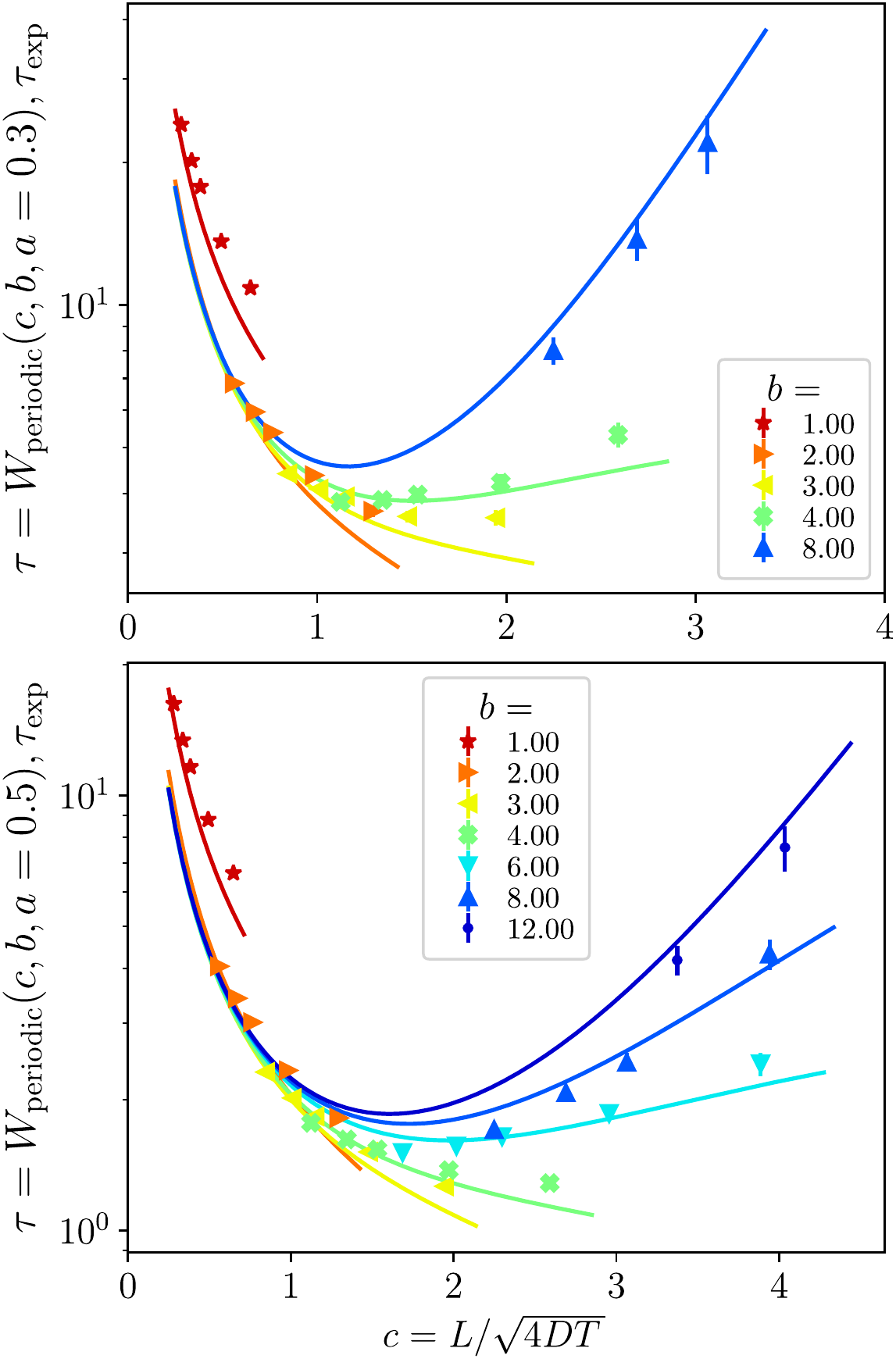}
	\caption{Experiment. $\tau$ versus $c$ for the periodic resetting protocol in 2D for finite $b$ and $a=0.3$ (top panel) and $a=0.5$ (bottom panel). Continuous lines are the the theoretical predictions of Eq.~\ref{tau_b_final.1} with $\sigma=\sigma_{x,\text{exp}}$ and the symbols corresponds to the results obtained from the experimental data.}
	\label{fig:2d_expe_period}
\end{figure}

Testing experimentally the theoretical predictions of the MFPT in 2D presents two main chalenges. The first is that a very good statistics is needed to get results with a rather good accuracy for several values of $a,b$ and $c$. The second is due to the unavoidable asymmetries in the $x$ and $y$ direction especially on the variances of the initial position, due to laser beam anisotropy ($\sigma_{x,\text{exp}} /\sigma_{y,\text{exp}} = 0.82$). {In order to compare with the theory (which has not been extended to anisotropic initial distributions), we must choose a representative $\sigma$. We could define $\sigma_\text{mean}=\sqrt{\smash{(\sigma_{x,\text{exp}}^2+\sigma_{y,\text{exp}}^2)}/2} = \SI{38}{\nano\m}$. However, this does not do justice to the fact that when the target is on the $x$ axis, a larger $\sigma_x$ \emph{decreases} the MFPT, while a larger $\sigma_y$ \emph{increases} the MFPT. Thus, $\sigma=\sigma_{x,\text{exp}}=\SI{34}{\nano\m}$ is a better choice to compare with the theory. The results are reported in Fig.~(\ref{fig:2d_expe_period}) where we observe a rather good agreement with the theoretical predictions in spite of the approximated $\sigma$ and the rather low statistics (about $10^4$ trajectories). The same results are reported against $\sigma_\text{mean}$ in Fig.~(\ref{fig:2d_expe_period_alt}), and the effect of anisotropy is discussed in annex \ref{annex:aniso}. While it is important ($\sim$10\% error on $\tau$), it is not enough to explain the mismatch. Again, the fact that the difference is significant only for low MFPTs and high $c$ indicates that it may come from the limited experimental acquisition rate.}
\subsection{Poissonian resetting in 2 dimensions}

\begin{figure}
	\centering
	\includegraphics[width=0.9\columnwidth]{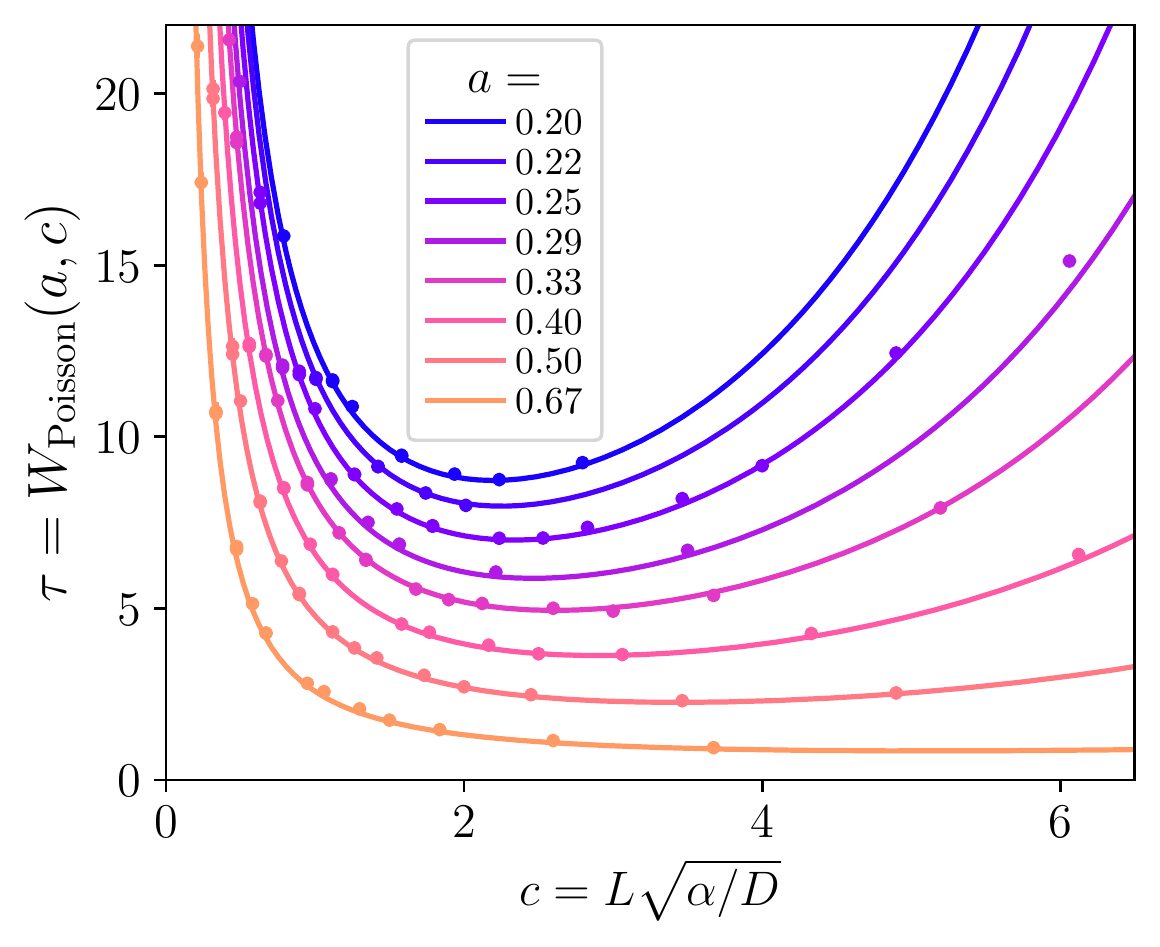}
	\caption{MFPT versus $c$ for the Poissonian resetting protocol in 2D for  $b\rightarrow \infty$ and several values of $a$. Continuous lines are the the theoretical predictions of Eq.~\ref{tau_poisson_origin.1} and the symbols corresponds to the  results obtained from the numerical integration of the Langvin equation.}
	\label{fig:b_infty_2d_numerical_pois}
\end{figure}

\begin{figure}
	\centering
	\includegraphics[width=0.9\columnwidth]{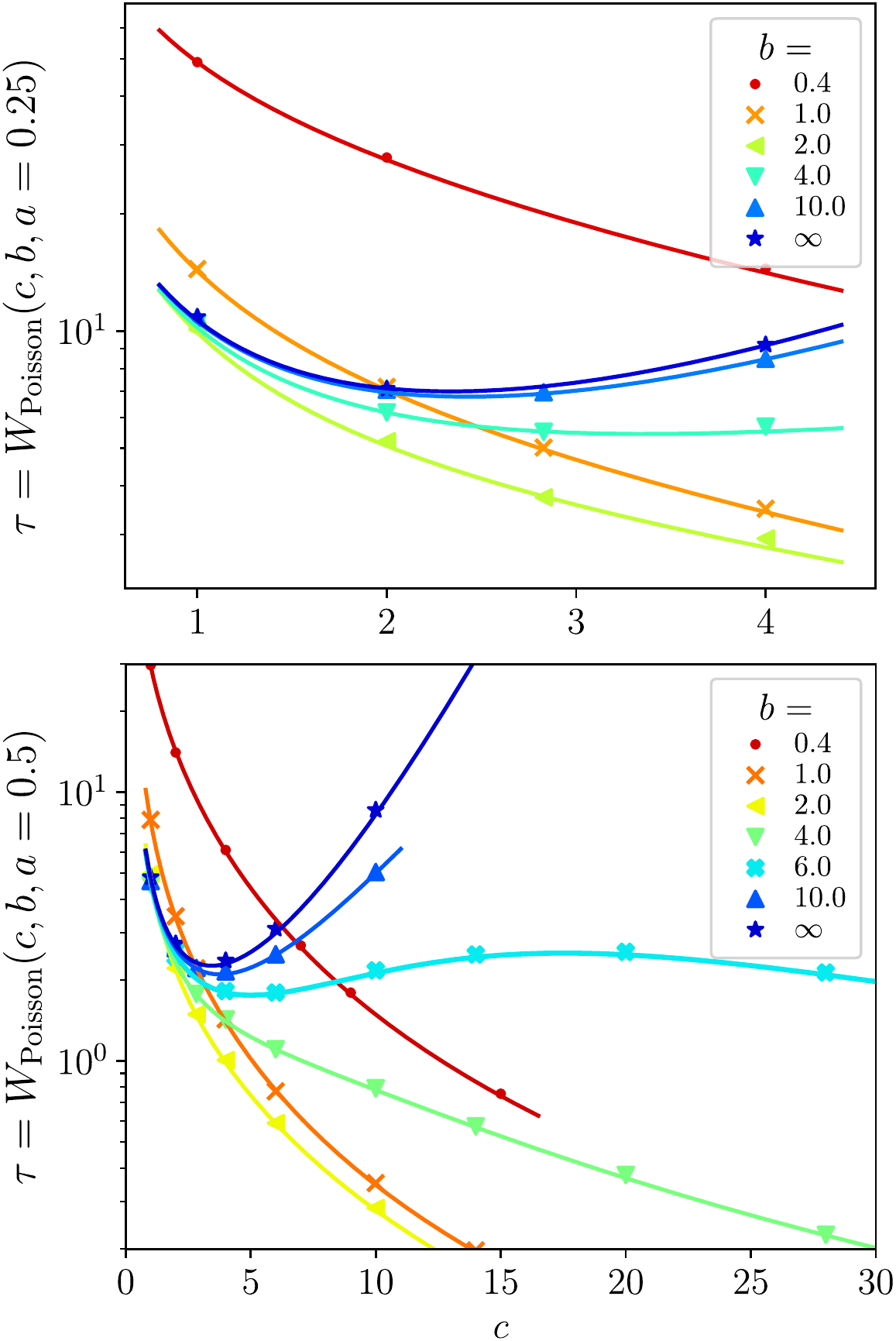}
	\caption{$\tau$ versus $c$ for the Poissonian resetting protocol in 2D for finite $b$ and $a=0.25$ (top panel) and $a=0.5$ (bottom panel). Continuous lines are the the theoretical predictions of Eq.~\ref{tau_poisson.1} and the symbols corresponds to the results obtained from the numerical integration of the Langvin equation. In the bottom panel, curves for $b=6$ and $b=4$ illustrate the two different regimes.}
	\label{fig:b_2d_numerical_pois}
\end{figure}

We discuss here the results with the Poissonian resetting. In this case the 2D MFPT can be analytically computed explicitly (see Eqs.~\ref{tau_poisson.1} and \ref{tau_poisson_origin.1} in Appendix \ref{apx:E_2d}). {Our experimental data statistics is to low to compare the theoretical 2D MFPT with experimental results for the Poissonian resetting protocol. Indeed in that case one needs higher statistics to sample correctly the exponential distribution of free diffusion times and then have a good estimate of the MFPT for the three parameters $(a,b,c)$. }
We thus compare the analytical results with a numerical integration of the Langevin equation for different values of $b$ and $a$. The numerical results for $b\rightarrow \infty$ (i.e.~$\sigma \rightarrow 0$) are compared in Fig.~(\ref{fig:b_infty_2d_numerical_pois}) with theoretical prediction (Eq.~\ref{tau_poisson_origin.1}) for several values of $a$.  
The theoretical predictions at finite $b$ are compared with the numerical results in Fig.~(\ref{fig:b_2d_numerical_pois}) for two values of $a$. The agreement is very good and one can notice that, as in the 1D case, the presence of a metastable minimum (e.g. see $b=6$).

{This demonstrates that, as in 1D, the 2D MFPT curve always features a metastable minimum for higher values of $b$, and a dynamical phase transition for closer targets.}

\section{Discussion and conclusions} \label{sec:conclusions}
We have studied theoretically, numerically  and experimentally, in one and two dimensions,
the statistics of first-passage times to a target by a Brownian particle subjected
to two different resetting protcols: (i) periodic resetting with period $T$
and (ii) Poissonian resetting with a constant $r$. In both cases, 
the distribution of the initial/reset position is not a delta function, but has
a Gaussian distribution with a finite width $\sigma$, corresponding
to the equilibrium Gibbs-Boltzmann position distribution of the particle
in a harmonic trapping potential. We have shown that the presence of a finite
width $\sigma$ in the resetting position has a profound and interesting effect
on the statistics of first-passage times. We have shown that, 
both in 1 and 2 dimensions and in both protocols, the curve of the rescaled mean 
first-passage time $\tau$ as a function of the parameter $c= L/\sqrt{4DT}$ for periodic 
protocol (and $c=\sqrt{r/D}\, L$ for Poissonian protocol) has very different behavior
for $b=L/\sigma>b_c$ and $b<b_c$. For $b>b_c$, the MFPT curve has two
minima, one at a finite value $c_1(b)$ and the other at $c\to \infty$, separated
by a maximum at $c=c_2(b)$. The first minimum turns out to be metastable and it
disappears for $b<b_c$ where the MFPT curve decreases monotonbically with
increasing $c$. We have argued that this phase transition at $b=b_c$ is dynamical in
the sense that it occurs when two different time scales $T_1\sim \mathcal{O}(L^2/D)$ and
$T_2\sim \mathcal{O}(\sigma^2/D)$ merge with each other.

We find the existence of the metastable minimum in MFPT for $b>b_c$ and the subsequent 
disappearence of it (when $b$ redcues below $b_c$) via a dynamical phase transition,
rather robust. It exists in both 1D and 2D and for two different types of 
resetting protocols.
For the periodic protocol, we also computed and measured the full PDF of the 
first-passage times and found that it 
displays striking spikes after each resetting event, a clear effect of the finiteness 
of the variance of the initial position. Both in one and two dimensions the numerical 
and experimental data agree well with theoretical predictions, but the experiment is 
not a mere test of the theory. Indeed, taking into account the experimental finite width $\sigma$ of optical trap resetting shed light on the dynamical phase transtion described here. We also detailed experimental 
difficulties encountered in first passage experiments. When the particle is free for a very long time (small $c$), one has to tackle with sedimentation or particle loss. On the very short times (large $c$), we have shown that finite sampling time affects the results, because some first-passages are not detected, leading to an overestimation of the 
first-passage time. In 2 dimensions the anisotropy of the optical trap and statistics limit the accuracy. Finally, it would be interesting to extend this work by taking into account the role of hydrodynamic memory effects and particle inertia in future theoretical developments and to compare with very fast sampling rate experiments.

\bigskip

\begin{acknowledgments}
	We thank A. Bovon for his participation in an early stage of the experiment and E. Trizac for useful discussions. This work has been partially supported by the FQXi Foundation, Grant No. FQXi-IAF19-05, "Information as a fuel in colloids and superconducting quantum circuits". SNM wants to thank the warm hospitality of ICTS (Bangalore)  where this work was completed.
\end{acknowledgments}

\bibliographystyle{apsrev4-1}
\bibliography{first_passage_biblio}

\begin{thebibliography}{49}%
\makeatletter
\providecommand \@ifxundefined [1]{%
 \@ifx{#1\undefined}
}%
\providecommand \@ifnum [1]{%
 \ifnum #1\expandafter \@firstoftwo
 \else \expandafter \@secondoftwo
 \fi
}%
\providecommand \@ifx [1]{%
 \ifx #1\expandafter \@firstoftwo
 \else \expandafter \@secondoftwo
 \fi
}%
\providecommand \natexlab [1]{#1}%
\providecommand \enquote  [1]{``#1''}%
\providecommand \bibnamefont  [1]{#1}%
\providecommand \bibfnamefont [1]{#1}%
\providecommand \citenamefont [1]{#1}%
\providecommand \href@noop [0]{\@secondoftwo}%
\providecommand \href [0]{\begingroup \@sanitize@url \@href}%
\providecommand \@href[1]{\@@startlink{#1}\@@href}%
\providecommand \@@href[1]{\endgroup#1\@@endlink}%
\providecommand \@sanitize@url [0]{\catcode `\\12\catcode `\$12\catcode
  `\&12\catcode `\#12\catcode `\^12\catcode `\_12\catcode `\%12\relax}%
\providecommand \@@startlink[1]{}%
\providecommand \@@endlink[0]{}%
\providecommand \url  [0]{\begingroup\@sanitize@url \@url }%
\providecommand \@url [1]{\endgroup\@href {#1}{\urlprefix }}%
\providecommand \urlprefix  [0]{URL }%
\providecommand \Eprint [0]{\href }%
\providecommand \doibase [0]{http://dx.doi.org/}%
\providecommand \selectlanguage [0]{\@gobble}%
\providecommand \bibinfo  [0]{\@secondoftwo}%
\providecommand \bibfield  [0]{\@secondoftwo}%
\providecommand \translation [1]{[#1]}%
\providecommand \BibitemOpen [0]{}%
\providecommand \bibitemStop [0]{}%
\providecommand \bibitemNoStop [0]{.\EOS\space}%
\providecommand \EOS [0]{\spacefactor3000\relax}%
\providecommand \BibitemShut  [1]{\csname bibitem#1\endcsname}%
\let\auto@bib@innerbib\@empty
%</preamble>
\bibitem [{\citenamefont {Redner}(2001)}]{Redner}%
  \BibitemOpen
  \bibfield  {author} {\bibinfo {author} {\bibfnamefont {S.}~\bibnamefont
  {Redner}},\ }\href@noop {} {\emph {\bibinfo {title} {A guide to first-passage
  processes}}}\ (\bibinfo  {publisher} {Cambridge University Press},\ \bibinfo
  {year} {2001})\BibitemShut {NoStop}%
\bibitem [{\citenamefont {Bray}\ \emph {et~al.}(2013)\citenamefont {Bray},
  \citenamefont {Majumdar},\ and\ \citenamefont {Schehr}}]{pers_review}%
  \BibitemOpen
  \bibfield  {author} {\bibinfo {author} {\bibfnamefont {A.~J.}\ \bibnamefont
  {Bray}}, \bibinfo {author} {\bibfnamefont {S.~N.}\ \bibnamefont {Majumdar}},
  \ and\ \bibinfo {author} {\bibfnamefont {G.}~\bibnamefont {Schehr}},\
  }\href@noop {} {\bibfield  {journal} {\bibinfo  {journal} {Adv. in Phys.}\
  }\textbf {\bibinfo {volume} {62}},\ \bibinfo {pages} {225} (\bibinfo {year}
  {2013})}\BibitemShut {NoStop}%
\bibitem [{\citenamefont {Villen-Altramirano}\ and\ \citenamefont
  {Villen-Altramirano}(1991)}]{VAVA91}%
  \BibitemOpen
  \bibfield  {author} {\bibinfo {author} {\bibfnamefont {M.}~\bibnamefont
  {Villen-Altramirano}}\ and\ \bibinfo {author} {\bibfnamefont
  {J.}~\bibnamefont {Villen-Altramirano}},\ }\enquote {\bibinfo {title}
  {Restart: A method for accelerating rare event simulations},}\ in\ \href@noop
  {} {\emph {\bibinfo {booktitle} {Qeueing Performance and Control in ATM}}},\
  \bibinfo {editor} {edited by\ \bibinfo {editor} {\bibfnamefont {J.~W.}\
  \bibnamefont {Cohen}}\ and\ \bibinfo {editor} {\bibfnamefont {C.~D.}\
  \bibnamefont {Pack}}}\ (\bibinfo  {publisher} {North-Holland, Amsterdam},\
  \bibinfo {year} {1991})\BibitemShut {NoStop}%
\bibitem [{\citenamefont {Luby}\ \emph {et~al.}(1993)\citenamefont {Luby},
  \citenamefont {Sinclair},\ and\ \citenamefont {Zuckerman}}]{LSZ93}%
  \BibitemOpen
  \bibfield  {author} {\bibinfo {author} {\bibfnamefont {M.}~\bibnamefont
  {Luby}}, \bibinfo {author} {\bibfnamefont {A.}~\bibnamefont {Sinclair}}, \
  and\ \bibinfo {author} {\bibfnamefont {D.}~\bibnamefont {Zuckerman}},\
  }\href@noop {} {\bibfield  {journal} {\bibinfo  {journal} {Inf. Proc. Lett.}\
  }\textbf {\bibinfo {volume} {47}},\ \bibinfo {pages} {4391} (\bibinfo {year}
  {1993})}\BibitemShut {NoStop}%
\bibitem [{\citenamefont {Tong}\ \emph {et~al.}(2008)\citenamefont {Tong},
  \citenamefont {Faloutsos},\ and\ \citenamefont {Pan}}]{TFP08}%
  \BibitemOpen
  \bibfield  {author} {\bibinfo {author} {\bibfnamefont {H.}~\bibnamefont
  {Tong}}, \bibinfo {author} {\bibfnamefont {C.}~\bibnamefont {Faloutsos}}, \
  and\ \bibinfo {author} {\bibfnamefont {J.-Y.}\ \bibnamefont {Pan}},\
  }\href@noop {} {\bibfield  {journal} {\bibinfo  {journal} {Knowl. Inf.
  Syst.}\ }\textbf {\bibinfo {volume} {14}},\ \bibinfo {pages} {327} (\bibinfo
  {year} {2008})}\BibitemShut {NoStop}%
\bibitem [{\citenamefont {Lorenz}(2018)}]{Lorenz18}%
  \BibitemOpen
  \bibfield  {author} {\bibinfo {author} {\bibfnamefont {J.~H.}\ \bibnamefont
  {Lorenz}},\ }\enquote {\bibinfo {title} {Runtime distributions and criteria
  for restarts},}\ in\ \href@noop {} {\emph {\bibinfo {booktitle} {SOFSEM
  2018:Theory and Practice of Computer Science}}},\ \bibinfo {series} {Lecture
  Notes in Computer Science}, Vol.\ \bibinfo {volume} {10706},\ \bibinfo
  {editor} {edited by\ \bibinfo {editor} {\bibfnamefont {T.}~\bibnamefont
  {A.}}, \bibinfo {editor} {\bibfnamefont {B.}~\bibnamefont {L.}}, \bibinfo
  {editor} {\bibfnamefont {v.~L.~J.}\ \bibnamefont {Biffl~S.}}, \ and\ \bibinfo
  {editor} {\bibfnamefont {W.}~\bibnamefont {J.}}}\ (\bibinfo  {publisher}
  {Springer, Berlin},\ \bibinfo {year} {2018})\BibitemShut {NoStop}%
\bibitem [{\citenamefont {Evans}\ \emph {et~al.}(2020)\citenamefont {Evans},
  \citenamefont {Majumdar},\ and\ \citenamefont {Schehr}}]{reset_review}%
  \BibitemOpen
  \bibfield  {author} {\bibinfo {author} {\bibfnamefont {M.~R.}\ \bibnamefont
  {Evans}}, \bibinfo {author} {\bibfnamefont {S.~N.}\ \bibnamefont {Majumdar}},
  \ and\ \bibinfo {author} {\bibfnamefont {G.}~\bibnamefont {Schehr}},\
  }\href@noop {} {\bibfield  {journal} {\bibinfo  {journal} {J. Phys. A: Math.
  Theor.}\ }\textbf {\bibinfo {volume} {53}},\ \bibinfo {pages} {193001}
  (\bibinfo {year} {2020})}\BibitemShut {NoStop}%
\bibitem [{\citenamefont {O}\ \emph {et~al.}(2011)\citenamefont {O},
  \citenamefont {Loverdo}, \citenamefont {Moreau},\ and\ \citenamefont
  {Voituriez}}]{BLMV11}%
  \BibitemOpen
  \bibfield  {author} {\bibinfo {author} {\bibfnamefont {O.~B.}\ \bibnamefont
  {O}}, \bibinfo {author} {\bibfnamefont {C.}~\bibnamefont {Loverdo}}, \bibinfo
  {author} {\bibfnamefont {M.}~\bibnamefont {Moreau}}, \ and\ \bibinfo {author}
  {\bibfnamefont {R.}~\bibnamefont {Voituriez}},\ }\href@noop {} {\bibfield
  {journal} {\bibinfo  {journal} {Rev. Mod. Phys.}\ }\textbf {\bibinfo {volume}
  {83}},\ \bibinfo {pages} {81} (\bibinfo {year} {2011})}\BibitemShut {NoStop}%
\bibitem [{\citenamefont {Evans}\ and\ \citenamefont
  {Majumdar}(2011{\natexlab{a}})}]{EM2011_1}%
  \BibitemOpen
  \bibfield  {author} {\bibinfo {author} {\bibfnamefont {M.~R.}\ \bibnamefont
  {Evans}}\ and\ \bibinfo {author} {\bibfnamefont {S.~N.}\ \bibnamefont
  {Majumdar}},\ }\href@noop {} {\bibfield  {journal} {\bibinfo  {journal}
  {Phys. Rev. Lett.}\ }\textbf {\bibinfo {volume} {106}},\ \bibinfo {pages}
  {160601} (\bibinfo {year} {2011}{\natexlab{a}})}\BibitemShut {NoStop}%
\bibitem [{\citenamefont {Evans}\ and\ \citenamefont
  {Majumdar}(2011{\natexlab{b}})}]{EM2011_2}%
  \BibitemOpen
  \bibfield  {author} {\bibinfo {author} {\bibfnamefont {M.~R.}\ \bibnamefont
  {Evans}}\ and\ \bibinfo {author} {\bibfnamefont {S.~N.}\ \bibnamefont
  {Majumdar}},\ }\href@noop {} {\bibfield  {journal} {\bibinfo  {journal} {J.
  Phys. A: Math. Theor.}\ }\textbf {\bibinfo {volume} {44}},\ \bibinfo {pages}
  {435001} (\bibinfo {year} {2011}{\natexlab{b}})}\BibitemShut {NoStop}%
\bibitem [{\citenamefont {Pal}\ \emph {et~al.}(2016)\citenamefont {Pal},
  \citenamefont {Kundu},\ and\ \citenamefont {Evans}}]{PKE16}%
  \BibitemOpen
  \bibfield  {author} {\bibinfo {author} {\bibfnamefont {A.}~\bibnamefont
  {Pal}}, \bibinfo {author} {\bibfnamefont {A.}~\bibnamefont {Kundu}}, \ and\
  \bibinfo {author} {\bibfnamefont {M.~R.}\ \bibnamefont {Evans}},\ }\href@noop
  {} {\bibfield  {journal} {\bibinfo  {journal} {J. Phys. A: Math. Theor.}\
  }\textbf {\bibinfo {volume} {49}},\ \bibinfo {pages} {225001} (\bibinfo
  {year} {2016})}\BibitemShut {NoStop}%
\bibitem [{\citenamefont {Bhat}\ \emph {et~al.}(2016)\citenamefont {Bhat},
  \citenamefont {Bacco},\ and\ \citenamefont {Redner}}]{BBR16}%
  \BibitemOpen
  \bibfield  {author} {\bibinfo {author} {\bibfnamefont {U.}~\bibnamefont
  {Bhat}}, \bibinfo {author} {\bibfnamefont {C.~D.}\ \bibnamefont {Bacco}}, \
  and\ \bibinfo {author} {\bibfnamefont {S.}~\bibnamefont {Redner}},\
  }\href@noop {} {\bibfield  {journal} {\bibinfo  {journal} {J. Stat. Mech.}\
  ,\ \bibinfo {pages} {083401}} (\bibinfo {year} {2016})}\BibitemShut {NoStop}%
\bibitem [{\citenamefont {Evans}\ \emph {et~al.}(2013)\citenamefont {Evans},
  \citenamefont {Majumdar},\ and\ \citenamefont {Mallick}}]{EMM13}%
  \BibitemOpen
  \bibfield  {author} {\bibinfo {author} {\bibfnamefont {M.~R.}\ \bibnamefont
  {Evans}}, \bibinfo {author} {\bibfnamefont {S.~N.}\ \bibnamefont {Majumdar}},
  \ and\ \bibinfo {author} {\bibfnamefont {K.}~\bibnamefont {Mallick}},\
  }\href@noop {} {\bibfield  {journal} {\bibinfo  {journal} {J. Phys. A: Math.
  Theor.}\ }\textbf {\bibinfo {volume} {46}},\ \bibinfo {pages} {185001}
  (\bibinfo {year} {2013})}\BibitemShut {NoStop}%
\bibitem [{\citenamefont {Montero}\ and\ \citenamefont
  {Villarroel}(2013)}]{MV13}%
  \BibitemOpen
  \bibfield  {author} {\bibinfo {author} {\bibfnamefont {M.}~\bibnamefont
  {Montero}}\ and\ \bibinfo {author} {\bibfnamefont {J.}~\bibnamefont
  {Villarroel}},\ }\href@noop {} {\bibfield  {journal} {\bibinfo  {journal}
  {Phys. Rev. E}\ }\textbf {\bibinfo {volume} {87}},\ \bibinfo {pages} {012116}
  (\bibinfo {year} {2013})}\BibitemShut {NoStop}%
\bibitem [{\citenamefont {Evans}\ and\ \citenamefont
  {Majumdar}(2014)}]{EM2014}%
  \BibitemOpen
  \bibfield  {author} {\bibinfo {author} {\bibfnamefont {M.~R.}\ \bibnamefont
  {Evans}}\ and\ \bibinfo {author} {\bibfnamefont {S.~N.}\ \bibnamefont
  {Majumdar}},\ }\href@noop {} {\bibfield  {journal} {\bibinfo  {journal} {J.
  Phys. A: Math. Theor.}\ }\textbf {\bibinfo {volume} {47}},\ \bibinfo {pages}
  {285001} (\bibinfo {year} {2014})}\BibitemShut {NoStop}%
\bibitem [{\citenamefont {Ku\'smierz}\ \emph {et~al.}(2014)\citenamefont
  {Ku\'smierz}, \citenamefont {Majumdar}, \citenamefont {Sabhapandit},\ and\
  \citenamefont {Schehr}}]{KMSS14}%
  \BibitemOpen
  \bibfield  {author} {\bibinfo {author} {\bibfnamefont {L.}~\bibnamefont
  {Ku\'smierz}}, \bibinfo {author} {\bibfnamefont {S.~N.}\ \bibnamefont
  {Majumdar}}, \bibinfo {author} {\bibfnamefont {S.}~\bibnamefont
  {Sabhapandit}}, \ and\ \bibinfo {author} {\bibfnamefont {G.}~\bibnamefont
  {Schehr}},\ }\href@noop {} {\bibfield  {journal} {\bibinfo  {journal} {Phys.
  Rev. Lett.}\ }\textbf {\bibinfo {volume} {113}},\ \bibinfo {pages} {220602}
  (\bibinfo {year} {2014})}\BibitemShut {NoStop}%
\bibitem [{\citenamefont {Reuveni}\ \emph {et~al.}(2014)\citenamefont
  {Reuveni}, \citenamefont {Urbakh},\ and\ \citenamefont {Klafter}}]{RUK14}%
  \BibitemOpen
  \bibfield  {author} {\bibinfo {author} {\bibfnamefont {S.}~\bibnamefont
  {Reuveni}}, \bibinfo {author} {\bibfnamefont {M.}~\bibnamefont {Urbakh}}, \
  and\ \bibinfo {author} {\bibfnamefont {J.}~\bibnamefont {Klafter}},\
  }\href@noop {} {\bibfield  {journal} {\bibinfo  {journal} {Proc. Natl. Acad.
  Sci. USA}\ }\textbf {\bibinfo {volume} {111}},\ \bibinfo {pages} {4391}
  (\bibinfo {year} {2014})}\BibitemShut {NoStop}%
\bibitem [{\citenamefont {Rotbart}\ \emph {et~al.}(2015)\citenamefont
  {Rotbart}, \citenamefont {Reuveni},\ and\ \citenamefont {Urbakh}}]{RRU15}%
  \BibitemOpen
  \bibfield  {author} {\bibinfo {author} {\bibfnamefont {T.}~\bibnamefont
  {Rotbart}}, \bibinfo {author} {\bibfnamefont {S.}~\bibnamefont {Reuveni}}, \
  and\ \bibinfo {author} {\bibfnamefont {M.}~\bibnamefont {Urbakh}},\
  }\href@noop {} {\bibfield  {journal} {\bibinfo  {journal} {Phys. Rev. E {\bf
  92} (2015).}\ }\textbf {\bibinfo {volume} {92}},\ \bibinfo {pages}
  {060101(R)} (\bibinfo {year} {2015})}\BibitemShut {NoStop}%
\bibitem [{\citenamefont {Christou}\ and\ \citenamefont
  {Schadschneider}(2015)}]{CS15}%
  \BibitemOpen
  \bibfield  {author} {\bibinfo {author} {\bibfnamefont {C.}~\bibnamefont
  {Christou}}\ and\ \bibinfo {author} {\bibfnamefont {A.}~\bibnamefont
  {Schadschneider}},\ }\href@noop {} {\bibfield  {journal} {\bibinfo  {journal}
  {J. Phys. A: Math. Theor.}\ }\textbf {\bibinfo {volume} {48}},\ \bibinfo
  {pages} {285003} (\bibinfo {year} {2015})}\BibitemShut {NoStop}%
\bibitem [{\citenamefont {Ku\'smierz}\ and\ \citenamefont
  {Gudowska-Nowak}(2015)}]{KGN15}%
  \BibitemOpen
  \bibfield  {author} {\bibinfo {author} {\bibfnamefont {L.}~\bibnamefont
  {Ku\'smierz}}\ and\ \bibinfo {author} {\bibfnamefont {E.}~\bibnamefont
  {Gudowska-Nowak}},\ }\href@noop {} {\bibfield  {journal} {\bibinfo  {journal}
  {Phys. Rev. E}\ }\textbf {\bibinfo {volume} {92}},\ \bibinfo {pages} {052127}
  (\bibinfo {year} {2015})}\BibitemShut {NoStop}%
\bibitem [{\citenamefont {Nagar}\ and\ \citenamefont {Gupta}(2016)}]{NG16}%
  \BibitemOpen
  \bibfield  {author} {\bibinfo {author} {\bibfnamefont {A.}~\bibnamefont
  {Nagar}}\ and\ \bibinfo {author} {\bibfnamefont {S.}~\bibnamefont {Gupta}},\
  }\href@noop {} {\bibfield  {journal} {\bibinfo  {journal} {Phys. Rev. E}\
  }\textbf {\bibinfo {volume} {93}},\ \bibinfo {pages} {060102 (R)} (\bibinfo
  {year} {2016})}\BibitemShut {NoStop}%
\bibitem [{\citenamefont {Reuveni}(2016)}]{Reuveni16}%
  \BibitemOpen
  \bibfield  {author} {\bibinfo {author} {\bibfnamefont {S.}~\bibnamefont
  {Reuveni}},\ }\href@noop {} {\bibfield  {journal} {\bibinfo  {journal} {Phys.
  Rev. Lett.}\ }\textbf {\bibinfo {volume} {116}},\ \bibinfo {pages} {170601}
  (\bibinfo {year} {2016})}\BibitemShut {NoStop}%
\bibitem [{\citenamefont {Pal}\ and\ \citenamefont {Reuveni}(2017)}]{PR17}%
  \BibitemOpen
  \bibfield  {author} {\bibinfo {author} {\bibfnamefont {A.}~\bibnamefont
  {Pal}}\ and\ \bibinfo {author} {\bibfnamefont {S.}~\bibnamefont {Reuveni}},\
  }\href@noop {} {\bibfield  {journal} {\bibinfo  {journal} {Phys. Rev. Lett.}\
  }\textbf {\bibinfo {volume} {118}},\ \bibinfo {pages} {030603} (\bibinfo
  {year} {2017})}\BibitemShut {NoStop}%
\bibitem [{\citenamefont {Chechkin}\ and\ \citenamefont
  {Sokolov}(2018)}]{CS18}%
  \BibitemOpen
  \bibfield  {author} {\bibinfo {author} {\bibfnamefont {A.}~\bibnamefont
  {Chechkin}}\ and\ \bibinfo {author} {\bibfnamefont {I.~M.}\ \bibnamefont
  {Sokolov}},\ }\href@noop {} {\bibfield  {journal} {\bibinfo  {journal} {Phys.
  Rev. Lett.}\ }\textbf {\bibinfo {volume} {121}},\ \bibinfo {pages} {050601}
  (\bibinfo {year} {2018})}\BibitemShut {NoStop}%
\bibitem [{\citenamefont {Belan}(2018)}]{Belan18}%
  \BibitemOpen
  \bibfield  {author} {\bibinfo {author} {\bibfnamefont {S.}~\bibnamefont
  {Belan}},\ }\href@noop {} {\bibfield  {journal} {\bibinfo  {journal} {Phys.
  Rev. Lett.}\ }\textbf {\bibinfo {volume} {120}},\ \bibinfo {pages} {080601}
  (\bibinfo {year} {2018})}\BibitemShut {NoStop}%
\bibitem [{\citenamefont {Evans}\ and\ \citenamefont {Majumdar}(2018)}]{EM18}%
  \BibitemOpen
  \bibfield  {author} {\bibinfo {author} {\bibfnamefont {M.~R.}\ \bibnamefont
  {Evans}}\ and\ \bibinfo {author} {\bibfnamefont {S.~N.}\ \bibnamefont
  {Majumdar}},\ }\href@noop {} {\bibfield  {journal} {\bibinfo  {journal} {J.
  Phys. A: Math. Theor.}\ }\textbf {\bibinfo {volume} {51}},\ \bibinfo {pages}
  {475003} (\bibinfo {year} {2018})}\BibitemShut {NoStop}%
\bibitem [{\citenamefont {A.~Mas\'o-Puigdellosas}\ and\ \citenamefont
  {M\'endez}(2019)}]{MPCM19a}%
  \BibitemOpen
  \bibfield  {author} {\bibinfo {author} {\bibfnamefont {D.~C.}\ \bibnamefont
  {A.~Mas\'o-Puigdellosas}}\ and\ \bibinfo {author} {\bibfnamefont
  {V.}~\bibnamefont {M\'endez}},\ }\href@noop {} {\bibfield  {journal}
  {\bibinfo  {journal} {Phys. Rev. E.}\ }\textbf {\bibinfo {volume} {99}},\
  \bibinfo {pages} {012141} (\bibinfo {year} {2019})}\BibitemShut {NoStop}%
\bibitem [{\citenamefont {Masoliver}\ and\ \citenamefont
  {Montero}(2019)}]{MM19}%
  \BibitemOpen
  \bibfield  {author} {\bibinfo {author} {\bibfnamefont {J.}~\bibnamefont
  {Masoliver}}\ and\ \bibinfo {author} {\bibfnamefont {M.}~\bibnamefont
  {Montero}},\ }\href@noop {} {\bibfield  {journal} {\bibinfo  {journal} {Phys.
  Rev. E}\ }\textbf {\bibinfo {volume} {100}},\ \bibinfo {pages} {042103}
  (\bibinfo {year} {2019})}\BibitemShut {NoStop}%
\bibitem [{\citenamefont {Durang}\ \emph {et~al.}(2019)\citenamefont {Durang},
  \citenamefont {Lee}, \citenamefont {Lizana},\ and\ \citenamefont
  {Jeon}}]{DLLJ19}%
  \BibitemOpen
  \bibfield  {author} {\bibinfo {author} {\bibfnamefont {X.}~\bibnamefont
  {Durang}}, \bibinfo {author} {\bibfnamefont {S.}~\bibnamefont {Lee}},
  \bibinfo {author} {\bibfnamefont {L.}~\bibnamefont {Lizana}}, \ and\ \bibinfo
  {author} {\bibfnamefont {J.-H.}\ \bibnamefont {Jeon}},\ }\href@noop {}
  {\bibfield  {journal} {\bibinfo  {journal} {J. Phys. A: Math. Theor.}\
  }\textbf {\bibinfo {volume} {52}},\ \bibinfo {pages} {224001} (\bibinfo
  {year} {2019})}\BibitemShut {NoStop}%
\bibitem [{\citenamefont {Pal}\ and\ \citenamefont {Prasad}(2019)}]{PP19}%
  \BibitemOpen
  \bibfield  {author} {\bibinfo {author} {\bibfnamefont {A.}~\bibnamefont
  {Pal}}\ and\ \bibinfo {author} {\bibfnamefont {V.~V.}\ \bibnamefont
  {Prasad}},\ }\href@noop {} {\bibfield  {journal} {\bibinfo  {journal} {Phys.
  Rev. E}\ }\textbf {\bibinfo {volume} {99}},\ \bibinfo {pages} {032123}
  (\bibinfo {year} {2019})}\BibitemShut {NoStop}%
\bibitem [{\citenamefont {Tal-Friedman}\ \emph {et~al.}(2020)\citenamefont
  {Tal-Friedman}, \citenamefont {Pal}, \citenamefont {Sekhon}, \citenamefont
  {Reuveni},\ and\ \citenamefont {Roichman}}]{Tal2020}%
  \BibitemOpen
  \bibfield  {author} {\bibinfo {author} {\bibfnamefont {O.}~\bibnamefont
  {Tal-Friedman}}, \bibinfo {author} {\bibfnamefont {A.}~\bibnamefont {Pal}},
  \bibinfo {author} {\bibfnamefont {A.}~\bibnamefont {Sekhon}}, \bibinfo
  {author} {\bibfnamefont {S.}~\bibnamefont {Reuveni}}, \ and\ \bibinfo
  {author} {\bibfnamefont {Y.}~\bibnamefont {Roichman}},\ }\href@noop {}
  {\bibfield  {journal} {\bibinfo  {journal} {J. Phys. Chem. Lett.}\ }\textbf
  {\bibinfo {volume} {11}},\ \bibinfo {pages} {7350} (\bibinfo {year}
  {2020})}\BibitemShut {NoStop}%
\bibitem [{\citenamefont {Besga}\ \emph {et~al.}(2020)\citenamefont {Besga},
  \citenamefont {Bovon}, \citenamefont {Petrosyan}, \citenamefont {Majumdar},\
  and\ \citenamefont {Ciliberto}}]{us_1D_Phys_rech}%
  \BibitemOpen
  \bibfield  {author} {\bibinfo {author} {\bibfnamefont {B.}~\bibnamefont
  {Besga}}, \bibinfo {author} {\bibfnamefont {A.}~\bibnamefont {Bovon}},
  \bibinfo {author} {\bibfnamefont {A.}~\bibnamefont {Petrosyan}}, \bibinfo
  {author} {\bibfnamefont {S.~N.}\ \bibnamefont {Majumdar}}, \ and\ \bibinfo
  {author} {\bibfnamefont {S.}~\bibnamefont {Ciliberto}},\ }\href {\doibase
  10.1103/PhysRevResearch.2.032029} {\bibfield  {journal} {\bibinfo  {journal}
  {Phys. Rev. Res.}\ }\textbf {\bibinfo {volume} {2}},\ \bibinfo {pages}
  {032029} (\bibinfo {year} {2020})}\BibitemShut {NoStop}%
\bibitem [{\citenamefont {Besga}\ \emph {et~al.}()\citenamefont {Besga},
  \citenamefont {Faisant}, \citenamefont {Petrosyan}, \citenamefont
  {Ciliberto},\ and\ \citenamefont {Majumdar}}]{FP_2021}%
  \BibitemOpen
  \bibfield  {author} {\bibinfo {author} {\bibfnamefont {B.}~\bibnamefont
  {Besga}}, \bibinfo {author} {\bibfnamefont {F.}~\bibnamefont {Faisant}},
  \bibinfo {author} {\bibfnamefont {A.}~\bibnamefont {Petrosyan}}, \bibinfo
  {author} {\bibfnamefont {S.}~\bibnamefont {Ciliberto}}, \ and\ \bibinfo
  {author} {\bibfnamefont {S.~N.}\ \bibnamefont {Majumdar}},\ }\href@noop {}
  {\bibinfo  {journal} {arXiv preprint: 2102.07232}\ }\BibitemShut {NoStop}%
\bibitem [{\citenamefont {B\`erut}\ \emph {et~al.}(2016)\citenamefont
  {B\`erut}, \citenamefont {Imparato}, \citenamefont {Petrosyan},\ and\
  \citenamefont {Ciliberto}}]{Berut2016}%
  \BibitemOpen
\bibfield  {journal} {  }\bibfield  {author} {\bibinfo {author} {\bibfnamefont
  {A.}~\bibnamefont {B\`erut}}, \bibinfo {author} {\bibfnamefont
  {A.}~\bibnamefont {Imparato}}, \bibinfo {author} {\bibfnamefont
  {A.}~\bibnamefont {Petrosyan}}, \ and\ \bibinfo {author} {\bibfnamefont
  {S.}~\bibnamefont {Ciliberto}},\ }\href@noop {} {\bibfield  {journal}
  {\bibinfo  {journal} {J. Stat. Mech.}\ ,\ \bibinfo {pages} {P054002}}
  (\bibinfo {year} {2016})}\BibitemShut {NoStop}%
\bibitem [{\citenamefont {Evans}\ and\ \citenamefont
  {Majumdar}(2019)}]{EM2019}%
  \BibitemOpen
  \bibfield  {author} {\bibinfo {author} {\bibfnamefont {M.~R.}\ \bibnamefont
  {Evans}}\ and\ \bibinfo {author} {\bibfnamefont {S.}~\bibnamefont
  {Majumdar}},\ }\href@noop {} {\bibfield  {journal} {\bibinfo  {journal} {J.
  Phys. A: Math. Theor.}\ }\textbf {\bibinfo {volume} {52}},\ \bibinfo {pages}
  {01LT01} (\bibinfo {year} {2019})}\BibitemShut {NoStop}%
\bibitem [{\citenamefont {Pal}\ \emph {et~al.}(2019)\citenamefont {Pal},
  \citenamefont {Kusmierz},\ and\ \citenamefont {Reuveni}}]{PKR2019}%
  \BibitemOpen
  \bibfield  {author} {\bibinfo {author} {\bibfnamefont {A.}~\bibnamefont
  {Pal}}, \bibinfo {author} {\bibfnamefont {L.}~\bibnamefont {Kusmierz}}, \
  and\ \bibinfo {author} {\bibfnamefont {S.}~\bibnamefont {Reuveni}},\
  }\href@noop {} {\bibfield  {journal} {\bibinfo  {journal} {Phys. Rev. E}\
  }\textbf {\bibinfo {volume} {100}},\ \bibinfo {pages} {040101(R)} (\bibinfo
  {year} {2019})}\BibitemShut {NoStop}%
\bibitem [{\citenamefont {Gupta}\ \emph {et~al.}(2020)\citenamefont {Gupta},
  \citenamefont {Plata},\ and\ \citenamefont {Pal}}]{GPP2019}%
  \BibitemOpen
  \bibfield  {author} {\bibinfo {author} {\bibfnamefont {D.}~\bibnamefont
  {Gupta}}, \bibinfo {author} {\bibfnamefont {C.~A.}\ \bibnamefont {Plata}}, \
  and\ \bibinfo {author} {\bibfnamefont {A.}~\bibnamefont {Pal}},\ }\href@noop
  {} {\bibfield  {journal} {\bibinfo  {journal} {Phys. Rev. Lett.}\ }\textbf
  {\bibinfo {volume} {124}},\ \bibinfo {pages} {110608} (\bibinfo {year}
  {2020})}\BibitemShut {NoStop}%
\bibitem [{\citenamefont {Bodrova}\ and\ \citenamefont
  {Sokolov}(2020)}]{BS2020}%
  \BibitemOpen
  \bibfield  {author} {\bibinfo {author} {\bibfnamefont {A.~S.}\ \bibnamefont
  {Bodrova}}\ and\ \bibinfo {author} {\bibfnamefont {I.~M.}\ \bibnamefont
  {Sokolov}},\ }\href@noop {} {\bibfield  {journal} {\bibinfo  {journal} {Phys.
  Rev. E}\ }\textbf {\bibinfo {volume} {101}},\ \bibinfo {pages} {052130}
  (\bibinfo {year} {2020})}\BibitemShut {NoStop}%
\bibitem [{\citenamefont {Mercado-V\'asquez}\ \emph {et~al.}(2020)\citenamefont
  {Mercado-V\'asquez}, \citenamefont {Boyer}, \citenamefont {Majumdar},\ and\
  \citenamefont {Schehr}}]{MBMS20}%
  \BibitemOpen
  \bibfield  {author} {\bibinfo {author} {\bibfnamefont {G.}~\bibnamefont
  {Mercado-V\'asquez}}, \bibinfo {author} {\bibfnamefont {D.}~\bibnamefont
  {Boyer}}, \bibinfo {author} {\bibfnamefont {S.~N.}\ \bibnamefont {Majumdar}},
  \ and\ \bibinfo {author} {\bibfnamefont {G.}~\bibnamefont {Schehr}},\
  }\href@noop {} {\bibfield  {journal} {\bibinfo  {journal} {J. Stat. Mech.}\
  ,\ \bibinfo {pages} {113203}} (\bibinfo {year} {2020})}\BibitemShut {NoStop}%
\bibitem [{\citenamefont {Gupta}\ \emph {et~al.}(2021)\citenamefont {Gupta},
  \citenamefont {Plata}, \citenamefont {Kundu},\ and\ \citenamefont
  {Pal}}]{GPKP21}%
  \BibitemOpen
  \bibfield  {author} {\bibinfo {author} {\bibfnamefont {D.}~\bibnamefont
  {Gupta}}, \bibinfo {author} {\bibfnamefont {C.~A.}\ \bibnamefont {Plata}},
  \bibinfo {author} {\bibfnamefont {A.}~\bibnamefont {Kundu}}, \ and\ \bibinfo
  {author} {\bibfnamefont {A.}~\bibnamefont {Pal}},\ }\href@noop {} {\bibfield
  {journal} {\bibinfo  {journal} {J. Phys. A: Math. Theor.}\ }\textbf {\bibinfo
  {volume} {54}},\ \bibinfo {pages} {025003} (\bibinfo {year}
  {2021})}\BibitemShut {NoStop}%
\bibitem [{\citenamefont {Mart{\'\i}nez}\ \emph {et~al.}(2016)\citenamefont
  {Mart{\'\i}nez}, \citenamefont {Petrosyan}, \citenamefont {Gu{\'e}ry-Odelin},
  \citenamefont {Trizac},\ and\ \citenamefont {Ciliberto}}]{ESE1}%
  \BibitemOpen
  \bibfield  {author} {\bibinfo {author} {\bibfnamefont {I.~A.}\ \bibnamefont
  {Mart{\'\i}nez}}, \bibinfo {author} {\bibfnamefont {A.}~\bibnamefont
  {Petrosyan}}, \bibinfo {author} {\bibfnamefont {D.}~\bibnamefont
  {Gu{\'e}ry-Odelin}}, \bibinfo {author} {\bibfnamefont {E.}~\bibnamefont
  {Trizac}}, \ and\ \bibinfo {author} {\bibfnamefont {S.}~\bibnamefont
  {Ciliberto}},\ }\href@noop {} {\bibfield  {journal} {\bibinfo  {journal}
  {Nature physics}\ }\textbf {\bibinfo {volume} {12}},\ \bibinfo {pages} {843}
  (\bibinfo {year} {2016})}\BibitemShut {NoStop}%
\bibitem [{\citenamefont {Chupeau}\ \emph {et~al.}(2018)\citenamefont
  {Chupeau}, \citenamefont {Besga}, \citenamefont {Guery-Odelin}, \citenamefont
  {Trizac}, \citenamefont {Petrosyan},\ and\ \citenamefont {Ciliberto}}]{ESE2}%
  \BibitemOpen
  \bibfield  {author} {\bibinfo {author} {\bibfnamefont {M.}~\bibnamefont
  {Chupeau}}, \bibinfo {author} {\bibfnamefont {B.}~\bibnamefont {Besga}},
  \bibinfo {author} {\bibfnamefont {D.}~\bibnamefont {Guery-Odelin}}, \bibinfo
  {author} {\bibfnamefont {E.}~\bibnamefont {Trizac}}, \bibinfo {author}
  {\bibfnamefont {A.}~\bibnamefont {Petrosyan}}, \ and\ \bibinfo {author}
  {\bibfnamefont {S.}~\bibnamefont {Ciliberto}},\ }\href@noop {} {\bibfield
  {journal} {\bibinfo  {journal} {Phys. Rev. E}\ }\textbf {\bibinfo {volume}
  {98}},\ \bibinfo {pages} {010104(R)} (\bibinfo {year} {2018})}\BibitemShut
  {NoStop}%
\bibitem [{\citenamefont {Guery-Odelin}\ \emph {et~al.}(2019)\citenamefont
  {Guery-Odelin}, \citenamefont {Ruschhaupt}, \citenamefont {A.~Kiely},
  \citenamefont {Martinez-Garaot},\ and\ \citenamefont {Muga}}]{RMP19}%
  \BibitemOpen
  \bibfield  {author} {\bibinfo {author} {\bibfnamefont {D.}~\bibnamefont
  {Guery-Odelin}}, \bibinfo {author} {\bibfnamefont {A.}~\bibnamefont
  {Ruschhaupt}}, \bibinfo {author} {\bibfnamefont {E.~T.}\ \bibnamefont
  {A.~Kiely}}, \bibinfo {author} {\bibfnamefont {S.}~\bibnamefont
  {Martinez-Garaot}}, \ and\ \bibinfo {author} {\bibfnamefont {J.~G.}\
  \bibnamefont {Muga}},\ }\href@noop {} {\bibfield  {journal} {\bibinfo
  {journal} {Rev. Mod. Phys.}\ }\textbf {\bibinfo {volume} {91}},\ \bibinfo
  {pages} {045001} (\bibinfo {year} {2019})}\BibitemShut {NoStop}%
\bibitem [{\citenamefont {C.~A.~Plata}\ and\ \citenamefont
  {Prados}(2019)}]{PGTP19}%
  \BibitemOpen
  \bibfield  {author} {\bibinfo {author} {\bibfnamefont {E.~T.}\ \bibnamefont
  {C.~A.~Plata}, \bibfnamefont {D.~Guery-Odelin}}\ and\ \bibinfo {author}
  {\bibfnamefont {A.}~\bibnamefont {Prados}},\ }\href@noop {} {\bibfield
  {journal} {\bibinfo  {journal} {Physical Review E}\ }\textbf {\bibinfo
  {volume} {99}},\ \bibinfo {pages} {01240} (\bibinfo {year}
  {2019})}\BibitemShut {NoStop}%
\bibitem [{\citenamefont {Majumdar}(1999)}]{pers_current}%
  \BibitemOpen
  \bibfield  {author} {\bibinfo {author} {\bibfnamefont {S.~N.}\ \bibnamefont
  {Majumdar}},\ }\href@noop {} {\bibfield  {journal} {\bibinfo  {journal}
  {Curr. Sci.}\ }\textbf {\bibinfo {volume} {77}},\ \bibinfo {pages} {370}
  (\bibinfo {year} {1999})}\BibitemShut {NoStop}%
\bibitem [{\citenamefont {Majumdar}(2010)}]{Louven_review}%
  \BibitemOpen
  \bibfield  {author} {\bibinfo {author} {\bibfnamefont {S.~N.}\ \bibnamefont
  {Majumdar}},\ }\href@noop {} {\bibfield  {journal} {\bibinfo  {journal}
  {Physica A}\ }\textbf {\bibinfo {volume} {389}},\ \bibinfo {pages} {4299}
  (\bibinfo {year} {2010})}\BibitemShut {NoStop}%
\bibitem [{\citenamefont {Barzykin}\ and\ \citenamefont
  {Tachiya}(1993)}]{BT1993}%
  \BibitemOpen
  \bibfield  {author} {\bibinfo {author} {\bibfnamefont {A.~V.}\ \bibnamefont
  {Barzykin}}\ and\ \bibinfo {author} {\bibfnamefont {M.}~\bibnamefont
  {Tachiya}},\ }\href@noop {} {\bibfield  {journal} {\bibinfo  {journal} {J.
  Chem. Phys.}\ }\textbf {\bibinfo {volume} {99}},\ \bibinfo {pages} {9591}
  (\bibinfo {year} {1993})}\BibitemShut {NoStop}%
\bibitem [{\citenamefont {Blythe}\ and\ \citenamefont {Bray}(2003)}]{BB2003}%
  \BibitemOpen
  \bibfield  {author} {\bibinfo {author} {\bibfnamefont {R.~A.}\ \bibnamefont
  {Blythe}}\ and\ \bibinfo {author} {\bibfnamefont {A.~J.}\ \bibnamefont
  {Bray}},\ }\href@noop {} {\bibfield  {journal} {\bibinfo  {journal} {Phys.
  Rev. E}\ }\textbf {\bibinfo {volume} {67}},\ \bibinfo {pages} {041101}
  (\bibinfo {year} {2003})}\BibitemShut {NoStop}%
\bibitem [{\citenamefont {Gradshteyn}\ and\ \citenamefont {Ryzhik}(1965)}]{GR}%
  \BibitemOpen
  \bibfield  {author} {\bibinfo {author} {\bibfnamefont {I.~S.}\ \bibnamefont
  {Gradshteyn}}\ and\ \bibinfo {author} {\bibfnamefont {I.~M.}\ \bibnamefont
  {Ryzhik}},\ }\href@noop {} {\emph {\bibinfo {title} {Table of integrals,
  series, and products}}}\ (\bibinfo  {publisher} {Academic press},\ \bibinfo
  {year} {1965})\BibitemShut {NoStop}%
\end{thebibliography}%

\newpage
\begin{widetext}
\appendix

\section{Preliminaries} \label{apx:A}

In this Appendix and in Appendices \ref{apx:B} and \ref{apx:C} we summarize, for the 1D case, 
how the theoretical predictions 
plotted in Figs. 2), 3) and 4) of the main text have been obtained, both for periodic and random 
resetting protocols.  Much of these 1D results can be found in Supplemental Material 
of \cite{us_1D_Phys_rech}, but we prefer to recall them here 
to have a full description of the resetting protocol in 1D and 2D. 
The theoretical results for 2D are entirely new and are presented in Appendices \ref{apx:D_2d}
and \ref{apx:E_2d} respectively.

As already pointed out in the main text we consider here a realistic situation 
in which the initial position at the begining of each free diffusion period (between successive 
resetting events)
has a Gaussian distribution of width $\sigma$.  The mean first-passage time (MFPT),
in the presence of resetting for either of the two protocols, was previously computed
only for the restting to a fixed initial position $x_0$~\cite{EM2011_1,EM2011_2,PKE16,BBR16,reset_review}. 
In practice this is not possible to realize experimentally 
because of the finite stiffness $\kappa$ of the trap which resets the particle to the initial position. 
In an optical trap at temperature $T$, the resetting position is always Gaussian distributed with a finite 
variance $\sigma^2=k_BT/\kappa$ given by equipartition. As $\kappa$ is 
proportional to the laser intensity starting with $\sigma=0$ would imply to trap with an infinite power 
which is of course not possible. Thus we study here the influence of a finite nonzero $\sigma$ on MFPT.

Consider a searcher undergoing a generic stochastic dynamics starting, say, at the initial position $x_0$.
The immobile target is located at $x=L$.
To compute the MFPT of a generic stochastic process, it is most convenient to first compute the survival probability or 
persistence~\cite{pers_current,Redner,Louven_review,pers_review} $S(t|x_0)$, starting from $x_0$.
This is simply the probability
that the searcher, starting its dynamics at $x_0$ at $t=0$, does not find the target up to time $t$. 
The first-passage probability density
$F(t|x_0)$ denotes the probability density to find the target for the first time at $t$.
The two quantities $F(t|x_0)$ and $S(t|x_0)$ are simply related to each other via $S(t|x_0)=\int_t^{\infty} F(t'|x_0)\, d{t'}$, because if the target is not found up to $t$, the first-passage time must occur after $t$. 
Taking a derivative with respect to $t$
gives
{
	\begin{equation}
	F(t|x_0)= - \dv{S(t|x_0)}{t} \, .
	\label{FQ.1}
	\end{equation}
Hence, if we can compute $S(t|x_0)$ (which is often easier to compute), we obtain $F(t|x_0)$ simply from Eq. (\ref{FQ.1}).}
Once we know $F(t|x_0)$, the MFPT is just its first moment
\begin{equation}
\langle t_f\rangle (x_0)= \int_0^{\infty} t\, F(t|x_0) \dd{t} = \int_0^{\infty} S(t|x_0) \dd{t}\, ,
\label{MFPT.0}
\end{equation}
where in arriving at the last equality, we substituted Eq. (\ref{FQ.1}) and did integration by parts. 
Below, we will first compute the survival probability $S(t|x_0)$ for fixed $x_0$
and then average over the distribution of $x_0$. We will consider the two protocols separately.

\section{Protocol-1: Periodic Resetting to a random initial position }\label{apx:B}

We consider an overdamped diffusing particle that starts at an initial
position $x_0$, which is drawn from a distribution ${\cal P}(x_0)$.
The particle diffuses for a fixed period $T$ and then its position is
instantaneously reset to a new position $z$, also drawn from the same
distribution ${\cal P}(z)$. Then the particle diffuses again for a period 
$T$, followed by a reset to a new position $z'$ drawn from ${\cal P}(z)$
and the process continues. We assume that after each resetting, the reset 
position $z$ is drawn {\rm independently} from cycle to cyle from the
same distribution ${\cal P}(z)$. We also have a fixed target at a
location $L$. For fixed $L$, $T$ and ${\cal P}(z)$, We want to 
first compute the mean first-passage time
$\langle t_f\rangle$ to find the target and then optimize (minimize) this
quantity with respect to $T$ (for fixed $L$ and ${\cal P}(z)$).
We first compute the MFPT for arbitrary ${\cal P}(z)$ and then focus on the
experimentally relevant Gaussian case.

We first recall that for a free diffusing particle, starting at an initial 
position $x_0$, the survival probability that it stays below the level $L$ up to 
time $t$ is given by~\cite{pers_current,Redner,Louven_review,pers_review}
\begin{equation}
S(t|x_0)= \erf\left(\frac{|L-x_0|}{\sqrt{4Dt}}\right)\, ,
\label{surv_diff.1}
\end{equation}
where $\erf(z)= (2/\sqrt{\pi})\, \int_0^{z} e^{-u^2} \dd{u}$ is the
error function and $D$ is the diffusion constant. If the initial
position is chosen from a distribution ${\cal P}(x_0)$, then
the survival probability, averaged over the initial position, is given by
\begin{equation}
Q_1(t)= \int_{-\infty}^{\infty} \dd{x_0} {\cal P}(x_0)\, S(t|x_0)=
\int_{-\infty}^{\infty} \dd{x_0} {\cal P}(x_0)\, {\rm 
	erf}\left(\frac{|L-x_0|}{\sqrt{4Dt}}\right)\, .
\label{surv_diff.2}
\end{equation}

Now, consider our protocol. Let us compute the survival probability  
$Q(t)$ up to time $t$, averaged over the distribution of the starting position at the begining of each cycle.
Then it is easy to see the following.
\begin{itemize}
	
	\item[{\bf a)}] $0<t\le T$: If the measurement time $t$ lies in the first cycle of 
	diffusion, then the survival probability upto time $t$ is simply
	$Q_1(t)$ given in Eq. (\ref{surv_diff.2}), since no resetting has taken 
	place up to $t$ yet. Note that at the end of the period $[0,T]$, the
	survival probability is simply $Q_1(T)$.
	
	\item[{\bf b)}] $T<t\le 2T$: In this case, the particle has to first survive the 
	period 
	$[0,T]$ with free diffusion: this occurs with probability $Q_1(T)$.
	Then from time $T$ till $T<t\le 2T$, it also undergoes a free diffusion 
	but starting from a new reset position $z$ drawn from ${\cal P}(z)$.
	Hence, the survival probability up to $T<t\le 2T$ is given by the product
	of these two events
	\begin{equation}
	Q_2(t)= Q_1(T)\, \int_{-\infty}^{\infty} \dd{z} {\cal P}(z)\, {\rm
		erf}\left(\frac{|L-z|}{\sqrt{4D(t-T)}}\right)\, .
	\label{cycle.2}
	\end{equation}
	Note that at the end of the cycle
	\begin{equation}
	Q_2(2T)= Q_1^2(T)
	\label{cycle_end.2}
	\end{equation}
	where $Q_1(T)$ is given by Eq. (\ref{surv_diff.2}).

	\item[{\bf c)}] $2T<t\le 3T$: By repeating the above argument
	\begin{equation}
	Q_3(t)= Q_1(2T)\, \int_{-\infty}^{\infty} \dd{z} {\cal P}(z)\, {\rm
		erf}\left(\frac{|L-z|}{\sqrt{4D(t-2T)}}\right)
	= Q_1^2(T)\, \int_{-\infty}^{\infty} \dd{z} {\cal P}(z)\, {\rm
		erf}\left(\frac{|L-z|}{\sqrt{4D(t-2T)}}\right)
	\label{cycle.3}
	\end{equation}
	At the end of the 3rd cycle
	\begin{equation}
	Q_3(3T)= Q_1^3(T)
	\label{cycle_end.3}
	\end{equation}
	where $Q_1(T)$ is given by Eq. (\ref{surv_diff.2}).
	
	\item[{\bf d)}] $(n-1)T<t\le n\,T$: For the $n$-th cycle, we have then
	\begin{equation}
	Q_n(t)= \left[Q_1(T)\right]^{n-1}\, 
	\int_{-\infty}^{\infty} \dd{z} {\cal P}(z)\, {\rm
		erf}\left(\frac{|L-z|}{\sqrt{4D(t- (n-1)\,T)}}\right)    
	\label{cycle.n}
	\end{equation}
	
\end{itemize}
\vskip 0.5cm

{Hence, the survival probability $Q(t)$ is just $Q_n(t)$ if $t$ belongs to the $n$-th period, i.e.,
	if $(n-1)T<t\le n\,T$, where $n=1,\,2,\,\ldots$. In other words, for fixed $t$, we need to find
	the cycle number $n$ associated to $t$ and then use the formula for $Q_n(t)$ in \eqref{cycle.n}.
	Mathematically speaking
	\begin{equation}
	Q(t)= \sum_{n=1}^{\infty} Q_n(t)\, \operatorname{I}_{ (n-1)T< t\le n\, T}
	\label{Qt_def}
	\end{equation}
	where the indicator function $\operatorname{I}_{A}=1$ if the clause $A$ in the subscript
	is satisfied and is zero otherwise.}

The mean first-passage time $\langle t_f\rangle$ to location $L$, average over the distribution of $x_0$, is then
given from Eq. (\ref{MFPT.0}) as
\begin{equation}
\langle t_f\rangle = \int_0^{\infty} \dd{t} Q(t)
\label{mfpt.1}
\end{equation}
where $Q(t)$ is the survival probability up to time $t$ in \eqref{Qt_def}. Using the
results above for different cycles, we can evaluate this integral
in Eq. (\ref{mfpt.1}) by dividing the time integral over different cycles.
This gives
\begin{eqnarray}
\langle t_f\rangle = \sum_{n=1}^{\infty} \int_{(n-1)T}^{nT} \dd{t} Q_n(t)
&= & \sum_{n=1}^{\infty} \left[Q_1(T)\right]^{n-1}\,
\int_{(n-1)T}^{nT} \dd{t} \int_{-\infty}^{\infty} \dd{z} {\cal P}(z)\, 
\erf\left(\frac{|L-z|}{\sqrt{4D(t- (n-1)\,T)}}\right) \nonumber \\
&=& \sum_{n=1}^{\infty} \left[Q_1(T)\right]^{n-1}\, 
\int_0^T \dd{\Delta}
\int_{-\infty}^{\infty} \dd{z} {\cal P}(z)\,
\erf\left(\frac{|L-z|}{\sqrt{4D\,\Delta}}\right)\, ,
\label{mfpt.2}
\end{eqnarray}
where in the last line we made a change of variable $\tau= t-(n-1)T$.
The sum on the right hand side (rhs) of Eq. (\ref{mfpt.2}) can be
easily performed as a geometric series. This gives 
\begin{equation}
\langle t_f\rangle = \frac{ \int_0^T \dd{\tau}
	\int_{-\infty}^{\infty} \dd{z} {\cal P}(z)\,
	\erf\left(\frac{|L-z|}{\sqrt{4D\,\tau}}\right)}{1-Q_1(T)}\, .
\label{mfpt.3}
\end{equation} 
The denominator can be further simplified as
\begin{equation}
1-Q_1(T)= 1- \int_{-\infty}^{\infty} \dd{z} {\cal P}(z)\, {\rm
	erf}\left(\frac{|L-z|}{\sqrt{4DT}}\right)
= \int_{-\infty}^{\infty} \dd{z} {\cal P}(z)\, {\rm
	erfc}\left(\frac{|L-z|}{\sqrt{4DT}}\right)
\label{denom.1d}
\end{equation}
where $\erfc(z)= 1-\erf(z)= (2/\sqrt{\pi})\int_z^{\infty} 
e^{-u^2} \dd{u}$ denotes the complementary error function. Note that
we have used the normalization: $\int_{-\infty}^{\infty} {\cal 
	P}(z) \dd{z}=1$. Substituting the result of Eq. \eqref{denom.1d} in
Eq. (\ref{mfpt.3}) we
get our final formula, valid for arbitrary reset/initial 
distribution ${\cal P}(z)$
\begin{equation}
\langle t_f\rangle = \frac{\int_0^T \dd{\tau}
	\int_{-\infty}^{\infty} \dd{z} {\cal P}(z)\,
	\erf\left(\frac{|L-z|}{\sqrt{4D\,\tau}}\right)}{\int_{-\infty}^{\infty} 
	\dd{z} {\cal P}(z)\, \erfc\left(\frac{|L-z|}{\sqrt{4DT}}\right)}\, .
\label{mfpt_final}
\end{equation}
The full first-passage probability density $F(t)$, averaged over the distribution of $x_0$, is then given by
\begin{equation}
F(t)= -\dv{Q}{t}\, ,
\label{FP.2}
\end{equation}
where $Q(t)$ is given in Eq. (\ref{Qt_def}).

%\vskip 0.3cm 

\subsection{ Reset to a Gaussian initial distribution} \label{sec:gaus_periodic}

As an example, let us 
consider the Gaussian distribution for the reset/initial position
\begin{equation}
{\cal P}(x_0) = \frac{1}{\sqrt{2\,\pi\,\sigma^2}}\, e^{- {x_0}^2/{2\sigma^2}}
\label{gauss.1}
\end{equation}
where $\sigma$ denotes the width. In this case, the survival probability $Q_1(T)$ at the end of the first cycle,
is given from Eq. (\ref{surv_diff.2}) upon setting $t=T$
\begin{equation}
Q_1(T)= \int_{-\infty}^{\infty} \frac{\dd{x_0}}{\sigma\, \sqrt{2\pi}}\, e^{-{x_0}^2/{2\sigma^2}}\, 
\erf\left(\frac{L-x_0}{\sqrt{4\, D\, T}}\right)\, .
\label{Q1T_Gauss.1}
\end{equation}
Let us introduce two dimensionless constants $b$ and $c$
\begin{equation}
b= \frac{L}{\sigma}\, ; \quad\quad\quad\quad c=
\frac{L}{\sqrt{4DT}}\, .
\label{bc.1}
\end{equation}
Then in terms of these two constants, Eq. (\ref{Q1T_Gauss.1}), after suitable rescaling, can be simplified to 
\begin{equation}
Q_1(T)= \frac{b}{\sqrt{2\pi}}\, \int_{-\infty}^{\infty} \dd{y} e^{-b^2\, y^2/2}\, \erf\left(c\, |1-y|\right)\, .
\label{Q1T_Gauss.2}
\end{equation}
Note that in the limit of $b\to \infty$, i.e., when $L\ll \sigma$ (corresponding to resetting to
the origin) one gets 
\begin{equation}
Q_1(T)\Big |_{b\to \infty}= \erf(c) \, ,
\label{Q_1.blarge.1}
\end{equation}
where we used $ (b/\sqrt{2\pi})\, e^{-b^2\, y^2/2} \to \delta(y)$ when $b\to \infty$. 

\subsubsection{ Mean first-passage time}
Substituting the Gaussian 
${\cal P}(z)$ in Eq. (\ref{mfpt_final}), and rescaling $\tau= v\, T$ we 
can write everything in dimensionless form
\begin{equation}
\tau=\frac{4D\langle t_f\rangle}{L^2} = \frac{\int_0^1 \dd{v}
	\int_{-\infty}^{\infty} \dd{u} e^{-u^2/2}\,
	{\rm
		erf}\left(   
	\frac{c}{\sqrt{v}}\,|1-u/b|
	\right)}{c^2\, \int_{-\infty}^{\infty} \dd{u} 
	e^{-u^2/2}\, 
	\erfc\left(c\,|1-u/b|\right)}\, \equiv w(b,c)
\label{mfpt_gauss.1}
\end{equation}
where the dimensionless constants $b$ and $c$ are given in Eq. (\ref{bc.1}).
It is hard to obtain a more explicit expression for the function $w(b,c)$ in
Eq. (\ref{mfpt_gauss.1}). But it can be easily evaluated numerically.

To make further analytical progress, we first consider the limit 
$\sigma\to 0$, i.e., $b=L/\sigma\to \infty$. 
In this limit, the reset distribution
${\cal P}(z)\to \delta(z)$. Hence, Eq. (\ref{mfpt_final}) 
or equivalently Eq. (\ref{mfpt_gauss.1}) simplifies 
considerably and the integrals can be evaluated
explicitly. We obtain an exact 
expression
\begin{equation}
\tau= \frac{4D\langle t_f\rangle}{L^2} = \frac{{\rm 
		erf}(c)+2c\left(e^{-c^2}/\sqrt{\pi}-c\, \erfc(c)\right)}{c^2\, 
	\erfc(c)}\equiv w(b\to \infty, c)\equiv w(c)
\label{delta_reset.1}
\end{equation}
where we recall $c= L/\sqrt{4DT}$. In Fig. 2 of the main text, we plot the 
function 
$w(c)$
vs. c, which  has a  minimum at 
\begin{equation}
c_{\rm opt}= 0.738412...
\label{copt.1}
\end{equation}
At this optimal value, $\tau_{\rm opt}= w(c_{\rm opt})= 5.34354\ldots$.
Hence, the optimal mean first-passage time to find the target located at 
$L$ is given by
\begin{equation}
\langle t_f\rangle_{\rm opt}= (5.34354\ldots)\, \frac{L^2}{4D}\, .
\label{tfopt.1}
\end{equation}
Note that this result is true in the limit $b=L/\sigma\to \infty$, i.e., 
when the target is very far away from the starting/resetting position.
In this limit, our result coincides with Ref. \cite{PKE16} where 
the authors studied
periodic resetting to the fixed initial position $x_0=0$ by a different method.
This is expected since the limit $b=L/\sigma \to \infty$ limit
is equivalent to $\sigma\to 0$ (with fixed $L$) and one would expect to
recover the fixed initial position results.

But the most interesting and unexpected result occurs for finite $b$.
In this case we can evaluate the rhs of Eq. 
(\ref{mfpt_gauss.1}) numerically. This has been done to plot the continuous lines 
in Fig. 3 of the main text, which shows the existence of a metastable minimum 
for $b>b_c\simeq 2.3$

\subsubsection{Full first-passage probability density \texorpdfstring{$F(t)$}{F(t)}}

The first-passage probability density can be computed from Eq. (\ref{FP.2}) and is plotted in Fig. 2 of the main text.
For a finite $b=L/\sigma$ we see those spectacular spikes at the beginning of each cycle. 
In this subsection, we analyse the origin of these spikes. For this let us analyse the survival probability
$Q_n(t)$ in Eq. (\ref{cycle.n}) close to the epoch $n\, T$, i.e., when the $n$-th period ends.
We set $t= n\, T+\Delta$ where $\Delta$ is small. For $\Delta>0$, we are in the $(n+1)$-th cycle, while for $\Delta<0$
we are in the $n$-th cycle. We consider the $\Delta>0$ and $\Delta<0$ cases separately.

\vskip 0.3cm

{\noindent {\bf {The case $\Delta>0$:}}} Since $t= n\, T+\Delta$ with $\Delta>0$ small,
$t$ now belongs to the $(n+1)$-th cycle, so we replace $n$ by $n+1$ in Eq. (\ref{cycle.n}) and get
\begin{equation}
Q_{n+1}(t= n\, T+ \Delta)= \left[Q_1(T)\right]^n\, \int_{-\infty}^{\infty} 
\frac{\dd{z}}{\sigma\, \sqrt{2\pi}}\, e^{-z^2/{2\sigma^2}}\,
\erf\left(\frac{|L-z|}{\sqrt{4\, D\, \Delta}}\right)\, ,
\label{right_cycle_n.1}
\end{equation}
where $Q_1(T)$ is given in Eq. (\ref{Q1T_Gauss.2}). It is convenient first to use the relation
$\erf(z)= 1-\erfc(z)$ and rewrite this as
\begin{equation}
Q_{n+1}(t= n\, T+ \Delta)= \left[Q_1(T)\right]^n\, \left[1- \int_{-\infty}^{\infty}
\frac{\dd{z}}{\sigma\, \sqrt{2\,\pi}}\, e^{-z^2/{2\sigma^2}}\,
\erfc\left(\frac{|L-z|}{\sqrt{4\, D\, \Delta}}\right)\,\right]\,  
\label{right_cycle_n.2}
\end{equation}
To derive the small $\Delta$ asymptotics, we make a change of variable $(L-z)/\sqrt{4\,D\,\Delta}=y$ on the rhs
of Eq. (\ref{right_cycle_n.2}). This leads to, using $b=L/\sigma$,
\begin{equation}
Q_{n+1}(t= n\, T+ \Delta)= \left[Q_1(T)\right]^n\, \left[1-\frac{\sqrt{4\,D\, \Delta}}{L}\, \frac{b}{\sqrt{2\,\pi}}\,
\int_{-\infty}^{\infty} \dd{y} e^{-\frac{b^2}{2}\, (1- \sqrt{4\,D\,\Delta}\, y/L)^2}\, \erfc(|y|)\, \right]\, .
\label{right_cycle_n.3}
\end{equation}
We can now make the limit $\Delta \to 0$. To leading order in $\Delta$, we can set $\Delta=0$ inside the exponential
in the integrand on the rhs and using $\int_{-\infty}^{\infty} \dd{y} \erfc(|y|)=2/\sqrt{\pi}$ we get
\begin{equation}
Q_{n+1}(t= n\, T+ \Delta)\simeq \left[Q_1(T)\right]^n\, \left[1- A_{+}(b)\, \sqrt{\Delta}\, \right]\, , \quad\quad
A_+(b)= \sqrt{\frac{8\,D}{\pi^2\, L^2}}\, b\, e^{-b^2/2}\, . 
\label{right_cycle_n.4}
\end{equation}
Taking derivative with respect to time, i.e., with respect $\Delta$, we find that as $\Delta\to 0^+$,
the first-passage probability denity $F(t)=-dQ/dt$ diverges for any finite $b$ as 
\begin{equation}
F(t= n\, T+ \Delta) \simeq \frac{A_n(b)}{ \sqrt{\Delta}}
\label{fp_n+.1}
\end{equation}
where the amplitude $A_n(b)$ of the inverse square root divergence is given by the exact formula
\begin{equation}
A_n(b)= \sqrt{\frac{2D}{\pi^2\, L^2}}\, b\, e^{-b^2/2}\, \left[Q_1(T)\right]^{n}\, 
\label{Ab.1}
\end{equation}
where $Q_1(T)$ (which also depends on $b$) is given in Eq. (\ref{Q1T_Gauss.2}). This inverse square root 
divergence at the
begining of any cycle, for any finite b,  describes the spikes in Fig. 2 of the main text. 
Note that the amplitude $A_n(b)$
of the $n$-th spike in Eq. (\ref{Ab.1}) decreases exponentially with $n$, as seen also in Fig. 2 of the main text.

Interestingly, we see from Eq. (\ref{Ab.1}) that the amplitudes of the spikes
decrease extremely fast as $b$ increases and the spikes disappear in the $b\to \infty$ limit.
Since $b=L/\sigma$, we see that the limit $b\to \infty$ corresponds to $\sigma\to 0$ limit (for fixed $L$).
This is indeed the case where one resets always to the origin. In fact, if we first take the limit $b\to \infty$
keeping $\Delta$ fixed and then take the limit $\Delta\to 0$, we get a very different reesult. Taking $b\to \infty$
limit first in Eq. (\ref{right_cycle_n.3}) we get, 
\begin{equation}
Q_{n+1}(t= n\, T+ \Delta)\Big|_{b\to \infty}= \left[Q_1(T)\right]^n 
\left[1- \erfc\left(\frac{L}{\sqrt{4\,D\,\Delta}}\right)\right]\, ,
\label{Q+blarge_n.1}
\end{equation}
where $Q_1(T)$ now is given in Eq. (\ref{Q_1.blarge.1}). Taking a derivative with respect $t$, i.e.,
with respect to $\Delta$, 
we get the first-passage probability density
\begin{equation}
F(t= n\, T+ \Delta)\Big|_{b\to \infty}= \left[Q_1(T)\right]^n\, \frac{L}{\sqrt{4\, \pi\, D\, \Delta^3}}\, 
e^{-L^2/(4\,D\,\Delta)}\, .
\label{fp+blarge_n.1}
\end{equation}
Thus, in the $b\to \infty$ limit, the first pasage probability actually vanishes extremely rapidly as $\Delta\to 0$
due to the essential singular term $\sim e^{-L^2/(4\, D \, \Delta)}$.

To summarize, the behavior of the first-passage probability density at the begining of the $n$-th 
period is strikingly different for $\sigma=0$ ($b\to \infty)$
and $\sigma>0$ ($b$ finite): in the former case it vanishes extremely rapidly as $t\to n\, T$ from above, 
while in the latter case it diverges as an inverse sqaure root leading to a spike at the begining of each period.
Thus the occurrence of spikes for $\sigma>0$ is a very clear signature of the finiteness of $\sigma$.
Physically, a spike occurs because if $\sigma$ is finite, immedaitely after each resetting the searcher
may be very close to the target and has a finite probability of finding the target immediately without further
diffusion.

\vskip 0.3cm

{\noindent {\bf {The case $\Delta<0$:}}} When $t= n\, T+ \Delta$ with $\Delta<0$, the effects are not so prominent
as in the $\Delta>0$ case. For $\Delta<0$, since $t$ belongs to the $n$-th branch, we consider the formula
in Eq. (\ref{cycle.n}) with $n$ and get
\begin{equation}
Q_{n}(t= n\, T+ \Delta)= \left[Q_1(T)\right]^{n-1}\, \int_{-\infty}^{\infty}
\frac{\dd{z}}{\sigma\, \sqrt{2\pi}}\, e^{-z^2/{2\sigma^2}}\,
\erf\left(\frac{|L-z|}{\sqrt{4\, D\, (T+\Delta)}}\right)\, .
\label{left_cycle_n.1}
\end{equation}
Now taking the $\Delta\to 0$ limit in Eq. (\ref{left_cycle_n.1}) is straightforward since there is no singular
behavior. Just Taylor expanding the error function for small $\Delta$ up to $O(\Delta)$, and making the
change of variable $(L-z)/\sqrt{4\, D\, T}=y$,  we obtain
\begin{equation}
Q_{n}(t= n\, T+ \Delta)= \left[Q_1(T)\right]^{n} - \Delta\, B_n(b) + O(\Delta^2)\, ,
\label{left_cycle_n.2}
\end{equation}
where the constant
\begin{equation}
B_n(b)= \frac{b \,\left[Q_1(T)\right]^{n-1}}{c\, T\, \pi\, \sqrt{2}}\,
\int_{-\infty}^{\infty} \dd{y} |y| \, e^{-y^2} \, e^{-\frac{b^2}{2}\,(1-c\, y)^2}\,  ,
\label{B_nb.1}
\end{equation}
with $c=L/\sqrt{4\, D\, T}$. 
Taking derivative with respect to $\Delta$ in Eq. (\ref{left_cycle_n.2}) 
gives the leading behavior of the first-passage probability density: it approaches
a constant as $\Delta\to 0$
\begin{equation}
F(t= n\, T+ \Delta) \simeq B_{n}(b) \, .
\label{left_fp.1}
\end{equation}
One can easily check that even in the $b\to \infty$ limit, this constant remains nonzero and is given by
\begin{equation}
F(t= n\, T+ \Delta)\Big|_{b\to \infty} \simeq \frac{\left[Q_1(T)\right]^{n-1}}{\sqrt{\pi}\, 
	T\, c^3}\, e^{-1/c^2}\, ,
\label{left_fp_blarge.1}
\end{equation}
where $Q_1(T)$ is now given in Eq. (\ref{Q_1.blarge.1}).
Thus, for any $\sigma\ge 0$, as one approaches
the epoch $t=n\, T$ from the left, i.e., at the end of the $(n-1)$-th cycle, 
the first-passage probability density approaches a constant $B_n(b)$ given
in Eq. (\ref{B_nb.1}). {This constant remains finite even when $b\to \infty$.
	On this side, there is no spectacular
	effect distinguishing $\sigma=0$ ($b\to \infty$) and $\sigma>0$ (finite $b$), as is the 
	case on the right side of the epoch $n\, T$.
	On the right, for finite $b$ we have an inverse square root divergence as in \eqref{fp_n+.1}, while
	for $b\to \infty$, it vanishes rapidly as $\Delta\to 0$. Thus in the large $b$ limit, we have
	a discontuity or drop from left to right (not a spike) as we see in the inset of Fig. 2
	in the main text.}    

\section{Protocol-2: Random resetting to  a random initial position}\label{apx:C}

We consider an overdamped diffusing particle that starts at an initial
position $x_0$, which is drawn from a distribution ${\cal P}(x_0)$.
The particle diffuses for a random interval $T$ drawn from an 
exponential distribution $P(T)= r\, e^{-r\,T}$  and then its position is
instantaneously reset to a new position $z$, also drawn from the same
initial distribution ${\cal P}(z)$. Then the particle diffuses again 
for another random interval
$T$ (again drawn from the exaponential distribution $P(T)$), 
followed by a reset to a new position $z'$ drawn from ${\cal P}(z)$
and the process continues. We assume that after each resetting, the reset 
position $z$ is drawn {\rm independently} from cycle to cyle from the
same distribution ${\cal P}(z)$. Similarly, the interval $T$ is
also drawn independently from cycle to cyle from the same distribution
$P(T)= r\, e^{-r\, T}$.
We also have a fixed target at a
location $L$. For fixed $L$, $r$ and ${\cal P}(z)$, We want to 
first compute the mean first-passage time
$\langle t_f\rangle$ to find the target and then optimize (minimize) this
quantity with respect to $r$ (for fixed $L$ and ${\cal P}(z)$). Recall
that in protocol-1, we just had a fixed $T$, but in protocol-2
$T$ is exponentially distributed.

\vskip 0.4cm

In this case, the mean first-passage time $\langle t_f\rangle$ has a
simple closed formula. This formula can be derived folllowing
steps similar to those in Ref. \cite{EM2011_2}. Skipping further
details, we find
\begin{equation}
\langle t_f\rangle= \frac{1}{r}\,\left[\frac{1}{\int_{-\infty}^{\infty}
	\dd{z} {\cal P}(z)\, \exp\left(- \sqrt{\frac{r}{D}}\, |L-z|\right)}-1\right]\, .
\label{mfpt.1p2}
\end{equation}

\vskip 0.4cm

Consider the special case of Gaussian distribution of Eq. (\ref{gauss.1})
Substituting this Gaussian 
${\cal P}(z)$ in Eq. (\ref{mfpt.1p2}), and performing the integral explicitly, we get
the dimensionless mean first-passage time
\begin{equation}
\tau=\frac{4D\langle t_f\rangle}{L^2} = 
\frac{4}{c^2}\, \left[\frac{2\,e^{-c^2/{2b^2}}}{e^c \, 
	\erfc\left(\frac{1}{\sqrt{2}}\left(\frac{c}{b}+b\right)\right)
	+ e^{-c} \, \erfc\left(\frac{1}{\sqrt{2}}\left(\frac{c}{b}-b\right)\right)}-1\right]\, \equiv w_2(b,c)
\label{mfpt.2p2}
\end{equation}
where the dimensionless constants $b$ and $c$
are given by
\begin{equation}
b= \frac{L}{\sigma}\, ; \quad\quad\quad\quad c= 
\sqrt{\frac{r}{D}}\, L\, .
\label{bc.1p2}
\end{equation}
Recall that in periodic protocol (with a fixed $T$ discussed in section \ref{sec:gaus_periodic}) , the parameter $b=L/\sigma$ is the same, but the
parameter $c=L/\sqrt{4DT}$ is  different.
In protocol-2, the reset rate $r$ effectively plays the role of $1/T$ in protocol-1.

\vskip 0.4cm

The formula for $w_2(b,c)$ in Eq. (\ref{mfpt.2p2}) for protocol-2 is simpler and more explicit, 
compared to the equivalent formula Eq.\ref{mfpt_gauss.1} for protocol-1.   

\vskip 0.4cm

In Eq. (\ref{mfpt.2p2}), consider first the limit $b\to \infty$, i.e.,
resetting to a fixed initial condition. In this case, Eq. (\ref{mfpt.2p2}) reduces precisely to the
formula derived by Evans and Majumdar~\cite{EM2011_1}
\begin{equation}
\tau= w_2(b\to \infty, c)= \frac{4}{c^2}\, \left[e^c-1\right]\, .
\label{b_infty.p2}
\end{equation}
In this $b\to \infty$ limit, $\tau$ as a function of $c$, has a unique minimum at $c=c^*=1.59362\ldots$,
where $\tau(c=c^*)=6.17655\ldots$.

\vskip 0.4cm

For finite fixed $b$, it is very easy to plot the function $w_2(b,c)$ vs. $c$ given in
Eq. (\ref{mfpt.2p2}) and one finds
that as in protocol-1, $w_2(b,c)$ first decreases with increasing $c$, achieves a minimum
at certain $c_1(b)$, then starts increasing again--becomes a maximum at $c_2(b)$ and finally decreases
for large $c$ as $\sim 1/c$ (this asymptotics for large $c$ can be computed exactly from
Eq. (\ref{mfpt.2p2})). Note that the actual values of $c_1(b)$band $c_2(b)$ are of course different in
the two protocols. Interestingly, this metastable optimum time at $c=c_1(b)$ 
disappears when $b$ becomes smaller
than a critical value $b_c= 2.53884$. For $b<b_c$, the function $w_2(b,c)$ decreases
monotonically with increasing $c$. Hence, for $b<b_c$, the best solution is $c=\infty$ (i.e.,
repeated resetting). The curves plotted in Fig. 4 of the main text correspond to this theoretical prediction
and they have been checked experimentally and numerically. This numerical test allows us to clearly 
state that the small disagreement at small $b$ and large $c$ of the experimental data  
with the theoretical predictions is due to the maximum sampling rate of our analog to digital 
converter which was not fast enough to detect very short MFPT.

\section{General Setting: the survival probability of a Brownian walker in arbitrary 
dimensions} \label{apx:D_2d}

We first consider the general problem of the first-passage probability of a Brownian 
particle to a fixed target in $d$-dimensions (without any resetting). Resetting will be 
incorporated later. Consider a Brownian searcher in $d$-dimensions, starting initially at 
the position ${\vec r}_0$. An immobile point target is located at ${\vec R}_{\rm tar}$ with 
a tolerance radius $R_{\rm tol}$ (see Fig. \ref{fig:capture_2d}).

If the searcher comes within the distance $R_{\rm tol}$ from the target at ${\vec R}_{\rm 
tar}$, the target is detected. Let $S_0(t|{\vec r}_0)$ denote the survival probability, 
i.e., the probability that the target is not detected up to time $t$, given that the 
searcher starts at the initial position ${\vec r}_0$. Note that $S_0(t|{\vec r}_0)$ depends 
implicitly on ${\vec R}_{\rm tar}$ and $R_{\rm tol}$, but we do not write them explicitly 
for the simplicity of notations. The subscript $0$ is $S_0|(t|{\vec r}_0)$ denotes the fact 
that we are considering the problem {\em without} resetting. Suppose now that the initial 
position is random and drawn from a normalised probability distribution ${\cal P}({\vec 
r}_0)$. Then the survival probability $Q_0(t)$, averaged over this distribution of the 
starting point of the searcher, is simply given by
\begin{equation}
Q_0(t)= \int S_0(t|{\vec r}_0)\, {\cal P}({\vec r}_0)\, \dd{{\vec r}_0} \, .
\label{Q0t_def.1}
\end{equation}
For later purposes, let us also define the Laplace transform of $Q_0(t)$ with respect to $t$ as follows
\begin{equation}
{\tilde Q}_0(s)= \int_0^{\infty} Q_0(t)\, \e^{-s\, t} \dd{t} \, .
\label{Q0s_def.1}
\end{equation}
Let us suppose that we know the survival probability $Q_0(t)$ without resetting explicitly. Then
we show below that when we switch on the resetting, the survival probability in the presence
of resetting can be computed in terms of $Q_0(t)$ quite generally both for (I) the 
Poissonian resetting and (II) the periodic resetting.

\vskip 0.4cm

{\noindent {\bf {Poissonian resetting:}}} In this case, we choose the
initial position ${\vec r}_0$ of the searcher from the distribution ${\cal P}({\vec r}_0)$ 
at $t=0$ and let the particle (searcher) diffuse for a random time $T$ distributed 
exponentially, $P(T)= r\, \e^{-r\, T}$ where $r\ge 0$ is the resetting rate. At the end of 
the random period $T$, we reset the position of the searcher, i.e., we choose afresh the 
initial position from the same distribution ${\cal P}({\vec r}_0)$ and again let it diffuse 
for a random period $T$ drawn afresh from $P(T)= r\, \e^{-r\, T}$ and so on. Let $Q_r(t)$ 
denote the survival probability, i.e., the probability that the target is not detected up to 
time $t$. The subscript $r$ denotes the presence of a nonzero resetting rate $r$. 
Considering several possibilities up to time $t$: (i) there is no resetting in $[0,t]$ (ii) 
there is exactly one resetting in $[0,t]$ (iii) there are exactly two resettings in $[0,t]$ 
etc, we can express $Q_r(t)$ as a series
\begin{equation}
Q_r(t)= \e^{-r\, t} Q_0(t) + r\, \int_0^{\infty} \dd{t_1} \int_0^{\infty} \dd{t_2}
\e^{-r\, t_1} Q_0(t_1)\, \e^{-r\, t_2} Q_0(t_2)\, \delta(t_1+t_2-t) + \ldots \, .
\label{series.1}
\end{equation}
The first term denotes the event when there is no resetting in $[0,t]$ which occurs with
probability $\e^{-r\, t}$ and the particle survives with probability $Q_0(t)$ during the
time interval. The second term denotes the event that exactly one resetting happens at time $t_1$
in $[0,t]$: this means a resetting occurs at $t_1$ and then there is no setting in the 
remaining interval $t_2=t-t_1$: this happens with probability density $r\, \e^{-r\, t_1} 
\dd{t_1}\, \e^{-r\, t_2}\, \delta(t_1+t_2-t)$. In addition the particle has to survive in 
both intervals $[0,t_1]$ and $[t_1, t]$ which occurs with probability $Q_0(t_1)\, Q_0(t_2)$ 
where $t_2=t-t_1$. We have used the renewal property that makes the intervals between 
resettings statistically independent. Similarly, one can work out the probability of two 
resettings in $[0,t]$ etc. This somewhat formidable looking series in \eqref{series.1} 
simplifies enormously in the Laplace space. We define the Laplace transform
\begin{equation}
{\tilde Q}_r(s)= \int_0^{\infty} Q_r(t)\, \e^{-s\, t} \dd{t}
\label{Qrs_def.1}
\end{equation}
and we recall that the Laplace transform ${\tilde Q}_0(s)$ is defined in \eqref{Q0s_def.1}. Taking 
Laplace transform
of \eqref{series.1} with respect to $t$ one obtains
\begin{equation}
{\tilde Q}_r(s) =  {\tilde Q}_0(r+s)+ r\, \left[ {\tilde Q}_0(r+s)\right]^2 +
r^2\, \left[{\tilde Q}_0(r+s)\right]^3 +\ldots 
= \frac{{\tilde Q}_0(r+s)}{1- r\, {\tilde Q}_0(r+s)}\, ,
\label{LT.2}
\end{equation}

The first-passage probability density $F_r(t)$, in the presence of the resetting, is given by
\begin{equation}
F_r(t)= - \dv{Q_r(t)}{t}\, .
\label{fpp_poisson.1}
\end{equation}
Consequently, its Laplace transform, using \eqref{fpp_poisson.1} and \eqref{LT.2}, is simply
\begin{equation}
{\tilde F_r}(s)=  \int_0^{\infty} F_r(t)\, \e^{-s\, t} \dd{t} 
= 1- s\, {\tilde Q}_r(s) 
= \frac{1- (r+s)\, {\tilde Q}_0(r+s)}{1- r\, {\tilde Q}_0(r+s)}\,
\label{ffp_lap.1}
\end{equation}
where ${\tilde Q}_0(s)$ is the Laplace transform of the survival probability without resetting
defined in \eqref{Q0s_def.1}. Finally, the mean first-passage time (MFPT), using the relation in 
\eqref{fpp_poisson.1}, is given by
\begin{equation}
\langle t_f\rangle= \int_0^{\infty} t\, F_r(t) \dd{t} = \int_0^{\infty} Q_r(t) \dd{t} 
= {\tilde Q}_r(s=0) 
= \frac{{\tilde Q}_0(r)}{1- r\,{\tilde Q}_0(r)} 
\label{mfpt_poisson.1}
\end{equation}
where we used \eqref{LT.2} in arriving at the last result. Thus, to summarize, in the presence of
the Possonian resetting with rate $r$, we can in principle
compute both $\langle t_f\rangle$ as well as the full distribution 
$F_r(t)$ (rather its Laplace transform),
provided we know ${\tilde Q}_0(s)$, i.e., the Laplace transform of 
the survival probability $Q_0(t)$ without resetting.

\vskip 0.4cm

{\noindent {\bf {Periodic resetting:}}} As in the case of the Poissonian resetting, we choose the
initial position ${\vec r}_0$ of the searcher from the distribution 
${\cal P}({\vec r}_0)$ at $t=0$ and
let the particle (searcher) diffuse for a fixed period $T$. At the end of the period $T$, we reset
the position of the searcher, i.e., we choose afresh the initial position from the
same distribution ${\cal P}({\vec r}_0)$ and again let it diffuse for a period $T$ and so on.
We then ask: what is the survival probability $Q(t|T)$ that the target is not detected up to 
time $t$?
Clearly if the measurement time $t$ belongs to the $n$-th period, i.e., 
$(n-1)T<t\le n\,T$ (with $n=1,2,\ldots$) then the survival probability is given by
\begin{equation} 
Q(t|T)= \left[Q_0(T)\right]^{n-1}\, Q_0\left(t- (n-1)T\right)\, ;\quad {\rm for}\,\, 
\quad (n-1)T<t\le n\,T
\label{tn.1}
\end{equation}
where $Q_0(t)$ is defined in \eqref{Q0t_def.1}. 
The formula \eqref{tn.1} is easy to derive: (i) the
target has to survive being detected for the full 
$(n-1)$ cycles prior to the running one at $t$ and
(ii) in the running period since the time passed is $t-(n-1)T$ 
since the begining of the period, its
survival probability is $Q_0\left(t- (n-1)T\right)$. Since the cycles are independent, the total 
survival probability is just the product over different periods till 
time $t$. Hence, for arbitrary $t$,
we can write the survival probability as
\begin{equation}
Q(t|T)= \sum_{n=1}^{\infty}  \left[Q_0(T)\right]^{n-1}\, Q_0\left(t- (n-1)T\right)\, 
\operatorname{I}_{ (n-1)T< t\le n\, T}
\label{Qt_def.1}
\end{equation}
where the indicator function $\operatorname{I}_{A}=1$ if the clause $A$ in the subscript
is satisfied and is zero otherwise. The first-passage probability density $F(t|T)$ (given the
period of resetting $T$) is then given by
\begin{equation}
F(t|T)= -\dv{Q(t|T)}{t}\, ,
\label{FP.2D}
\end{equation}
where $Q(t|T)$ is given in Eq. (\ref{Qt_def.1}). The MFPT is given by
the first moment of $F(t|T)$
\begin{equation}
\langle t_f\rangle = \int_0^{\infty} t\, F(t|T) \dd{t} = \int_0^{\infty} Q(t|T) \dd{t}
\label{mfpt.1d2}
\end{equation}
where we used \eqref{FP.2D} and integrated by parts using $Q(\infty|T)=0$. Plugging in the formula
in \eqref{Qt_def.1} and integrating we get
\begin{equation}
\langle t_f\rangle = \sum_{n=1}^{\infty} \left[Q_0(T)\right]^{n-1} \int_0^{T}Q_0(\tau) \dd{\tau}
= \frac{\int_0^T Q_0(t) \dd{t}}{1- Q_0(T)}\, ,
\label{mfpt_periodic.1}
\end{equation}
where we recall that $Q_0(t)$ is the survival probability without resetting 
as defined in \eqref{Q0t_def.1}. Thus, as in the case of the Poissonian resetting,
for periodic resetting with a period $T$ also, we can compute the MFPT 
$\langle t_f\rangle$ and the full first-passage time distribution $F(t|T)$, provided
we know the survival probability $Q_0(t)$ without resetting. 

Hence, to summarize, the central quantity of interest is the survival probability $Q_0(t)$ 
without resetting. If we know this object, then we can compute, in principle, the 
first-passage properties in the presence of resetting, both for the Poissonian as well as 
for the periodic protocol. Below, we will focus in $d=2$ (where experimental data is 
available) and first compute the survival probability $Q_0(t)$ without resetting, and then 
apply the general results above to compute the MFPT as well as the distribution of 
first-passage times for the two protocols. It turns out that the final formulae are a bit 
more explicit in the case of Poissonian resetting than the periodic resetting.

\section{Two dimensions}\label{apx:E_2d}

We want to first compute the survival probability $Q_0(t)$ without resetting, defined in
\eqref{Q0t_def.1}, in dimension $d=2$. For this we need to first compute the survival
probability $S_0(t|{\vec r}_0)$, i.e., the probability that the Brownian searcher, 
starting at the fixed initial position ${\vec r}_0$, does not enter the 
circle of tolerance
(of radius $R_{\rm tol}$) around the target located at 
${\vec R}_{\rm tar}$ up to time $t$ (see
Fig. \ref{fig:capture_2d}). The result for 
$S_0(t|{\vec r}_0)$ is actually well known~\cite{BT1993,
BB2003,EM2014} and can be easily computed as follows. We first make a change of variable
$\vec {r_p}= \vec r_0- {\vec R}_{\rm tar}$ so that the distance is measured 
with respect to the target
center and we denote, for the simplicity of notations, 
$S_0(t|{\vec r}_0)= S_0(t|\vec r_p+ {\vec R}_{\rm tar})$
by $S(\vec r_p,t)$. The subscript $p$ in $\vec {r_p}$ simply stands for `particle'.

Then $S(\vec r_p,t)$ satisfies the diffusion equation 
\begin{equation}
\partial_t S(\vec r_p, t)= D\, \nabla_{\vec r_p}^2 S(\vec r_p,t)\, 
\label{diff.1}
\end{equation}
in the region $r_p=|\vec r_p|> R_{\rm tol}$, with the absorbing boundary 
condition $S(r_p=R_{\rm tol},t)=0$.
Using spherical symmetry, $S(\vec r_p,t)$ depends only on 
$r_p= |\vec r_p|$ and hence \eqref{diff.1} reduces
to a one dimensional partial differential equation in the radial coordinate

\begin{equation}
\partial_t S(r_p,t)= D\left[ \partial_{r_p}^2 S + \frac{2}{r_p}\, 
\partial_{r_p} S\right]\, ; \quad {\rm for} \quad r_p>R_{\rm tol}\, .
\label{diff.2}
\end{equation}
with the boundary condition $S(r_p=R_{\rm tol},t)=0$ and also $S(r_p\to \infty,t)=1$. 
The latter condition
follows from the fact that the particle must survive with probability $1$ 
up to a finite time $t$, if it
starts infinitely far away from the target. The initial condition is $S(r_p,0)=1$ for
all $r_p> R_{\rm tol}$. To solve this partial differential equation, it is first
convenient to take the Laplace transform with respect to $t$ and define
${\tilde S}(r_p,s)= \int_0^{\infty} S(r_p,t)\, \e^{-s\, t} \dd{t}$. Taking
Laplace transform of \eqref{diff.2} and 
using the initial condition, we arrive at an ordinary second order
differential equation
\begin{equation}
D\, \left[ \dv[2]{\tilde S}{r_p}  + \frac{2}{r_p}\, \dv{\tilde S}{r_p}\right]= -1 + s \, {\tilde S}\, 
\label{diff.3}
\end{equation}
valid for $r_p>R_{\rm tol}$ with the boundary conditions: (i) ${\tilde S}(r_p=R_{\rm tol},s)=0$
and (ii) ${\tilde S}(r_p\to \infty,s)=1/s$. The latter condition follows from the
boundary condition $S(r_p\to \infty,t)=1$. The inhomogeneous term on the right hand side
of \eqref{diff.3} can be eliminated by a shift ${\tilde S}(r_p,s)\to {\tilde S}(r_p,s) +1/s$. The
solution of the resulting ordinary homogeneous differential equation is 
given by the limear combination of 
modified Bessel functions. Finally, retaining only the solution that satisfies the correct 
boundary conditions, one arrives at~\cite{BT1993,BB2003,EM2014} 
\begin{equation}
{\tilde S}(r_p,s)= \frac{1}{s}\,\left[1- \frac{ K_0\left(\sqrt{\frac{s}{D}}\, r_p\right)}
{K_0\left( \sqrt{\frac{s}{D}}\, 
R_{\rm tol}\right)}\right]\, ; \quad {\rm for}\,\, r_p\ge R_{\rm tol}\, 
\label{rad_sol.1}
\end{equation}
where $K_\nu(x)$ is the modified Bessel function of the 
second kind with index $\nu$~\cite{GR}. Translating back
to the original coordinate and using 
$S_0(t|{\vec r}_0)= S(r_p=|\vec r_0- \vec R_{\rm tar}|,\, t)$ 
we then have the exact solution in the Laplace space
\begin{equation}
\int_0^{\infty} S_0(t|{\vec r}_0)\, \e^{-s\, t} \dd{t}= \frac{1}{s}\, \left[1- 
\frac{ K_0\left(\sqrt{\frac{s}{D}}\, |{\vec r}_0- {\vec R}_{\rm tar}|\right)}
{K_0\left( \sqrt{\frac{s}{D}}\, R_{\rm tol}\right)}\right]\,
\theta\left(|{\vec r}_0- {\vec R}_{\rm tar}|-R_{\rm tol}\right)
\label{S0t_sol.1}
\end{equation}
where $\theta(z)$ is the Heaviside step function: $\theta(z)=1$ if $z>0$ 
and $\theta(z)=0$ for $z<0$.
Interestingly, the Laplace transform can be inverted~\cite{BT1993} and one obtains
\begin{equation}
S_0(t|{\vec r}_0)= \frac{2}{\pi}\, \int_0^{\infty} \frac{\dd{x}}{x}\, 
\e^{-D\,t\, x^2/R_{\rm tol}^2}\,
\frac{ \left[Y_0\left(\frac{x\, |{\vec r}_0- {\vec R}_{\rm tar}|}{R_{\rm tol}}\right)\, J_0(x)-
J_0\left(\frac{x\, |{\vec r}_0- {\vec R}_{\rm tar}|}{R_{\rm tol}}\right)\, 
Y_0(x)\right] }{J_0^2(x)+Y_0^2(x)}\, 
\label{lap_inv.1}
\end{equation}
where $J_0(x)$ and $Y_0(x)$ are the Bessel functions of the first and 
the second kind respectively~\cite{GR}.
It can be checked that the integral is covergent for large $x$, and also for small $x$. 
To see this,
one can use the asymptotic behaviors of the Bessel functions
for small and large $x$~\cite{GR}. For small $x$
\begin{eqnarray}
J_0(x) & =& 1- \frac{x^2}{4} + O(x^3) \label{BesselJ_x0.1} \\
Y_0(x) & =& \frac{2}{\pi}\, \ln x+ \frac{2 (\gamma_E-\ln 2)}{\pi} + O(x^2) \label{BesseY_x0.1}
\end{eqnarray}
where $\gamma_E$ is the Euler constant. Similarly for large $x$
\begin{eqnarray}
J_0(x) & =& \sqrt{\frac{2}{\pi x}}\, 
\cos\left(x-\frac{\pi}{4}\right)+ \ldots \label{BesselJ_xlarge.1} \\
Y_0(x) & =& \sqrt{\frac{2}{\pi x}}\, 
\sin\left(x-\frac{\pi}{4}\right)+ \ldots \label{BesseY_xlarge.1}
\end{eqnarray}

Finally, averaging over the distribution of ${\vec r}_0$, drawn from ${\cal P}({\vec r}_0)$
as in \eqref{Q0t_def.1}, we obtain the Laplace transform ${\tilde Q}_0(s)$
\begin{equation}
{\tilde Q}_0(s)= \int_0^{\infty} Q_0(t)\, \e^{-s\, t} \dd{t}= 
\frac{1}{s} \int {\cal P}({\vec r}_0)\,
\dd{\vec{r}_0} 
\theta\left(|{\vec r}_0- {\vec R}_{\rm tar}|-R_{\rm tol}\right)
\left[1-\frac{ \int K_0\left(\sqrt{\frac{s}{D}}\, 
|{\vec r}_0- {\vec R}_{\rm tar}|\right)} 
{K_0\left( \sqrt{\frac{s}{D}}\, R_{\rm tol}\right)} \right]\, . 
\label{Q0s_sol.1}
\end{equation}
Using \eqref{lap_inv.1}, we can even invert this Laplace transform
and get
\begin{equation}
Q_0(t)= \frac{2}{\pi}\, \int_0^{\infty} \frac{\dd{x}}{x}\, \frac{\e^{-D\,t\, x^2/R_{\rm tol}^2}}
{J_0^2(x)+Y_0^2(x)}\,
\int_{<} \dd{\vec{r}_0} {\cal P}({\vec r}_0)\, 
\left[Y_0\left(\frac{x\, |{\vec r}_0- {\vec R}_{\rm tar}|}{R_{\rm tol}}\right)
\, J_0(x)-J_0\left(\frac{x\, |{\vec r}_0- 
{\vec R}_{\rm tar}|}{R_{\rm tol}}\right)\, Y_0(x)\right]\, ,
\label{Q0t_2d.1}
\end{equation}
where $\int_{<}$ is a shorthand notation to
indicate that the integral is over the restricted
space where $|{\vec r}_0- {\vec R}_{\rm tar}|>R_{\rm tol}$.
By examining the expressions for the MFPT in the two protocols 
respectively in \eqref{mfpt_poisson.1}
and \eqref{mfpt_periodic.1}, we see that for the Poissonian resetting we just need the 
Laplace transform ${\tilde Q}_0(s=r)$ itself (which is given explicitly in 
\eqref{Q0s_sol.1}) and we do not need to invert this Laplace transform. Hence the expression 
for MFPT in the Poissonian case turns out to be a bit simpler. In contrast, for the periodic 
resetting, we need $Q_0(T)$ in \eqref{mfpt_periodic.1} to evaluate the MFPT and hence we 
need to use the result in \eqref{Q0t_2d.1} which is a bit more cumbersome.

\subsection{Poissonian Resetting}

In this case, substituting \eqref{Q0s_sol.1} directly into 
the expression for MFPT in \eqref{mfpt_poisson.1}
we get after simplifying
\begin{equation}
\langle t_f\rangle= \frac{1}{r}\,
\left[ \frac{1}{1- g + {\cal D}_1}-1\right]\, ,
\label{mfpt_poisson.2}
\end{equation}
where 
\begin{equation}
{\cal D}_1= \int_{<}{\cal P}({\vec r}_0) \dd{\vec{r}_0}
\frac{K_0\left( \sqrt{\frac{r}{D}}\, 
|{\vec r}_0-{\vec R}_{\rm tar}|\right)}{K_0\left(\sqrt{\frac{r}{D}}\, 
R_{\rm tol}\right)} \, , \quad {\rm and}\quad g= \int_{<} 
{\cal P}({\vec r}_0) \dd{\vec{r}_0}
\label{denom.2d}
\end{equation}

Let us first evaluate the integral ${\cal D}_1$ in \eqref{denom.2d},
where we recall that the integral over ${\vec r}_0$ is restricted to the 
outside of the
tolerance disk of radius $R_{\rm tol}$. 
We consider a Gaussian distribution for ${\vec r}_0$ in two dimensions
\begin{equation}
{\cal P}({\vec r}_0)= \frac{1}{2\,\pi\, \sigma^2}\, 
\e^{- \frac{{\vec r}_0^2}{2\, \sigma^2}}\, .
\label{Pr0_Gaussian.1}
\end{equation}
Substituting \eqref{Pr0_Gaussian.1} in \eqref{denom.2d} and making a change of variable
${\vec r}_0- {\vec R}_{\rm tar}= \vec R$ and shifting to polar coordinates we get
\begin{eqnarray}
{\cal D}_1 & = & \frac{1}{2\,\pi\, \sigma^2}\, 
\int_{R>R_{\rm tol}}\int_0^{2\pi}  \e^{- ({\vec R}_{\rm tar}+{\vec R})^2/
{2\sigma^2}}\,
K_0\left( \sqrt{\frac{r}{D}}\, R\right)\, R\, \dd{R} \dd{\theta} \nonumber \\
& = & \frac{\e^{- L^2/{2\sigma^2}}}{2\,\pi\, \sigma^2}\, 
\int_{R_{\rm tol}}^{\infty} \dd{R} R\, 
K_0\left( \sqrt{\frac{r}{D}}\, R\right)\, \e^{-R^2/{2\sigma^2}}\,
\int_0^{2\pi} \e^{- L\, R\, \cos(\theta)/\sigma^2} \, \dd{\theta} \nonumber \\
&=& \frac{\e^{- L^2/{2\sigma^2}}}{\sigma^2}\,
\int_{R_{\rm tol}}^{\infty} \dd{R} R\, 
K_0\left( \sqrt{\frac{r}{D}}\, R\right)\, \e^{-R^2/{2\sigma^2}}\,
I_0\left(\frac{R\, L}{\sigma^2}\right)\, ,
\label{denom.2}
\end{eqnarray} 
where $I_0(z)$ is the modified Bessel function of the first kind~\cite{GR}, and $L=|{\vec R}_{\rm tar}|$.
By rescaling $R= \sigma\, z$, it can be re-expressed as
\begin{equation}
{\cal D}_1= \e^{-L^2/{2\sigma^2}}\, 
\int_{R_\text{tol}/\sigma}^{\infty} \dd{z} z\, \e^{-z^2/2}\, 
K_0\left( \sqrt{\frac{r}{D}}\, \sigma\, z\right)\, 
I_0\left( \frac{L}{\sigma}\, z\right)\, .
\label{denom.3}
\end{equation}
Unfortunately, we could not perform the integral in \eqref{denom.3} 
explicitly.
Similarly, one can perform the second integral $g$ in \eqref{denom.2d} to get
\begin{equation}
g= \e^{-L^2/{2\sigma^2}}\, 
\int_{R_{\rm tol}}^{\infty} \dd{R} R\, \e^{-R^2/{2\sigma^2}}\, 
I_0\left(\frac{R\, L}{\sigma^2}\right)\, .
\label{def_g1}
\end{equation}

To simplify things, we now define the following dimensionlesss variables
\begin{equation}
\tau_\text{Poisson}= \frac{4\, D\, \langle t_f\rangle}{L^2}\, ; \quad\,\, 
a=\frac{R_{\rm tol}}{L}\, ;
\quad\,\, b= \frac{L}{\sigma}\, ; \quad \,\, c=\frac{\sqrt{r}\, L}{\sqrt{D}}\, .
\label{adim_def.1}
\end{equation}
In terms of these adimensional variables, the integral ${\cal D}_1$ 
in \eqref{denom.3} reads
\begin{equation}
{\cal D}_1= \e^{-b^2/2}\, \int_{ab}^{\infty} \dd{z} z\,
\e^{-z^2/2}\, K_0\left(\frac{c}{b}\, z\right)\, I_0(b\, z)\, 
\label{D1.1}
\end{equation}
and similarly, Eq. (\ref{def_g1}) reads
\begin{equation}
g(a,b)= \e^{-b^2/2} \int_{ab}^{\infty} \dd{z} z\, \e^{-z^2/2}\, I_0(b\, z)\, .
\label{def_g2}
\end{equation}
Note that $g(a,b)$ is independent of $c$ and is a function of only 
$a$ and $b$.
Finally, we can then re-express the scaled MFPT 
$\tau_\text{Poisson}= 4D\langle t_f\rangle/R_\text{tar}^2$, using 
\eqref{mfpt_poisson.2}, as
\begin{equation}
\tau_\text{Poisson}= 
W_\text{Poisson}(a,b,c) \equiv  
\frac{4}{c^2}\, \left[ \frac{1}{1-g(a,b) + \frac{\e^{-b^2/2}}{K_0(ac)}\, 
\int_{ab}^{\infty} \dd{z} z\, 
\e^{-z^2/2}\, K_0\left(\frac{c}{b}\, z\right)\, I_0(b\, z)}-1 \right]\, ,
\label{tau_poisson.1}
\end{equation}
with $g(a,b)$ given in \eqref{def_g2}.
We note that the parameter $0<a= R_{\rm tol}/{L}<1$, but the other two parameters 
$b$ and $c$ are positive but arbitrary. We recall that in $d=1$, we had already put $a=0$, 
since in $1$-d there is no need for a cut-off or tolerance radius, as the searcher can find 
the point target with a finite nonzero probability in $d=1$.

\subsubsection{The limit \texorpdfstring{$\sigma\to 0$}{σ=0}: resetting to the origin}

When $\sigma\to 0$, ${\cal P}({\vec r}_0)
\to \delta({\vec r}_0)$ and hence it corresponds to resetting to the origin~\cite{EM2011_1}.
In this case, only the parameter $b= L/\sigma\to \infty$, 
while $a$ and $c$ remain fixed.
To find the limiting behavior of \eqref{tau_poisson.1} when $b\to \infty$, 
let us investigate in this limit
the quantity ${\cal D}_1$ 
given in \eqref{D1.1}.
For large $b$ with $z$ fixed, we can approximate $I_0(b\, z)\approx \e^{bz}/\sqrt{2\pi bz}$.
Substituting this in the integral in \eqref{D1.1} and rescaling $z=b\, y$ we get
\begin{equation}
{\cal D}_1\approx \frac{b}{\sqrt{2\pi}}\, \int_a^{\infty} \dd{y} \sqrt{y}\, 
K_0(c\, y)\, \e^{-b^2(y-1)^2/2}\, .
\label{D1.2}
\end{equation}
In the limit $b\to \infty$, the integrand is sharply peaked at $y=1$. Hence we can further
approximate
\begin{eqnarray}
{\cal D}_1 &\approx & \frac{b}{\sqrt{2\pi}}\, 
K_0(c) \, \int_{a}^\infty \dd{y} \e^{-b^2(y-1)^2/2}\nonumber \\
&=& \frac{1}{\sqrt{\pi}}\, 
K_0(c) \, \int_{b (a-1)/\sqrt{2}}^{\infty} \dd{u} \e^{-u^2} \nonumber \\
&= & K_0(c)\, ,
\label{D1.3}
\end{eqnarray}
where we made a change of variable $b(y-1)/\sqrt{2}=u$ in the first line to arrive 
at the second line.
Further noting that for $a<1$ the lower limit can be pushed to $b(a-1)/\sqrt{2}\to -\infty$ when 
$b\to \infty$, we arrived at the third line. Similarly, one can analyse
the integral $g(a,b)$ given in \eqref{def_g2} in the limit $b\to \infty$.
Following exactly similar arguments we get, as $b\to \infty$
\begin{equation}
g(a,b\to \infty) \approx 1 \, .
\label{g_largeb}
\end{equation}

Substituting these results in \eqref{tau_poisson.1} we then
get in the limit $b\to \infty$
\begin{equation}
\tau_\text{Poisson}=
W_\text{Poisson}(a,\,c)= \frac{4}{c^2}\, \left[\frac{K_0(c\,a)}{K_0(c)}-1\right]\, ,
\label{tau_poisson_origin.1}
\end{equation}
which coincides, in these dimensionless units, exactly with the result derived in Ref. \cite{EM2014}
for resetting to the origin. For fixed $a$, one can easily find out the asymptotic behaviors of
$W_\text{Poisson}(a,c)$ in the two limits $c\to 0$ and $c\to \infty$. For this, one can use
the following leading asymptotic behaviors of the function $K_0(z)$
\begin{eqnarray}
K_0(z) \approx \begin{cases}
& -\ln (z/2) - \gamma_E  \quad\,\, {\rm as}\quad z\to 0  \\
\\
& \sqrt{\frac{\pi}{2\,z}}\, \e^{-z} \quad\,\, {\rm as}\quad z\to \infty   
\label{K0z_asymp}
\end{cases}
\end{eqnarray}
Substituting these asymptotics in \eqref{tau_poisson_origin.1} we get
\begin{eqnarray}
W_\text{Poisson}(a,c) \approx \begin{cases}
& \frac{4}{c^2}\, \frac{\ln (1/a)}{\ln (1/c)} \quad\,\, {\rm as}\quad 
c\to 0 \label{c0.1} \\
\\
& \frac{4}{\sqrt{a}\, c^2}\, \e^{(1-a)\, c} \quad\,\, {\rm as}\quad c\to \infty
\label{cinf.1}
\end{cases}
\end{eqnarray}
Thus the scaled MFPT $W_\text{Poisson}(a,c)$, for fixed $0<a<1$, as a function of $c$ diverges
in both limits $c\to 0$ and $c\to \infty$ and
also exhibits a unique minimum at $c=c^*(a)$ which depends on the parameter $a$ 
(see Fig. \ref{fig:tau_poisson_origin} for a plot).

\begin{figure}[H]
\centering
\includegraphics[width = 0.6\linewidth]{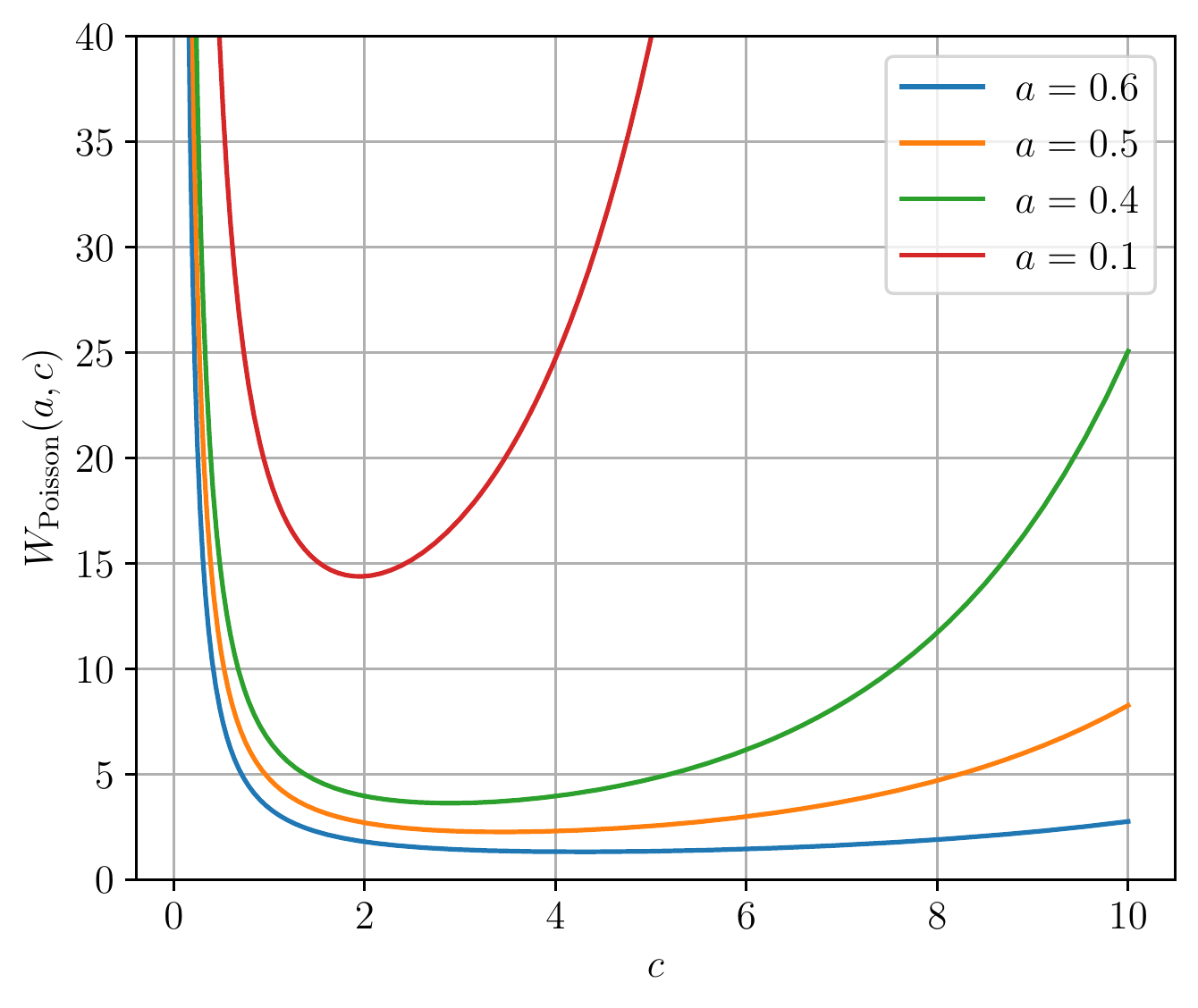}
\caption{The scaled MFPT $W_\text{Poisson}(a,c)$ in the limit 
$b\to \infty$ (or equivalently $\sigma\to 0$ limit)
plotted as a function of $c$ for different values of the parameter $a= R_{\rm tol}/L<1$.
The MFPT exhibits a unique minimum at $c^*(a)$ that depends on $a$.}
\label{fig:tau_poisson_origin}
\end{figure}

\subsubsection{Finite \texorpdfstring{$\sigma>0$}{σ}}

In this case, the parameter $b= L/\sigma$ is finite, and
we have to evaluate the function $W_\text{Poisson}(a,b,c)$ in \eqref{tau_poisson.1} numerically.
For finite $b$, the scaled MFPT $\tau_\text{Poisson}= W(a,b,c)$ exhibits very different behavior 
as a function of $c$, for fixed $a$ and $b$. One of the biggest difference is 
that for finite $b$, as
$c\to \infty$, the function $W_\text{Poisson}(a,b,c)$ now decays to zero, in stark contrast
to the $b\to \infty$ case where it diverges (see \eqref{cinf.1}). Indeed, for finite $b$, one can
extract the asymptotics behaviors of $W_\text{Poisson}(a,b,c)$ from \eqref{tau_poisson.1} 
in the two limits $c\to 0$ and $c\to \infty$, by using the asymptotic behaviors
of $K_0(z)$ given in \eqref{K0z_asymp}. 
Skipping the details of the derivation we find
\begin{eqnarray}
W_\text{Poisson}(a,b,c) \approx \begin{cases}
& \frac{4\, A_1(a,b)}{c^2\, \ln(1/c)} \quad\,\, {\rm as}\quad c\to 0 \label{c0b.1} \\
\\
& \frac{4\, g(a,b)}{1-g(a,b)}\, \frac{1}{c^2} \quad\,\, {\rm as}\quad c\to \infty
\label{cinfb.1}
\end{cases}
\end{eqnarray}
where $g(a,b)$ is given in \eqref{def_g2}.
The amplitude $A_1(a,b)$ can also be computed as
\begin{equation}
A_1(a,b)= \e^{-b^2/2}\,\int_0^{a\, b} \dd{z} z\, \e^{-z^2/2}\, \ln(z/{ab})\,
I_0(b\, z)
\label{A1_def.1}
\end{equation}
One can check that in the limit $b\to \infty$, one obtains
$A_1(a, b\to \infty)= \ln (1/a)$ and thus the first line
in \eqref{cinfb.1} coincides with the first line
of \eqref{cinf.1}.

In fact, as in the $d=1$ case, for finite but large $b$, $W_\text{Poisson}(a,b,c)$ as a 
function of $c$, first becomes a minimum at $c_1^*(a,b)$, then achieves a maximum at 
$c_2^*(a,b)$ and finally decays to zero as $1/c$ when $c\to \infty$. Thus, the minimum at 
$c_1^*(a,b)$ is metastable as in $d=1$. Finally, below a critical value $b<b_c(a)$, the 
curve $W_\text{Poisson}(a,b,c)$ decreases monotonically as a function of $c$, indicating 
that the metastable minimum disappears and the true optimal value is $c=\infty$ 
(corresponding to infinite resetting rate $r\to \infty$). In Fig. \ref{fig:tau_poisson} we 
plot $W_\text{Poisson}(a,b,c)$ vs $c$ (for fixed $a=1/2$ and for different values of $b$)

\begin{figure}[H]
\centering
\includegraphics[width = 0.6\linewidth]{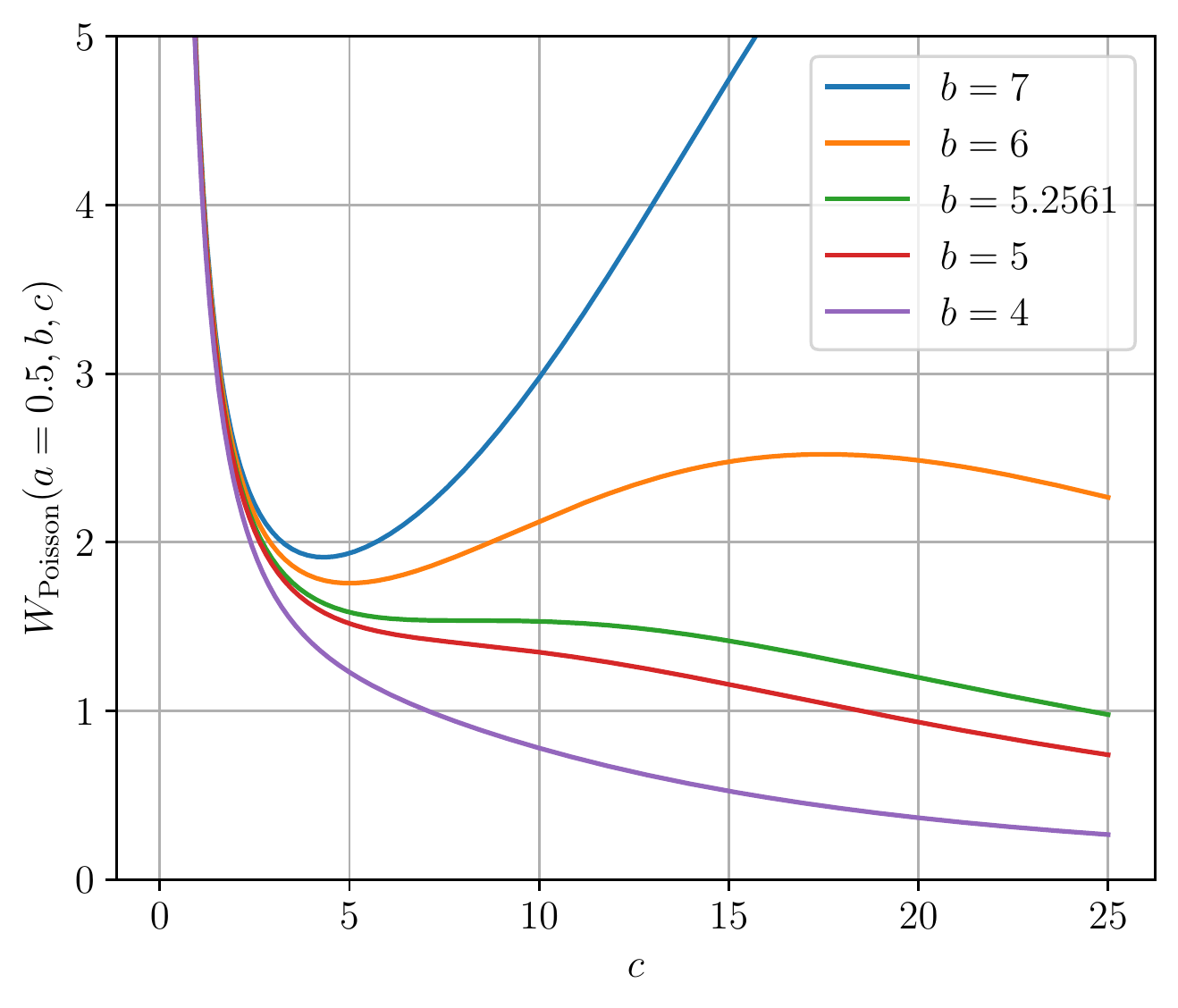}
\caption{The scaled MFPT $W_\text{Poisson}(a,b,c)$ 
as a function of $c$ for fixed $a=1/2$ and different values of $b$. For $b>b_c(a)$, the curve
shows a metastable minimum at $c_1^*(a,b)$, then a maximum at 
$c_2^*(a,b)$ and finally decreases
monotonically to $0$ as $c\to \infty$. 
For $b<b_c(a)$, the scaled MFPT decreaes monotonically with $c$
with the only minimum at $c\to \infty$. 
The critical value is $b_c(a)\approx 5.2561$ for $a=1/2$.} 
\label{fig:tau_poisson}
\end{figure}

\subsection{Periodic Resetting}

In this case, the MFPT is given by the exact formula \eqref{mfpt_periodic.1} that reads
\begin{equation}
\langle t_f\rangle = 
\frac{\int_0^T Q_0(t) \dd{t}}{1- Q_0(T)}\, ,
\label{mfpt_periodic.2}
\end{equation}
where $Q_0(t)$ is given explicitly in \eqref{Q0t_2d.1}. We consider the Gaussian distribution
for the resetting position: ${\cal P}({\vec r}_0)= \e^{- {\vec r}_0^2/{2\sigma^2}}/{2\pi \sigma^2}$
and again define the dimensionless variables
\begin{equation}
\tau_\text{periodic}= \frac{4\, D\, \langle t_f\rangle}{L^2}\, ; \quad\,\,
a=\frac{R_{\rm tol}}{L}<1\, ;
\quad\,\, b= \frac{L}{\sigma}\, ; \quad \,\, c=\frac{L}{\sqrt{4\,D\, T}}\, .
\label{adim_def.2}
\end{equation}

\subsubsection{The limit \texorpdfstring{$\sigma\to 0$}{σ=0}: resetting to the origin}

In this limit,
we have ${\cal P}({\vec r}_0)
\to \delta({\vec r}_0)$, corresponding to resetting to the origin. 
In terms of the dimensionless variables,
we then have from \eqref{mfpt_periodic.2}
\begin{equation}
\tau_\text{periodic}= W_\text{periodic}(a,c)=
\frac{\displaystyle\frac{8\, a^2}{\pi}
\int_0^{\infty} \frac{1-\e^{-x^2/{4\, a^2\,c^2}}}{x^3}\,
\frac{Y_0(x/a)\, J_0(x)-J_0(x/a)\, Y_0(x)}{J_0^2(x)+Y_0^2(x)}
\dd{x} }{\displaystyle 1- 
\frac{2}{\pi}\, \int_0^{\infty} \frac{\e^{-x^2/{4\, a^2\,c^2}}}{x}\,
\frac{Y_0(x/a)\, J_0(x)-J_0(x/a)\, Y_0(x)}{J_0^2(x)+Y_0^2(x)}
\dd{x}}
\label{W_periodic_ac.1}
\end{equation} 

If we plot $W_\text{periodic}(a,c)$ vs. $c$, for fixed $a<1$, we again expect to see a 
unique minimum at $c^*(a)$, as in Fig. (\ref{fig:tau_poisson_origin}) for the Poisonian 
resetting. Unfortunately, numerical integration in \eqref{W_periodic_ac.1} is rather hard as 
the intergrand is oscillatory and the integral is convergent but extremely slowly, in 
particular for large $c$. Indeed, it is easier to extract the behavior of 
$W_\text{periodic}(a,c)$ for small and large $c$ (for fixed $a$) directly from the Laplace 
transform, rather than from \eqref{W_periodic_ac.1}. We note that since $c=L/\sqrt{4DT}$, 
large (respectively small) $c$ corresponds to small (large) $T$. Hence we need to analyse 
$\langle t_f\rangle (T)$ in \eqref{mfpt_periodic.2} for small and large $T$. For this we 
start with the Laplace transform ${\tilde Q_0}(s)$ in \eqref{Q0s_sol.1}. For the resetting 
to the origin, using ${\cal P}({\vec r}_0) \to \delta({\vec r}_0)$, we get
\begin{equation}
{\tilde Q}_0(s)= \int_0^{\infty} Q_0(t)\, \e^{-s\, t} \dd{t}=
 \frac{1}{s}\left[1-
\frac{ K_0\left(\sqrt{\frac{s}{D}}\, L\right)}
{K_0\left( \sqrt{\frac{s}{D}}\, R_{\rm tol}\right)} \right]\, .
\label{Q0s_sol_origin.1}
\end{equation}

\vskip 0.4cm

{\noindent {\bf { Small $T$ behavior of $\langle t_f\rangle$:}}} To extract the small
$t$ asymptotics of $Q_0(t)$, we need to
analyse the Laplace transform ${\tilde Q}_0(s)$ in \eqref{Q0s_sol_origin.1} for large $s$.
Using $K_0(z) \approx \sqrt{\pi/{(2\,z)}}\, \e^{-z}$ for large $z$, we get for large $s$
\begin{equation}
{\tilde Q}_0(s)\approx \frac{1}{s}\left[1- \sqrt{a}\, \e^{- \sqrt{\frac{s}{D}}\, (L-
R_{\rm tol})}\right]\, ,
\label{small_T.1}
\end{equation}
where $a= R_{\rm tol}/L$. The Laplace transform can now be explicitly inverted to give
for small $t$
\begin{equation}
Q_0(t) \approx 1- \sqrt{a}\, \erfc\left(\frac{L-R_{\rm tol}}{\sqrt{4Dt}}\right)\, ,
\label{small_T.2}
\end{equation}
where $\erfc(z)= (2/{\sqrt{\pi}})\, \int_0^{z} \e^{-u^2}\, du$ is the 
complementary error function.
It has the following asymptotic behavior for large argument 
$\erfc(z) \approx \e^{-z^2}/{z\, \sqrt{\pi}}$.
Now, from \eqref{mfpt_periodic.2}, we get for small $T$ (using \eqref{small_T.2} and keeping
only leading order terms for small $T$) 
\begin{equation}
\langle t_f\rangle= 
\frac{\int_0^T Q_0(t) \dd{t}}{1- Q_0(T)}\approx 
\frac{T}{\sqrt{a}}\, \erfc\left(\frac{L-R_{\rm tol}}{\sqrt{4DT}}\right)\, .
\label{small_T.3}
\end{equation}
Finally, using the asymptotics of $\erfc(z)$ for large $z$ in \eqref{small_T.3}, we can
express, after a few steps of straightforward algebra, 
the scaled MFPT $\tau_\text{periodic}$ for small $T$, 
or equivalently for large $c$, in terms of adimensional variables in \eqref{adim_def.2}
\begin{equation}
\tau_\text{periodic}=W_\text{periodic}(a,c)= \frac{4\,D\, \langle t_f\rangle}{L^2}
\approx \frac{\sqrt{\pi}\, (1-a)}{\sqrt{a}\, c}\, 
\e^{(1-a)^2\, c^2}\, ; \quad {\rm as} \quad c\to \infty\, .
\label{large_c_origin.1}
\end{equation}
Thus, $W_\text{periodic}(a,c)$, for fixed $a$, diverges very fast as $c\to \infty$. 
Let us remark that deriving this asymptotic behavior directly from the expression in 
\eqref{W_periodic_ac.1}
is rather hard.

\vskip 0.4cm

{\noindent {\bf { Large $T$ behavior of $\langle t_f\rangle$:}}} To compute the large 
$t$ behavior of
$Q_0(t)$, we need to analyse the small $s$ behavior of 
${\tilde Q}_0(s)$ in \eqref{Q0s_sol_origin.1}.
Fortunately this large $t$ behavior is well known~\cite{BT1993} and one gets
\begin{equation}
Q_0(t) \approx \frac{2 \ln (1/a)}{ \ln \left( \frac{4Dt}{L^2\, a^2}\right)}\, ; \quad 
{\rm for}\quad t\gg L^2\,  
\label{large_t.1}
\end{equation}
where $a= R_{\rm tol}/L<1$. Substituting this result in \eqref{mfpt_periodic.2}, we get
for large $T$, i.e., small $c= L/\sqrt{4DT}$ the following result
\begin{equation}
\langle t_f\rangle \approx 2\, a^2\, \ln(1/a)\, \int_0^{1/{(c\,a)^2}} \frac{\dd{z}}{\ln z}\, .
\label{large_t.2}
\end{equation}
We can easily evaluate the integral for large $c$.
Finally the scaled MFPT for small $c$ is given by (keeping only the leading asymptotic behavior)
\begin{equation}
\tau_\text{periodic}=W_\text{periodic}(a,c)\approx \frac{\ln(1/a)}{c^2 \, \ln (1/{ca})}\, ; 
\quad {\rm as} \quad c\to 0
\label{small_c_origin.1}
\end{equation}

Thus, summarizing for the $\sigma=0$ (or $b\to \infty$) case of the periodic resetting, 
the scaled MFPT, as a
function of $c$ has the following limiting behaviors
\begin{eqnarray}
\tau_\text{periodic}=W_\text{periodic}(a,c) \approx\begin{cases}
& \frac{\ln(1/a)}{c^2 \, \ln (1/{ca})}\, ;
\quad {\rm as} \quad c\to 0 \label{smallc} \\
\\
& \frac{\sqrt{\pi}\, (1-a)}{\sqrt{a}\, c}\, \e^{(1-a)^2\, c^2}\, ; 
\quad {\rm as} \quad c\to \infty \label{largec} 
\end{cases}
\end{eqnarray}
Thus $W_\text{periodic}(a,c)$, as a function of $c$ for fixed $a<1$, diverges as the both ends $c\to 0$ 
and $c\to \infty$. It has a unique minimum at some $c=c^*(a)$. 
As mentioned before, plotting directly the full function $W(a,c)$ as given explicitly in \eqref{W_periodic_ac.1} 
turns out to be hard, due to the oscillatory and exponential nature of the integrands. Using a few numerical
tricks, we managed to plot $W_\text{periodic}(a,c)$ vs. $c$ for different values of $a$ as shown in Fig.~(\ref{fig:tau_per_origin}).

\begin{figure}[H]
\centering
\includegraphics[width = 0.6\linewidth]{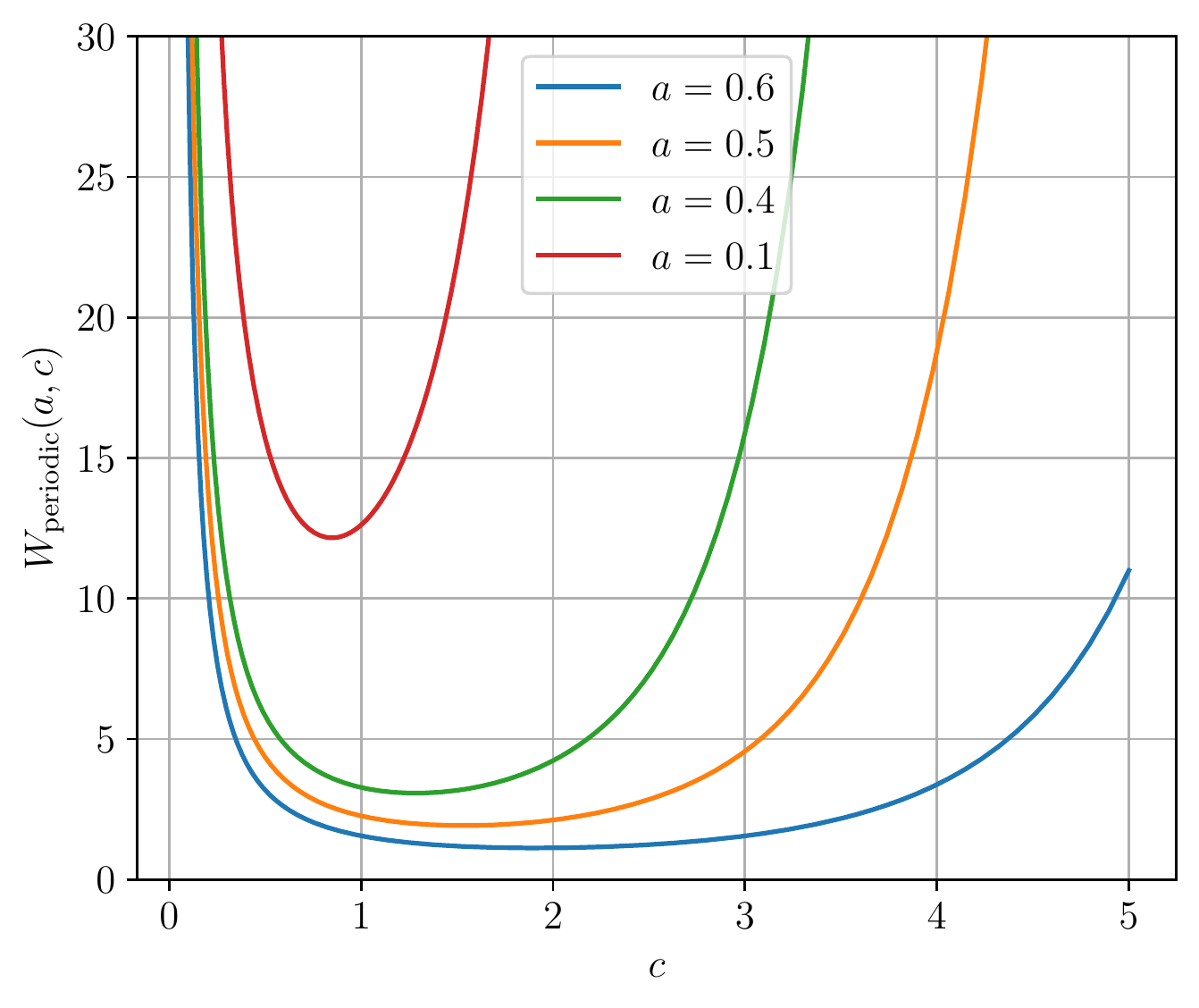}
\caption{The scaled MFPT $W_\text{periodic}(a,c)$ in the limit $b\to \infty$ (or equivalently $\sigma\to 0$ limit)
plotted as a function of $c$ for different values of the parameter $a= R_{\rm tol}/L<1$.
The MFPT exhibits a unique minimum at $c^*(a)$ that depends on $a$.}
\label{fig:tau_per_origin}
\end{figure}

To perform the numerical integration of \eqref{W_periodic_ac.1}, we must extract out the divergences of the integrands in order to improve precision. Indeed, the denominator becomes tends to zero exponentially quicky when $c$ increases, so an accurate integration of the denominator integral is crucial. To do that, we expand the integrand around $x=0$ as
\begin{equation*} f_1 (x) =
   \tfrac{Y_0 (x / a) J_0 (x) - Y_0 (x) J_0 (x / a)}{J_0 (x)^2 + Y_0 (x)^2} 
   \frac{1}{x} \e^{- x^2 / 4 a^2 c^2}
   \underset{x \rightarrow 0}{\simeq}
   \tfrac{2}{\pi} \ln \left( \tfrac{1}{a} \right)  \frac{1}{x}  \frac{1}{1 +
   \frac{4}{\pi^2}  \left( \gamma + \ln \tfrac{x}{2} \right)^2}
   =:
   f_2 (x)
\end{equation*}
Happily, the integral of $f_2$ is analytical. The denominator can be rewritten
\begin{equation*}
  1 - \frac{2}{\pi}  \int_0^{\infty} \dd{x}f_1 (x) = 1 - \frac{2}{\pi}  \int_0^{\infty} \dd{x} (f_1 (x) - f_2 (x)) - 2
  \ln \left( \frac{1}{a} \right)
\end{equation*}
where $\int_0^{\infty} \dd{x} (f_1 (x) - f_2 (x))$ is much easier to compute
numerically. Still, this is not enough, because the integrand now converges slowly when $x \rightarrow \infty$, preventing good precision. Fortunately, because $f_1$ decays exponentially, we can cut off the integral ($x_c =c^2$ seems to be a good cutoff), and because $\int_0^{x_c} \dd{x}f_2 (x)$ is still analytical, we can rewrite the denominator as
\begin{eqnarray*}
  1 - \frac{2}{\pi}  \int_0^{\infty} \dd{x}f_1 (x) & \approx & 1 -
  \frac{2}{\pi}  \int_0^{x_c} \dd{x}f_1 (x)\\
  & = & 1 - \frac{2}{\pi}  \int_0^{x_c} \dd{x} (f_1 (x) - f_2 (x)) -
  \frac{2}{\pi}  \int_0^{x_c} \dd{x}f_2 (x)\\
  & = & 1 - \frac{2}{\pi}  \int_0^{x_c} \dd{x} (f_1 (x) - f_2 (x)) - \ln
  \left( \tfrac{1}{a} \right)  \left( 1 + \tfrac{2}{\pi} \arctan \left(
  \tfrac{2}{\pi}  \left( \gamma + \ln \tfrac{x_c}{2} \right) \right) \right)
\end{eqnarray*}
The remaining integral of $f_1 - f_2$ is easy to compute accurately. The same regularization can be used for the numerator.

\subsubsection{Finite \texorpdfstring{$\sigma>0$}{σ}}

For finite $\sigma$, we have ${\cal P}({\vec r}_0)= \e^{- {\vec r}_0^2/{2\sigma^2}}/{2\pi \sigma^2}$
and our main objective is to evaluate $Q_0(t)$ in \eqref{Q0t_2d.1} and then
use \eqref{mfpt_periodic.2} to compute $\langle t_f\rangle$. Thus the main technical
challenge is to evaluate the integral over ${\vec r}_0$ in \eqref{Q0t_2d.1}. We consider
first the following integral
\begin{equation}
I_1= \int \dd{\vec{r}_0} {\cal P}({\vec r}_0)\, 
Y_0\left(x\, \frac{|{\vec r}_0- L|}{R_{\rm tol}}\right) \, .
\label{I1_def.1}
\end{equation}
We first make the change of variable ${\vec r}_0- {\vec R}_{\rm tar}= \vec R$, denote $L=|{\vec R}_{\rm tar}|$, and 
shift to the polar coordinates. This gives
\begin{eqnarray}
I_1 & = & \frac{1}{2\,\pi\, \sigma^2}\,
\int_{R>R_{\rm tol}}\int_0^{2\pi} \e^{- ({\vec R}_{\rm tar}+{\vec R})^2/
{2\sigma^2}}\, Y_0\left( x\, \frac{R}{R_{\rm tol}}\right)\, R\, \dd{R} \dd{\theta} \nonumber \\
& = & \frac{\e^{- L^2/{2\sigma^2}}}{2\,\pi\, \sigma^2}\, 
\int_{R_{\rm tol}}^{\infty} \dd{R} R\, \e^{-R^2/{2\sigma^2}}
Y_0\left(x\, \frac{R}{R_{\rm tol}}\right)\,
\int_0^{2\pi} \e^{- L\, R\, \cos(\theta)/\sigma^2} \, \dd{\theta} \nonumber \\
&=& \frac{\e^{- L^2/{2\sigma^2}}}{\sigma^2}\,
\int_{R_{\rm tol}}^{\infty} \dd{R} R\, \e^{-R^2/{2\sigma^2}}\, Y_0\left(x\, \frac{R}{R_{\rm tol}}\right)
\,I_0\left(\frac{R\, L}{\sigma^2}\right)\, .
\label{I1.2}
\end{eqnarray}
Similarly, we have
\begin{equation}
I_2= \int \dd{\vec{r}_0} {\cal P}({\vec r}_0)\,
J_0\left(x\, \frac{|{\vec r}_0-\vec{R}_\text{tar}|}{R_{\rm tol}}\right)
= \frac{\e^{- L^2/{2\sigma^2}}}{\sigma^2}\,
\int_{R_{\rm tol}}^{\infty} \dd{R} R\, \e^{-R^2/{2\sigma^2}}\, 
J_0\left(x\, \frac{R}{R_{\rm tol}}\right)\,
I_0\left(\frac{R\, L}{\sigma^2}\right)\, .
\label{I2.2}
\end{equation}
Putting all these results in \eqref{Q0t_2d.1}, and further rescaling $R=\sigma\,z$ we can write
$Q_0(t)$ in terms of dimensionless variables $a= R_{\rm tol}/L$ and $b=L/\sigma$ 
in a somewhat compact form
\begin{equation}
Q_0(t)= \frac{2}{\pi}\, \e^{-b^2/2}\, \int_0^{\infty} \dd{x} \frac{\e^{- D\, t\, x^2/R_{\rm tol}^2}}{x}\, 
\frac{ J_0(x)\,  F_1(x,a,b)- Y_0(x)\, F_2(x,a,b) }{ J_0^2(x)+Y_0^2(x) }\,
\label{Q0t_2d_b.1}
\end{equation}
where we have defined
\begin{eqnarray}
F_1(x,a,b) &=& \int_{a\, b}^{\infty} \dd{z} z\, \e^{-z^2/2}\, Y_0\left(\frac{x\, z}{a\,b}\right)\, I_0(b\, z)
\label{F1_def} \\
F_2(x,a,b) &=&  \int_{a\, b}^{\infty} \dd{z} z\, \e^{-z^2/2}\, J_0\left(\frac{x\, z}{a\,b}\right)\, I_0(b\, z)
\label{F2_def}  
\end{eqnarray} 
Finally, substituting the expression \eqref{Q0t_2d_b.1} of $Q_0(t)$ in \eqref{mfpt_periodic.2} we can obtain
the expression of $\langle t_f\rangle$ and eventually that of the scaled MFPT for finite $\sigma$, i.e.,
for finite $b$, in terms of $a$, $b$ and $c= L/\sqrt{4\, D\, T}$. We get the following 
exact expression
\begin{equation}
\tau_\text{periodic} =  \frac{4\, D\, \langle t_f\rangle}{L^2}
\equiv W_\text{periodic}(a,b,c) =
\frac{\displaystyle\frac{8\, a^2}{\pi} \, \e^{-b^2/2}
\int_0^{\infty} \frac{1-\e^{-x^2/{4\, a^2\,c^2}}}{x^3}\,
\frac{J_0(x)\,F_1(x,a,b)-Y_0(x)\,F_2(x,a,b)}{J_0^2(x)+Y_0^2(x)}
\dd{x} }{\displaystyle 1- 
\frac{2}{\pi}\, \e^{-b^2/2}\, \int_0^{\infty} \frac{\e^{-x^2/{4\, a^2\,c^2}}}{x}\,
\frac{J_0(x)\,F_1(x,a,b)-Y_0(x)\,F_2(x,a,b)}{J_0^2(x)+Y_0^2(x)}
\dd{x}}
\label{tau_b_final.1}
\end{equation}
where $F_1(x,a,b)$ and $F_2(x,a,b)$ are given respectively in \eqref{F1_def} and \eqref{F2_def}.
Note that the result for $b\to \infty$ limit, i.e., $\sigma= 0$ case is given 
in \eqref{W_periodic_ac.1}.
The result in \eqref{tau_b_final.1} is the finite $b$ analogue of \eqref{W_periodic_ac.1}.
Plotting $W_\text{periodic}(a,b,c)$ vs. $c$, for fixed $a<1$ and $b$, 
is challenging, and similar numerical tricks are used to numerically integrate it. In particular, the integral in the denominator can be re-written as
\begin{equation*}
  I_{\text{den}} \simeq \e^{- \frac{b^2}{2}} \int_0^{x_c} \dd{x}
  \int_{ab}^{\infty} \dd{z} z \e^{-\frac{z^2}{2}} I_0 (bz) (f_1 - f_2) + \Big( \tfrac{\pi}{2} + \arctan \big( \tfrac{2}{\pi}  \left( \gamma + \ln \tfrac{x_c}{2} \right) \big) \Big) A_1 (b, z)
\end{equation*}
where $A_1$ is defined bellow, $f_1$ and $f_2$ are as before but with $\ln(z/ab)$ in place of $\ln(1/a)$, and where the remaining double integral is relatively easy to compute accurately. To test the integration, we compared the results to the asymptotic behaviors in Eq. \eqref{largec.1}. The results are plotted in Fig.~(\ref{fig:tau_per}).

\begin{figure}[H]
\centering
\includegraphics[width = 0.6\linewidth]{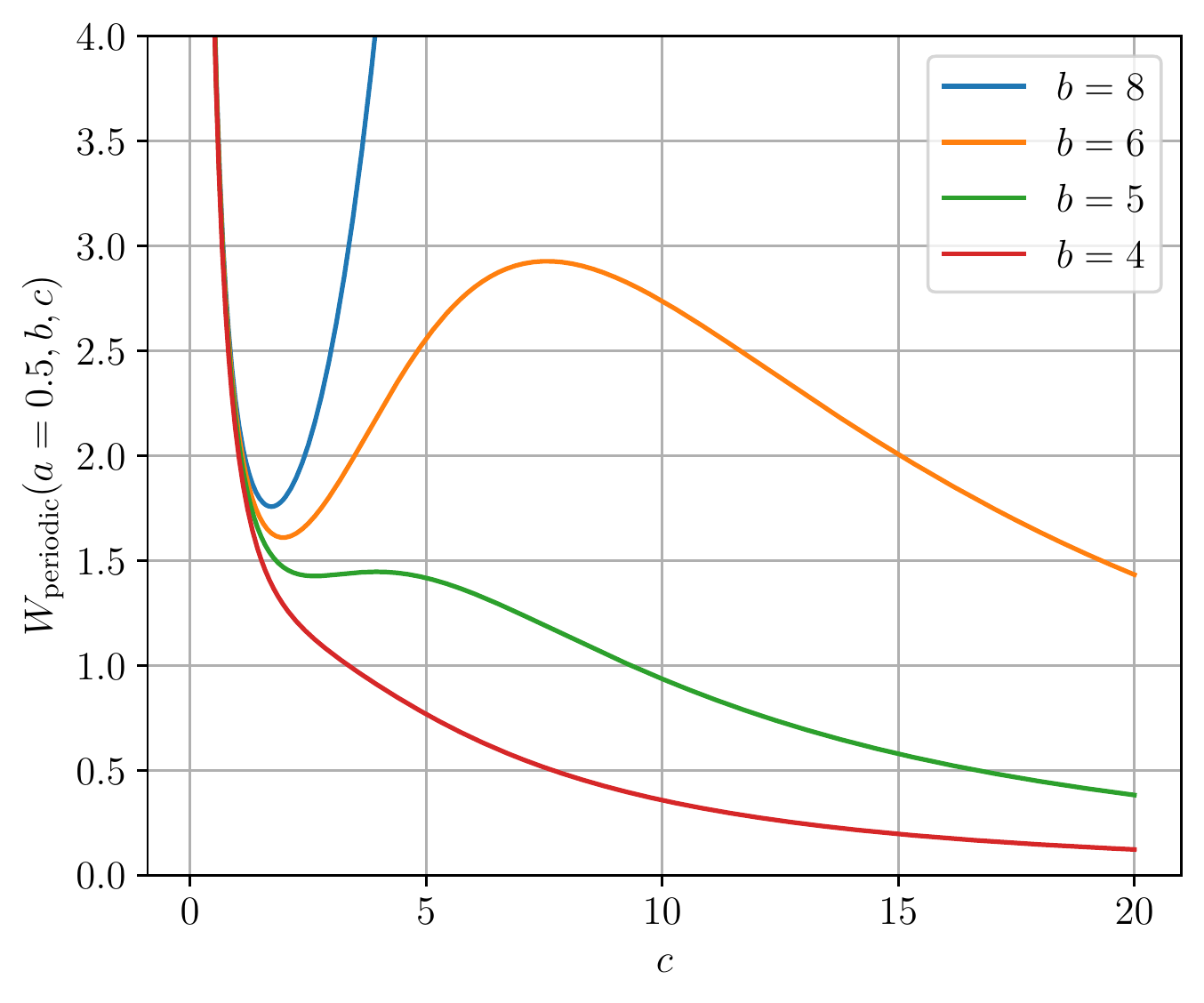}
\caption{The scaled MFPT $W_{\rm per}(a,b,c)$ 
as a function of $c$ for fixed $a=1/2$ and different values of $b$. For $b>b_c(a)$, the curve
shows a metastable minimum at $c_1^*(a,b)$, then a maximum at 
$c_2^*(a,b)$ and finally decreases
monotonically to $0$ as $c\to \infty$. 
For $b<b_c(a)$, the scaled MFPT decreaes monotonically with $c$
with the only minimum at $c\to \infty$. 
The critical value is $b_c(a)\approx 5$ for $a=1/2$.} 
\label{fig:tau_per}
\end{figure}

As in the Poissonian case, if we plot $W_\text{periodic}(a,b,c)$ vs. $c$, for fixed $a<1$ and $b$, we expect to see a
minimum at $c_1^*(a,b)$, followed by a maximum at $c_2^*(a,b)>c_1^*(a,b)$ and
then an absolute minimum at $c\to \infty$. The metstable minimum $c_1^*(a,b)$ disappears
below a critical value $b<b_c(a)$, as in the Poissonian case. 
Thus, for any finite $b$, the function $W_\text{periodic}(a,b,c)$ vs $c$ 
finally decays to $0$ as $c\to \infty$.
This is in stark contrast to
the $b\to \infty$ case, where we recall that  
$W_\text{periodic}(a,\infty,c)=W_\text{periodic}(a,c)$ diverges as $c\to \infty$ (see the
\eqref{largec}). It would be nice to see this decay at large $c$ by plotting
$W_\text{periodic}(a,b,c)$ vs. $c$ in \eqref{tau_b_final.1}. However, as mentioned earlier,
plotting this function is hard since the numerical integration in \eqref{tau_b_final.1} converges
very slowly, in particular for large $c$.

Indeed, it is easier to extract the behavior of $W_\text{periodic}(a,b,c)$ for small and large $c$
(with fixed $a$ and $b$) directly from the Laplace transform, rather than from \eqref{tau_b_final.1}.
We note that since $c=L/\sqrt{4DT}$, large (respectively small) $c$ corresponds to small (large) $T$.
Hence we need to analyse $\langle t_f\rangle (T)$ in \eqref{mfpt_periodic.2} for small and large $T$.
For this we start with the Laplace transform
${\tilde Q_0}(s)$ in \eqref{Q0s_sol.1} and analyse its small $s$ and large $s$ behaviors, as we have done
before in the $\sigma=0$ (or $b\to \infty$ limit) case. Skipping details, we just write the final results
\begin{eqnarray}
W_\text{periodic}(a,b,c) \approx \begin{cases}
& \frac{A_1(a,b)}{c^2 \, \ln \left(\frac{1}{c\,a}\right)} \quad\,\, {\rm as}\quad c\to 0 
\\
\\
& \frac{g(a,b)}{1-g(a,b)}\,\frac{1}{c^2}
\quad\,\, {\rm as}\quad c\to \infty \label{largec.1}
\end{cases}
\end{eqnarray}
where $I_0(z)$ is the modified Bessel function of the first kind and 
the amplitude $A_1(a,b)$ already appeared in \eqref{A1_def.1} and reads
\begin{equation}
A_1(a,b)= \e^{-b^2/2}\, \int_{a\, b}^{\infty} \dd{z} z\, \e^{-z^2/2}\, \ln\left(\frac{z}{a\, b}\right)\, 
I_0(b\, z) \, .
\label{A2_def.1}
\end{equation}
Thus, for fixed $b$ finite and fixed $a<1$, the scaled MFPT $W_\text{periodic}(a,b,c)$ decays
for large $c$ as $c\to \infty$, in contrast to the $b\to \infty$ case where it diverges as $c\to \infty$
(as in \eqref{largec}). Thus the true optimal value of $c$ occurs at $c\to \infty$, even though there
is a metastable minimum at an earlier value $c_1^*(a,b)$ for $b>b_c(a)$.

\section{Anisotropy of the initial position distribution} \label{annex:aniso}

To quantify the effect of initial position anisotropy, which is unfortunately important with our experimental setup, we performed simulations with an non-symetric gaussian initial position distribution where $\sigma_x \neq \sigma_y$. Fig.~\ref{fig:2d-aniso-effect} shows that for $a=0.5$, the mismatch induced is about $\pm 15\%$ for a $20\%$ anisotropy. Moreover, it is about $\pm 10\%$ and $\pm 20\%$ for $a=0.9$ and $a=0.3$ respectively. As expected, the MFPT is decreased when $\sigma_x > \sigma_y$. The effect is more pronounced at high $c$, i.e.~for frequent resetting, especially when $a$ approaches $1$. Many more figures at available at \url{https://github.com/xif-fr/BrownianMotion/tree/master/langevin-mfpt}.

\begin{figure}[H]
	\centering
	\includegraphics[width=1\textwidth]{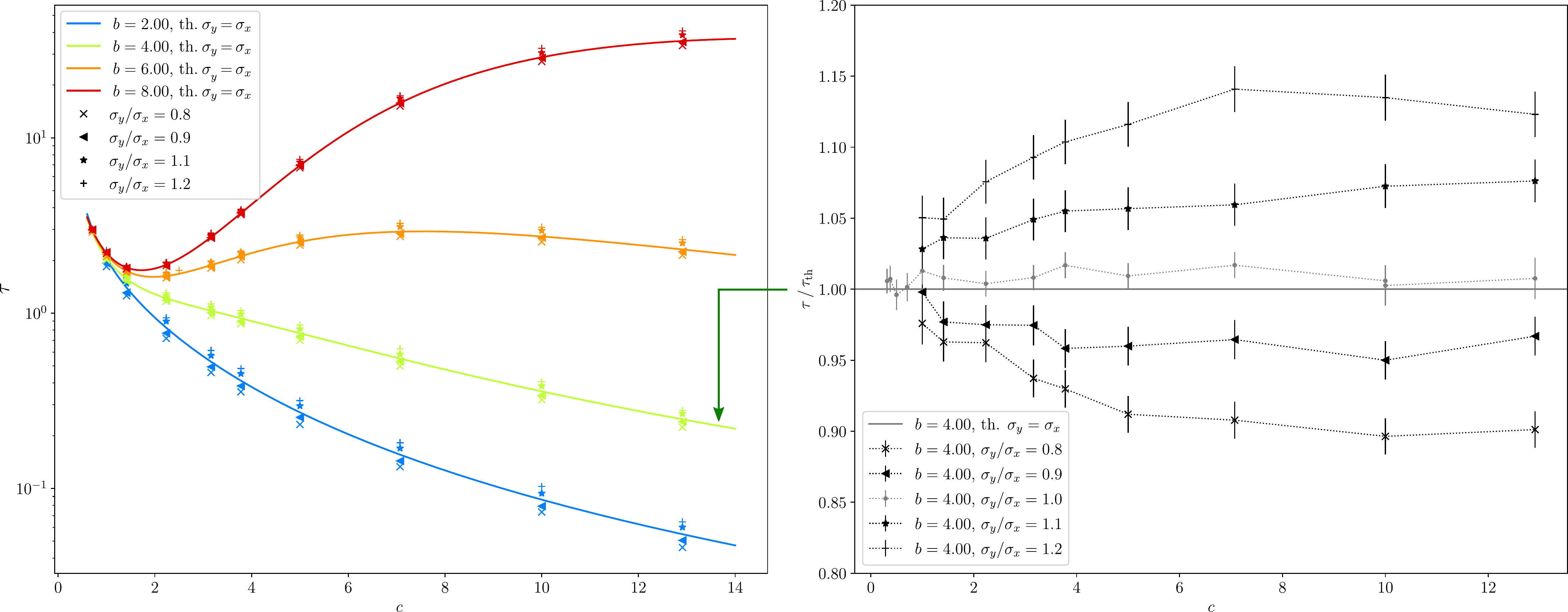}
	\caption{Simulation. Effect of anisotropy of the initial position distribution on the MFPT. Example with $a=0.5$ and periodic resetting, where $\sigma_y/\sigma_x$ is varied from 0.8 to 1.2, and $b$ from 8 to 2. Left : $\tau$ as a function of $c$ for several anisotropies (points) againt the theoretical MFPT (solid lines). Right : deviation (points) relative to theory (solid line) for several anisotropies.}
	\label{fig:2d-aniso-effect}
\end{figure}

Fig.~\ref{fig:2d_expe_period_alt} represents the 2D experimental data with periodic resetting against the theory using $$\sigma=\sigma_\text{mean}=\sqrt{\frac{\sigma_{x,\text{exp}}^2+\sigma_{y,\text{exp}}^2}{2}}\,.$$
The mismatch is more pronounced than with $\sigma=\sigma_{x,\text{exp}}$, confirming the choice made in Section \ref{sec:res-exp2d}.

\begin{figure}[H]
	\centering
	\includegraphics[width=0.8\columnwidth]{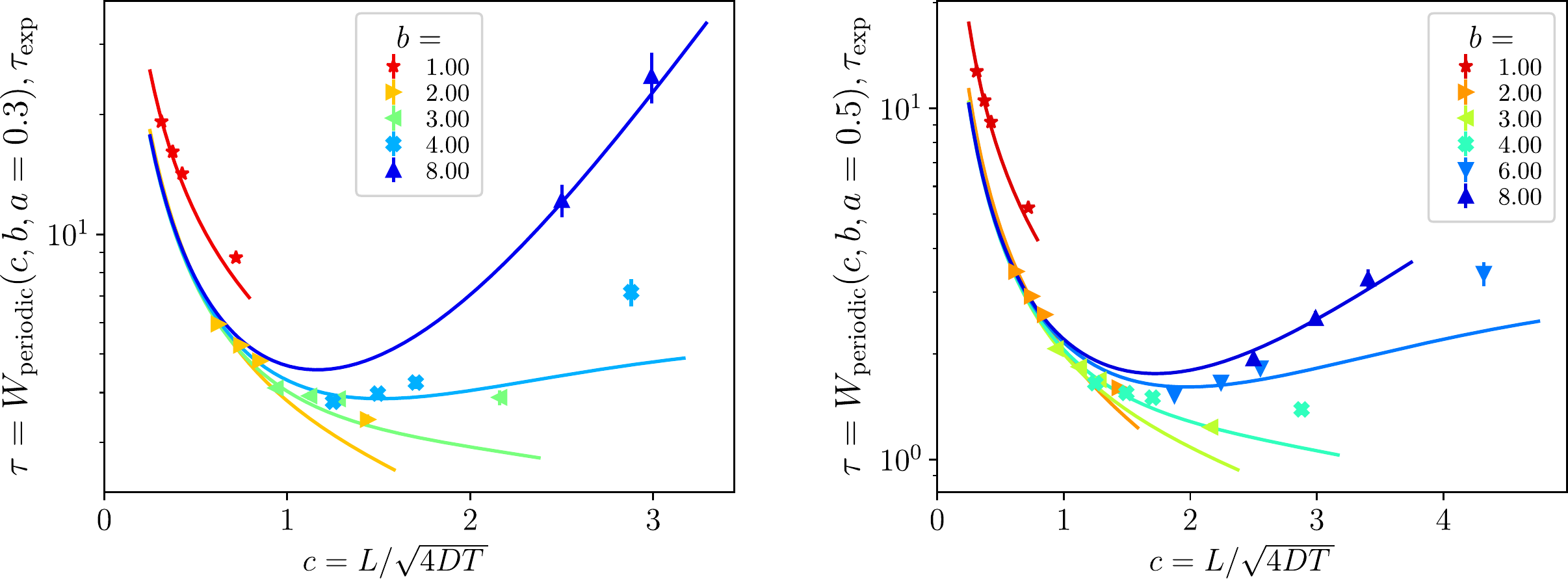}
	\caption{Experiment. $\tau$ versus $c$ for the periodic resetting protocol in 2D for finite $b$ and $a=0.3$ (top panel) and $a=0.5$ (bottom panel). Continuous lines are the the theoretical predictions of Eq.~\ref{tau_b_final.1} with $\sigma=\sigma_\text{mean}$ and the symbols corresponds to the results obtained from the experimental data.}
	\label{fig:2d_expe_period_alt}
\end{figure}

\section{Numerical integration of the Langevin equation} \label{apx:langevin}

To measure numerically the MFPT, we integrate the Langevin equation :
\begin{equation}
\dot x(t) = \sqrt{2\, D}\, \eta_x(t),\quad
\dot y(t) = \sqrt{2\, D}\, \eta_y(t) 
\end{equation}
where $D$ is the diffusion constant (which we set equal to 1) and $\eta_m$ ($m=x,y$) are delta correlated noises with $\langle \eta_m(t)\eta_m(t') \rangle = \delta(t-t')$ and $\langle \eta_m(t)\eta_{m'}(t) \rangle = 0$ for $m\ne m'$. The simulation starts with $x(0),y(0)$ randomly drawn from a Gaussian distribution of standard deviation $\sigma$. The particle freely diffuses till a time $T$ which can be either constant (periodic reset) or randomly distributed (poissonian reset), and $x,y$  are reset at a new random position, drawn from the same Gaussian distribution. The cycle is repeated utill the target is reached (a line at $x=L$ in 1 dimension, and a circle of radius $R_\text{tol}$ centered at $x=L$ in 2 dimensions). The first passage time $t_\text{f}$ is then recorded. The integration step is fixed at a small enough value so as the systematic error introduced by the time discretization on $\langle t_\text{f} \rangle$ is typically smaller or equal to 1\%. To compute the MFPT $\langle t_\text{f} \rangle$, between $N=20000$ and 200000 trajectories (representing about $2\,\mathrm{hour}\cdot\mathrm{core}$ for a typical 2015 CPU) are computed for each $(a,b,c)$. Multiple values of $(L,R_\text{tol},\sigma,T|\alpha)$ are possible for each $(a,b,c)$, and we checked that the resulting adimentionalized MFPT $\tau$ does not depend on these values, but only on $(a,b,c)$ (withing a 5\% error margin maximum, typically 1\%). The uncertainty on the MFPT ($\operatorname{std}(t_\text{f})/\sqrt{N}$) is typically $<1\%$, which is too small for the error bars in figures \ref{fig:b_2d_numerical_period},\ref{fig:b_infty_2d_numerical_pois},\ref{fig:b_2d_numerical_pois} to be visible.

The code is available at \url{https://github.com/xif-fr/BrownianMotion/}.

\end{widetext}
\end{document}